%% file: main.tex
\documentclass[11pt]{article}
\usepackage[letterpaper,margin=0.9in]{geometry}
\usepackage[T1]{fontenc}
\usepackage{lmodern,microtype}
\usepackage{amsmath,amssymb,graphicx,array,booktabs,tabularx,longtable}
\usepackage[numbers,sort&compress]{natbib}
\usepackage[font=small,labelfont=bf]{caption}
\usepackage{float,placeins,lineno,xcolor}
\usepackage{etoolbox}
\usepackage{listings}
\lstdefinestyle{yaml}{basicstyle=\ttfamily\scriptsize,breaklines=true,breakatwhitespace=true,columns=fullflexible,keepspaces=true,frame=single,framerule=0.4pt,rulecolor=\color{gray!60},xleftmargin=2pt,xrightmargin=2pt,literate={—}{{\textemdash}}1}

\usepackage{xcolor} 
\usepackage[normalem]{ulem} 

\definecolor{yubocolor}{HTML}{CC79A7}

\definecolor{joshcolor}{RGB}{0, 0, 255}      
\definecolor{cerencolor}{RGB}{255, 0, 0}     
\definecolor{ericcolor}{RGB}{0, 128, 0}      
\definecolor{lauracolor}{RGB}{148, 0, 211}   

\AtBeginEnvironment{table}{\centering\footnotesize\setlength{\tabcolsep}{3pt}}

\usepackage[hidelinks]{hyperref}
\usepackage{doi}
\hypersetup{pdftitle={The As-If Machine: Personalized AI counterfactuals make distant policy risks feel close},pdfauthor={Joshua Ashkinaze, Yubo Zhou, Pengyu Chen, Avi Bhattacharya, Hyebin Park, Yanman Zhang, Stephanie Preston, Eric Gilbert, Ceren Budak},bookmarksdepth=2}
\graphicspath{{static_figures/}{explain_divergence/plots/}{study_comparison/plots/}{sci_fund_real/plots/}{satellite/plots/}}

\input{study_comparison/tables/results.tex}
\title{The As-If Machine: Personalized AI counterfactuals make distant policy risks feel close}
\author{%
\begin{tabular}{c}
Joshua Ashkinaze, Yubo Zhou\textsuperscript{*}, Pengyu Chen\textsuperscript{*},\\
Avi Bhattacharya\textsuperscript{*}, Hyebin Park\textsuperscript{\dag}, Yanman Zhang\textsuperscript{\dag},\\
Stephanie Preston, Eric Gilbert, Ceren Budak\\[0.5em]
\small University of Michigan\\[0.3em]
\footnotesize\textsuperscript{*}These authors contributed equally. \textsuperscript{\dag}These authors contributed equally.
\end{tabular}}
\date{}
\begin{document}
\raggedbottom
\maketitle

\begin{abstract}
Life experiences can shape policy attitudes, but we only live one life. Citizens may therefore fail to support policies addressing risks they have never personally experienced. Narratives can make such risks immediate and personal, but personalized stories remain difficult to produce at scale. Here we ask whether AI can make distant scenarios feel concrete by simulating a participant's life as if that scenario occurred. We present the As-If Machine (AIM), a multi-agent system through which people co-write personalized counterfactual narratives—stories about their own lives under different policy futures. In two pre-registered and quota-matched experiments on science funding ($N=1{,}007$) and satellite cybersecurity ($N=894$), we compared AIM with two controls: reading an expert-written vignette on the same topic and co-writing with AIM about an unrelated topic. AIM increased psychological closeness—how close a risk feels to the self—by nearly identical margins in the two domains ($0.36$ and $0.38$ standard deviations (SDs) against the vignette, $0.80$ and $0.90$ against unrelated writing). AIM increased psychological closeness even among participants with unfavorable views of AI. For policy attitudes, AIM's effects were $0.16$ SDs against the vignette and $0.10$ SDs (n.s.) against unrelated writing for science funding; the corresponding effects for satellite cybersecurity were $0.54$ and $0.67$ SDs. Across $23{,}496$ human–AI autocomplete interactions, participants and the system played distinct roles: AIM supplied concrete detail that grounded the scenario, and participants supplied first-person perspective and emotion. Co-writing with the As-If Machine makes distant policy risks feel close.
\end{abstract}

\section{Introduction}

Life experiences can shape attitudes, behavior, and policy support~\cite{egan_turning_2012,naumann_how_2016,pud_personal_2004,weinstein_effects_1989}. People who experience extreme weather, for example, report stronger belief in climate change~\cite{egan_turning_2012}. But we only live one life. We cannot personally experience every consequence that should inform our attitudes. A patient may never see how decreased funding for scientific training grants affects specialist availability. A commuter may never realize how much navigation and communication systems depend on satellites until those systems fail. By the time such consequences become direct experience, the opportunity to prevent them may have passed.

 Unexperienced consequences can remain ``psychologically distant''---removed from the self in time, space, social connection, or likelihood~\cite{liberman_traversing_2014,williams_psychological_2014,baltatescu_psychological_2016,trope_construal_2007,liberman_construal-level_2008,trope_construal-level_2010}. Distant events tend to stay abstract, whereas closer events become concrete~\citep{soderberg_effects_2015}. Personal experience can reduce psychological distance. However, many important risks cannot be experienced in advance. Could co-writing with artificial intelligence (AI) approximate what experience provides---a concrete sense of what an unfamiliar future would mean for a person's own life?

Existing approaches face a tradeoff. Educational interventions explain why a risk matters; perspective-taking asks people to consider unfamiliar experiences; guided simulation asks people to imagine possible futures; and narratives can make imagined events resemble direct experience while reducing argument against the message~\cite{pham_thought_1999,lee_using_2020,green_interactive_2014,green_need_2020,green_transportation_2004,green_transportation_2012}. But scalable interventions generally present everyone with similar material, while tailored interventions are generally difficult to scale. A fixed story can show what \textit{someone} might experience. Yet it cannot show how the event would unfold in \textit{your} job, \textit{your} family, or \textit{your }commute.

We developed the As-If Machine (AIM) to overcome this tradeoff. AIM is a multi-agent system through which people co-write personalized counterfactual narratives---stories about their own lives under possible policy futures. We designed AIM around principles that support ``narrative transportation'' (immersion in a story)~\citep{green_transportation_2021}. These principles include: personal relevance, plausible circumstances, mental imagery, and coherent causal sequences. AIM uses information about a participant's circumstances to generate both how the event might begin for a person and how its consequences might unfold in that person's life. These outputs provide context that feeds into a large language model (LLM) which offers short continuations as the participant writes. Participants contribute their own text and can accept or reject each AI suggestion. Personalization therefore continues through the writing itself, since participants bring their own responses to the imagined situation. Most AIM participants (75\%) reported that these suggestions were helpful for continuing the story. Our linguistic analysis reveals distinct roles: AIM supplies causal structure and concrete detail, while participants supply first-person perspective and emotion (\autoref{sec:linguistic-results}).

Making a consequence feel close, however, is not the same as changing what someone thinks should be done about it. Experimental attempts to reduce psychological distance have produced inconsistent downstream effects on attitudes and behavior~\citep{van_valkengoed_psychological_2023,huang_disentangling_2024,langlais_does_2026}. We therefore separate two questions. Can personalized simulation make an unfamiliar consequence feel concrete and personally relevant? And when does that change extend to policy attitudes?

We evaluated AIM in two pre-registered randomized experiments on science funding and satellite cybersecurity. In each study, we compared AIM with (1) an expert-written narrative on the same issue and (2) the same interactive writing system applied to an unrelated topic. These comparisons allow us to benchmark AIM's effects against both conventional narrative persuasion and interactive writing in general. We measured psychological closeness and policy attitudes, and analyzed effects by prior issue familiarity.

AIM increased psychological closeness in both studies by nearly identical margins. Its effects on policy attitudes diverged: attitudes changed substantially for satellite cybersecurity, weakly for science funding, and more among participants with less prior familiarity. Across 23,496 human--AI autocomplete interactions, AIM supplied concrete detail while participants supplied perspective and emotion. The results show that AI can help people construct experiences they have never lived and reliably make distant risks feel closer. Whether that closeness changes what people support depends on the domain.

 \begin{figure}[h]
\centering
\includegraphics[width=\linewidth]{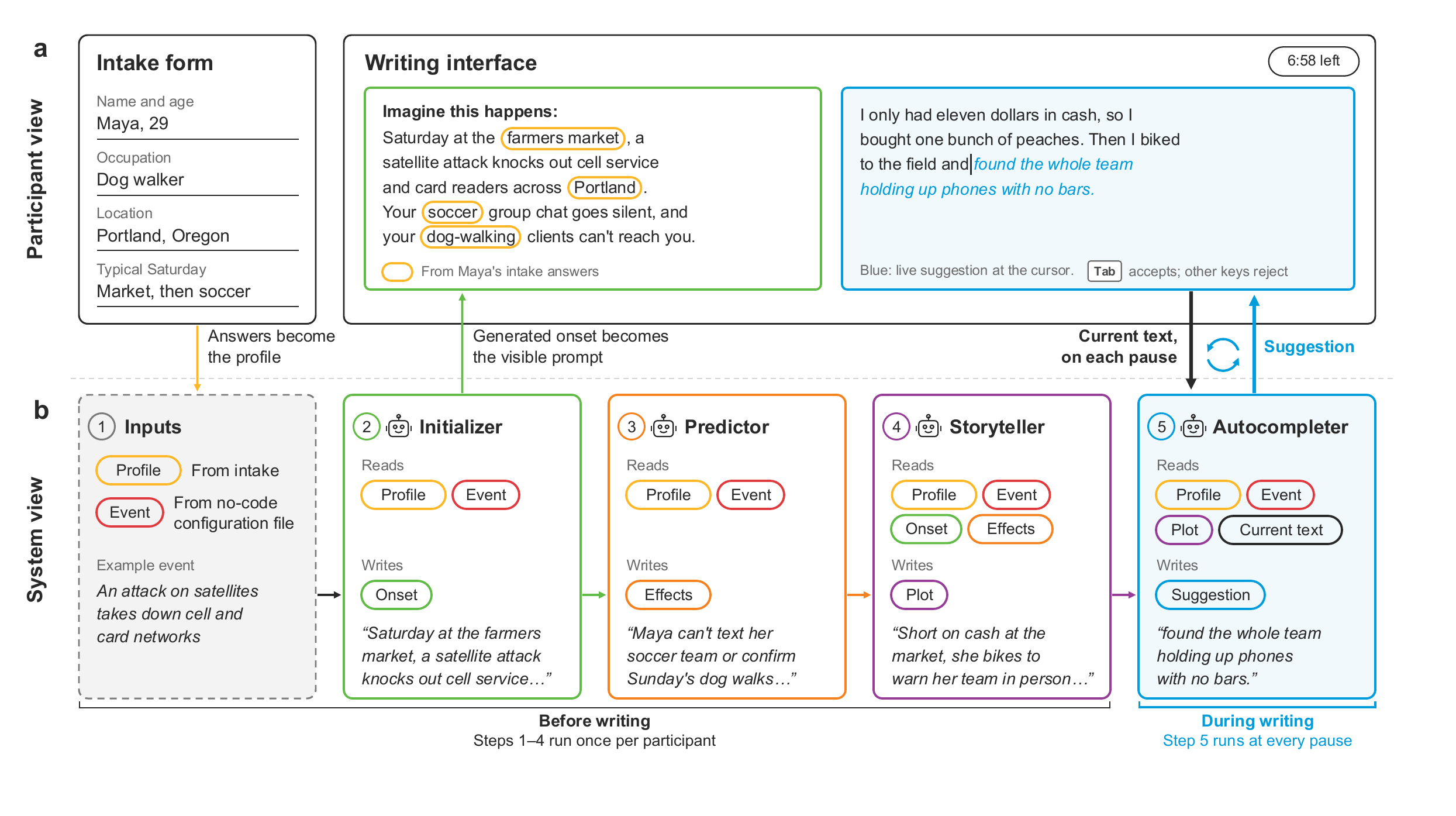}
\caption{\textbf{Panel a}, \textbf{Participant view:} Participants first complete an intake form. They then continue a scenario prompt that the system built from their answers, shown by orange outlines. Whenever they pause, the system suggests a continuation, which they can accept by pressing Tab.
\textbf{Panel b}, \textbf{System view:} Four steps run once before writing begins. The system turns the intake answers into a text profile (1). The Initializer writes a personalized event onset, which becomes the visible prompt for the participant (2). The Predictor forecasts how the event would affect the participant (3). The Storyteller combines the onset and effects into a plot (4). During writing, the Autocompleter reads the profile, event, plot, and current text and offers continuations (5).}
\label{fig:asif_diagram}
\end{figure}

\section{System description: An open-source engine of interventions}
\label{sec:system-description}
\label{sec:aim_system}
\label{sec:aim_architecture}

AIM is a web application with four specialized LLM agents (Figure~\ref{fig:asif_diagram}). The Initializer generates a plausible scenario onset from the participant's background and circumstances. The Predictor forecasts its consequences for that participant, and the Storyteller connects these events into a narrative arc. The Autocompleter offers short suggestions as the participant writes, maintaining continuity with the scenario, and participants can accept, reject, or modify them.

The architecture embodies principles of narrative transportation~\cite{green_transportation_2021}, such as connecting the story to the reader's life, beginning plausibly, and unfolding through a coherent causal sequence (Extended Data Table~\ref{best_practices} maps the principles onto AIM).  By separating these functions across agents, each stage produces reusable output. The Initializer’s personalized onset can serve as the participant’s starting prompt.  The Predictor’s consequences combine with that Initializer's onset to guide the Writer’s plot---which then provides context for the Autocompleter. Based on informal user testing, we also constrained the Autocompleter from attributing thoughts or emotions to participants (Methods, \autoref{sec:methods-formative-development}).

\label{sec:intervention_engine}
We designed AIM as reusable research infrastructure\footnote{\url{https://github.com/josh-ashkinaze/asif-machine}}. Researchers can vary the intervention while retaining the four-agent architecture, co-writing interface, and experimental workflow. For each experiment, a plain-text YAML file specifies the scenario that guides the agents, intake questions for personalization, deployment parameters, models, and agent instructions. From this configuration, AIM automatically renders the appropriate participant-facing web pages and creates a dedicated experiment URL. Extended Data Figure~\ref{fig:yaml_config} shows a complete configuration for a scenario in which gravity stops working, and Extended Data Figure~\ref{fig:asif_screens} shows the resulting interface. SI~\autoref{sec:si-event-configs} gives the event configurations for both experiments. Researchers can distribute the URL directly to participants or incorporate it into study workflows using platforms such as Qualtrics.

\begin{figure}[tbp]
\centering
\includegraphics[width=\linewidth]{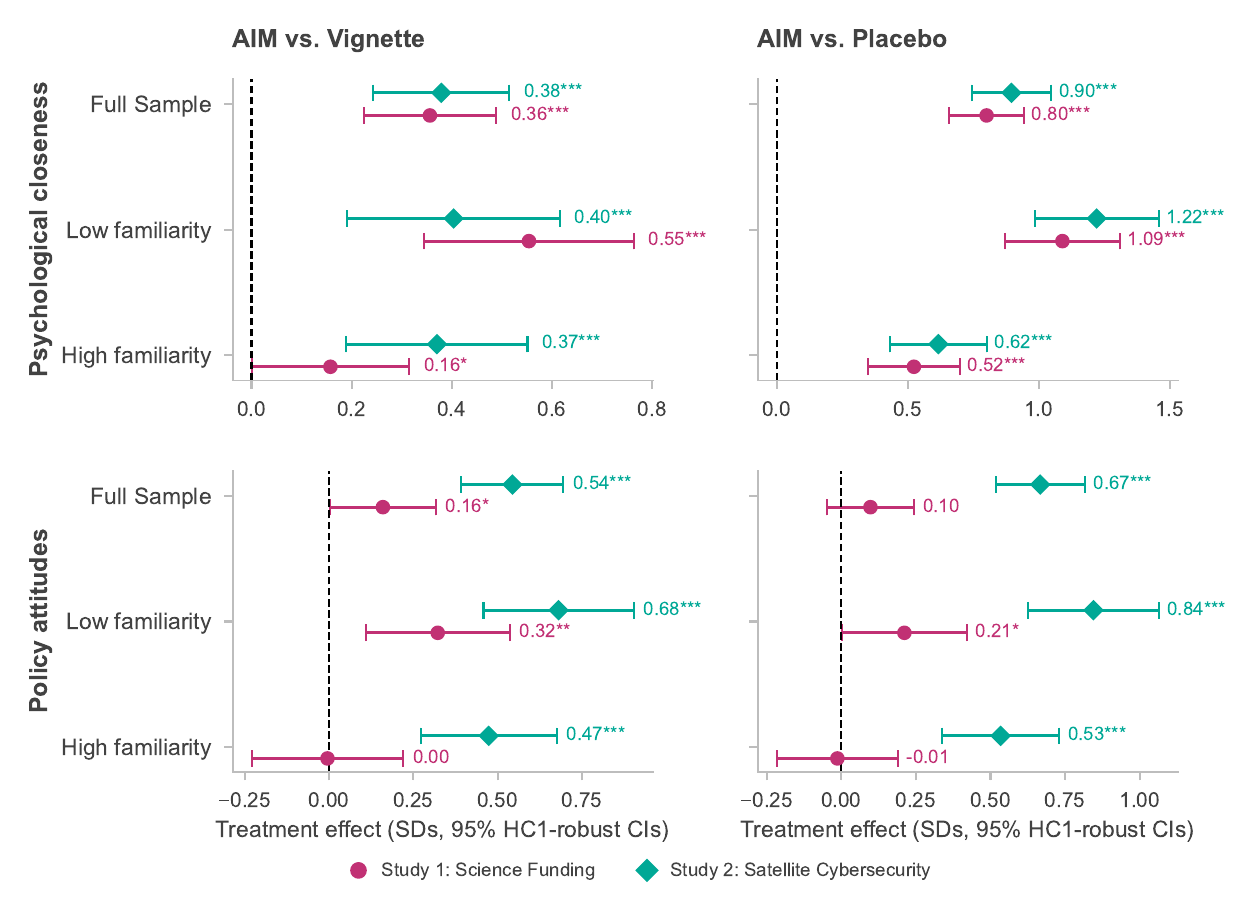}
\caption{Treatment effects by study and familiarity group. Points show AIM-minus-control contrasts in standard deviations with $95\%$ confidence intervals. Low familiarity is self-reported familiarity $\leq2$. Stars from multiple comparison corrected sharpened FDR q-values~\cite{anderson_multiple_2008}, treating each study as a family of hypotheses: * $q\leq.05$, ** $q\leq.01$, *** $q\leq.001$.}
\label{fig:forest_2}
\end{figure}

\section{Study 1: Science funding}
\label{sec:study-1-science-funding}
\label{sec:aim_experiment}

We conducted a pre-registered three-arm experiment on science funding. Most Americans support science~\cite{kennedy_americans_2026} in the abstract, but the consequences of funding cuts are not immediately visible. We focused on a concrete scenario: a participant's specialist appointment was 14 months away because of reductions in National Institutes of Health (NIH) T32 training grants, which train specialists.

We recruited adults through Prolific and quota-matched the sample on age, political party, and sex. We randomly assigned participants to AIM ($n=333$), an expert vignette ($n=331$), or an interactive placebo ($n=343$), yielding a final sample of $N=1{,}007$ (see SI \autoref{sec:si-descriptives} for descriptives by arm). AIM participants spent approximately six minutes co-writing a personalized account of delayed care. Vignette participants read a domain-relevant story about a girl whose cancer was cured through CAR-T immunotherapy, designed to encourage support for science and used in prior research~\cite{ophir_effects_2021}. Placebo participants used AIM to write about meeting a celebrity or sounds turning into colors. The vignette controls for demand effects, and the placebo preserves the co-writing format without policy content.

As pre-registered, we measured psychological closeness and policy attitudes as z-scored indices of two items each (Methods). The main pre-registered analyses shown in \autoref{fig:forest_2} use HC1 robust standard errors and correct for multiple comparisons within each study with sharpened FDR $q$ values~\cite{anderson_multiple_2008}.

We pre-registered four hypotheses and accompanying analyses: (H1) AIM would increase psychological closeness relative to both controls; (H2) AIM would increase support for science funding relative to both controls; (H3) closeness would mediate AIM's effect on attitudes; and (H4) AIM's effects would be larger among participants with lower prior familiarity (thresholds of $\leq 2$ and $\leq 3$ on the seven-point scale).

\subsection{Main effects}
\label{sec:study-1-science-funding-main-effects}

\paragraph{Closeness.} AIM increased psychological closeness relative to both controls (Figure~\ref{fig:forest_2}). Relative to the vignette, the difference was $\beta=0.36$ SDs, 95\% CI $[0.22,0.49]$, $t=5.27$, sharpened $q\leq.001$; relative to the placebo, it was $\beta=0.80$ SDs, 95\% CI $[0.66,0.94]$, $t=10.96$, sharpened $q\leq.001$.

\paragraph{Attitudes.} Attitude effects were smaller. Relative to the vignette, the difference was $\beta=0.16$ SDs, 95\% CI $[0.00,0.32]$, $t=2.00$, sharpened $q\leq.034$; relative to the placebo, it was $\beta=0.10$ SDs, 95\% CI $[-0.05,0.24]$, $t=1.31$, sharpened $q\leq.071$. AIM increased policy support relative to the vignette; the placebo contrast was in the hypothesized direction but not significant. Extended Data Tables~\ref{tab:main_effects_desc} and~\ref{tab:primary} report item-level descriptive statistics and regression estimates.

\subsection{Mediation}
\label{sec:study-1-science-funding-mediation}

The pre-registered mediation analysis yielded positive indirect effects of AIM on attitudes through closeness, against both the vignette ($b=0.17$, 95\% CI $[0.10,0.25]$) and the placebo ($b=0.33$, 95\% CI $[0.25,0.42]$). Because we did not randomize closeness, these estimates do not identify causal mediation (Extended Data Table~\ref{tab:mediation}). However, we tested the indirect effects' sensitivity to an unmeasured factor affecting both closeness and attitudes~\citep{imai_identification_2010}. Both remained positive unless such a confounder induced a correlation of approximately $\rho^{\star}=0.40$ between the unexplained components of closeness and attitudes (Supplementary Fig.~\ref{fig:mediation-sensitivity}).

\subsection{Familiarity}
\label{sec:study-1-science-funding-familiarity}
Across all four pre-registered contrasts (vignette and placebo, attitudes and closeness), AIM produced larger estimated effects among participants less familiar with science funding (Figure~\ref{fig:forest_2}). At the pre-registered cutoff of 2 of 7 (on a question asking about self-reported familiarity with science funding), 488 participants were in the low-familiarity group and 519 were in the high-familiarity group. Among low-familiarity participants, AIM increased closeness relative to the vignette ($\beta=0.55$ SDs, $t=5.20$, sharpened $q\leq.001$) and placebo ($\beta=1.09$ SDs, $t=9.76$, sharpened $q\leq.001$), and attitudes relative to the vignette ($\beta=0.32$ SDs, $t=2.96$, sharpened $q=.004$) and placebo ($\beta=0.21$ SDs, $t=1.99$, sharpened $q=.034$). Among higher-familiarity participants, closeness effects were smaller ($\beta=0.16$ SDs against the vignette and $0.52$ against the placebo) and attitude effects were near zero ($\beta=-0.004$ and $-0.01$ SDs, sharpened $q=.298$). A cutoff of 3 produced a similar but weaker pattern (Extended Data Tables~\ref{tab:familiarity2} and~\ref{tab:familiarity3}).

\subsection{Heterogeneous treatment effects}
\label{sec:study-1-science-funding-heterogeneous-treatment-effects}

To examine familiarity continuously and explore other moderators, we fit pre-registered causal random forests~\citep{wager_estimation_2018,athey_estimating_2019} and summarized each conditional average treatment effect (CATE) function by its best linear projection (BLP) onto the moderator set~\citep{semenova_debiased_2021}. Compared to the placebo, higher familiarity ($B=-0.36$, $p<.001$, $q\leq.001$), older age ($B=-0.21$, $p=.010$, $q=.084$), and greater personal distress ($B=-0.18$, $p=.018$, $q=.099$) predicted smaller closeness effects; older age also predicted a smaller attitude effect ($B=-0.18$, $p=.039$, $q=.913$). Compared to the vignette, higher income predicted a smaller closeness effect ($B=-0.15$, $p=.030$, $q=.451$), and greater mental imagery predicted a larger attitude effect ($B=0.17$, $p=.043$, $q=.674$). Only the familiarity effect survived correction for multiple comparisons within each forest. SI \autoref{sec:si-moderators} reports all coefficients, BLP plots, and CATE/LOESS panels.

\section{Study 2: Satellite cybersecurity}
\label{sec:study-2-satellite-cybersecurity}

To test robustness, we ran a follow-up experiment on satellite cybersecurity with the same co-writing procedure and control conditions, though see SI \autoref{sec:si-study-differences} for other changes between studies. We expected larger attitude effects in satellite cybersecurity, a less polarizing domain (Discussion). AIM participants wrote about a satellite cyberattack disrupting everyday life, with effects based on~\cite{van_camp_world_2022}. Vignette participants read a NATO workshop scenario~\cite{north_atlantic_treaty_organization_scenario_2026} describing the everyday consequences of a hypothetical satellite cyberattack. Placebo participants used AIM to write about the same unrelated scenarios as in Study 1. The final sample contained $N=894$ participants: 298 in AIM, 303 in the vignette condition, and 293 in the placebo condition (see SI \autoref{sec:si-descriptives} for descriptives by arm).

We changed the outcome batteries in Study 2. Because Study 1's two attitude items moved weakly, we could not tell whether AIM had little effect or affected an unmeasured dimension. We therefore used seven 0--100 sliders, chosen to reduce ceiling effects, spanning three subscales---importance of satellite cybersecurity, spending priorities, and willingness to oppose funding cuts. As pre-registered, a three-factor model fit better than a one-factor model (CFI = .973 versus .775; TLI = .948 versus .662; Supplementary Table~\ref{tab:satellite-cfa}). The pre-registered primary attitude index z-scored and equally weighted the three subscales. Closeness was a z-scored index of four 1--9 items (Methods). The pre-registered hypotheses and analyses follow Study 1.  Extended Data Tables~\ref{tab:main_effects_desc} and~\ref{tab:primary} report item-level descriptive statistics and regression estimates.

\subsection{Main effects}
\label{sec:study-2-satellite-cybersecurity-main-effects}

\paragraph{Closeness.} AIM increased psychological closeness relative to both controls (Figure~\ref{fig:forest_2}). Relative to the vignette, the difference was $\beta=0.38$ SDs, 95\% CI $[0.24,0.51]$, $t=5.47$, sharpened $q\leq.001$; relative to the placebo, it was $\beta=0.90$ SDs, 95\% CI $[0.74,1.05]$, $t=11.55$, sharpened $q\leq.001$.

\paragraph{Attitudes.} AIM increased policy support relative to both controls. Relative to the vignette, the difference was $\beta=0.54$ SDs, 95\% CI $[0.39,0.69]$, $t=7.06$, sharpened $q\leq.001$; relative to the placebo, it was $\beta=0.67$ SDs, 95\% CI $[0.52,0.82]$, $t=8.75$, sharpened $q\leq.001$. Unlike Study 1, AIM produced substantial attitude effects. Attitude effects were strong across all three sub-scales. Relative to the vignette, AIM increased importance, spending, and action by $0.50$, $0.51$, and $0.41$ SDs, respectively. Relative to the placebo, it increased them by $0.60$, $0.56$, and $0.56$ SDs (Extended Data Table~\ref{tab:subscales}).

\subsection{Mediation}
\label{sec:study-2-satellite-cybersecurity-mediation}
The pre-registered mediation analysis again yielded positive indirect effects through closeness: $b=0.24$, $95\%$ CI $[0.15,0.34]$ against the vignette ($n=601$) and $b=0.54$, $95\%$ CI $[0.43,0.66]$ against the placebo ($n=591$; Extended Data Table~\ref{tab:mediation}). The sensitivity thresholds were higher than in Study 1 ($\rho^{\star}=0.56$ and $0.62$; Supplementary Fig.~\ref{fig:mediation-sensitivity}).

\subsection{Familiarity}
\label{sec:study-2-satellite-cybersecurity-familiarity}

At the pre-registered cutoff of 2, 425 participants reported low familiarity and 469 reported higher familiarity. All eight subgroup contrasts had confidence intervals above zero. As in Study 1, lower-familiarity participants had larger estimated effects across all four outcome-control (placebo and vignette, attitudes and closeness) combinations. However, at $p<0.05$, the pre-registered moderation tests supported larger effects only in the placebo comparisons. The cutoff of 3 produced a similar but weaker pattern (Extended Data Tables~\ref{tab:familiarity2} and~\ref{tab:familiarity3}).

\subsection{Heterogeneous treatment effects}
\label{sec:study-2-satellite-cybersecurity-heterogeneous-treatment-effects}

Using the same pre-registered causal-forest analysis, we found limited heterogeneity in Study 2. Supplementary Information~\ref{sec:si-moderators} reports BLP estimates and LOESS plots. Only the familiarity effect survived correction for multiple comparisons. Compared to the placebo, higher familiarity ($B=-0.27$, $p=.001$, $q=.021$) and older age ($B=-0.21$, $p=.015$, $q=.134$) predicted smaller closeness effects. Compared to the vignette, greater personal distress predicted a larger closeness effect ($B=0.18$, $p=.012$, $q=.187$). No moderator of attitude effects reached $p<.05$. Note that familiarity and age predicted smaller closeness effects against the placebo in both studies. Their attitude coefficients had the same sign but were smaller and mostly nonsignificant.

\section{Exploratory distributional effects}
\label{sec:exploratory-distributional-effects}
\begin{figure}[h]
\centering
\includegraphics[width=0.8\linewidth]{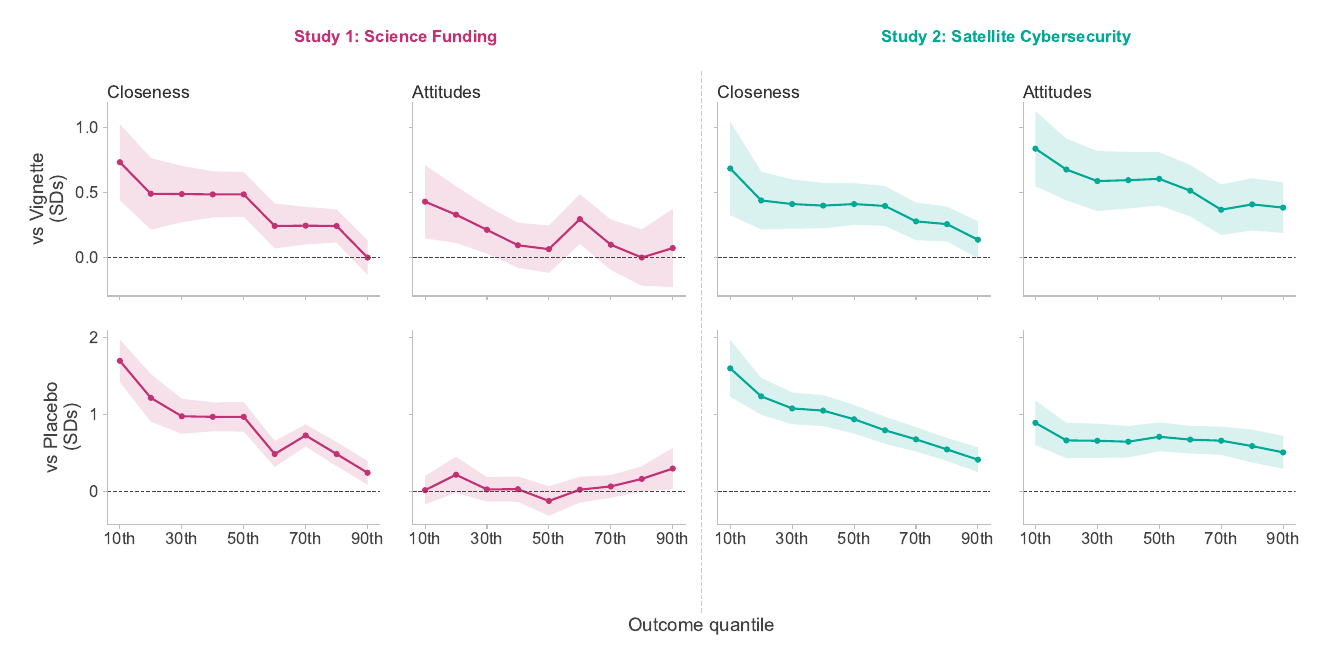}
\caption{Quantile treatment effects. Lines show estimated treatment effect on different outcome quantiles 0.1--0.9, with pointwise $95\%$ intervals. Rows distinguish controls; colors distinguish studies.}
\label{fig:quantile}
\end{figure}

While OLS models whether the treatment shifts the mean, the treatment could also have affected other parts of the distribution. We use quantile regression---which models how a variable (such as a treatment) shifts different quantiles of an outcome distribution---to explore this (Figure~\ref{fig:quantile}). In both studies, treatment effects on closeness were generally concentrated in the lower quantiles of closeness. That is: the treatment appeared to raise the lower end of the distribution, meaning fewer AIM participants than control participants felt that the issue had very little to do with their own lives.

\begin{figure}[tbp]
\centering
\includegraphics[width=\linewidth,height=0.9\textheight,keepaspectratio]{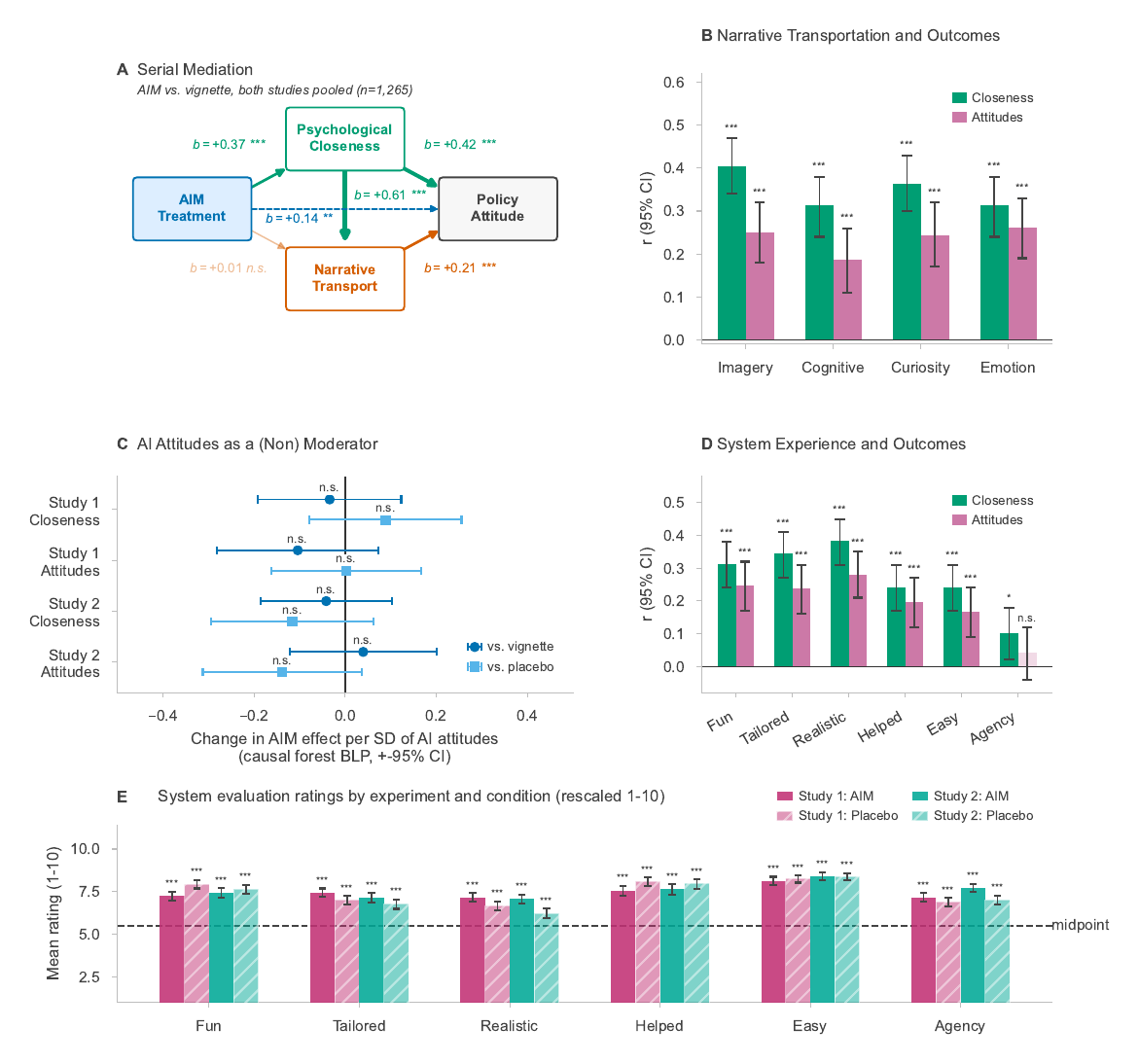}
\caption{Suggestive mechanisms. \textbf{A}, Serial multiple-mediator model (closeness predicts transportation), AIM versus vignette, both studies pooled ($n=1{,}265$). Line widths correspond to effect size. \textbf{B}, \textbf{D}, Pearson correlations of transportation items (B) and system ratings (D) with each outcome among AIM participants, with $95\%$ intervals; opacity indicates $p<.05$. \textbf{C}, Change in the AIM effect per standard deviation of AI attitudes, from each causal forest's best linear projection. \textbf{E}, Mean system ratings by study and condition, rescaled to 1--10; stars test each mean against the scale midpoint. For all plots, ${}^{*}p<.05$, ${}^{**}p<.01$, ${}^{***}p<.001$; n.s., not significant.}
\label{fig:mechanisms}
\label{fig:sys_eval_cor}
\label{fig:sys_eval}
\end{figure}

\section{Suggestive mechanisms and non-mechanisms}
\label{sec:suggestive-mechanisms}

AIM appeared to change attitudes mainly by making the issue feel relevant to participants' own lives (closeness) rather than by immersing them in the story (transportation). Closeness was more associated with AIM's effect on attitudes than narrative transportation. In a serial mediation model (SI~\ref{sec:si-mediation-model-selection} gives the rationale and the parallel alternative), pooling both studies and comparing AIM with the vignette control ($n=1{,}265$), AIM increased closeness ($b=0.37$, 95\% CI $[0.27,0.47]$), closeness increased transportation ($b=0.61$, 95\% CI $[0.54,0.67]$), and both closeness ($b=0.42$, 95\% CI $[0.35,0.49]$) and transportation ($b=0.21$, 95\% CI $[0.15,0.27]$) predicted attitudes (Figure~\ref{fig:mechanisms}A). AIM did not affect transportation beyond its effect through closeness ($b=0.01$, 95\% CI $[-0.09,0.12]$). Closeness therefore carried most of the indirect effect: the path through closeness alone was $b=0.15$ (95\% CI $[0.11,0.21]$), the path from closeness through transportation was $b=0.05$ (95\% CI $[0.03,0.07]$), and the path through transportation alone was near zero ($b=0.00$, 95\% CI $[-0.02,0.03]$). These paths mediated $59\%$ of AIM's effect on attitudes.

Among AIM treatment participants, outcomes were associated with how participants experienced the story and the system. Greater narrative transportation was associated with greater closeness and more supportive attitudes (Figure~\ref{fig:mechanisms}B). The same pattern appeared for five system ratings: participants who found AIM more enjoyable, tailored, realistic, helpful, or easy to use also reported greater closeness and more supportive attitudes (Figure~\ref{fig:mechanisms}D). These analyses are correlational, so we cannot say that a positive system experience caused the outcomes.

Three other plausible mechanisms received surprisingly little support (Figure~\ref{fig:mechanisms}C and D). First, AI adoption---the amount of AI text participants accepted---was unrelated to closeness ($b=-0.002$, 95\% CI $[-0.065,0.062]$, $p=.963$) or attitudes ($b=-0.001$, 95\% CI $[-0.081,0.079]$, $p=.979$). Second, participants' prior views of AI did not moderate AIM's effect on either outcome, against either control, in either study: All eight causal-forest best-linear-projection coefficients were small and nonsignificant (range $-0.14$ to $0.09$ SD per SD of AI attitudes, all $p\geq.12$; Figure~\ref{fig:mechanisms}C). Third, perceived agency over the story was the exception to the general system-rating pattern: it was unrelated to attitudes ($r=.04$, 95\% CI $[-.04,.12]$, $p=.28$) and only weakly related to closeness ($r=.10$, 95\% CI $[.02,.18]$, $p=.010$), the smallest association among the system ratings (Figure~\ref{fig:mechanisms}D).

Participants also evaluated AIM positively  (Figure~\ref{fig:mechanisms}E). Across arms and studies, participants rated AIM as easy to use and fun, its content as tailored and realistic, and its suggestions as helpful. Every rating exceeded its scale midpoint.

\section{Qualitative feedback reveals success modes and failure modes}
\label{sec:success-modes-and-failure-modes-in-study-1}
\label{sec:qualitative-analysis}

We asked AIM participants what worked and what did not. We then identified themes using BERTopic (Methods), an unsupervised method that clusters texts using neural embeddings. We excluded clusters containing low-quality answers (e.g., restatements of the question). Percentages use responses in clusters we classified as substantive. Supplementary Tables~\ref{tab:qualitative_themes} and~\ref{tab:qualitative_themes_satellite} report every retained theme and a representative quotation.

\textbf{Personal relevance.} Participants most often credited AIM with connecting an abstract issue to their own lives. Making science funding personal was the largest Study 1 success theme ($30.1\%$ of success responses): ``Most of the time, this sort of funding feels entirely separated from my day to day life. Creating a story about something that I could easily relate to was effective.'' In Study 2, the largest success theme was recognizing the scale of disruption ($31.4\%$): ``It made me realize how big a threat this is to my everyday life.'' These accounts fit the strong effects on closeness.

\textbf{Hidden dependencies.} The satellite scenario helped participants see which services depend on satellites, revealing hidden dependencies. Three Study 2 themes centered on everyday dependencies, personal reliance, or unnoticed vulnerability ($34.7\%$ of success responses). About a third of those responses concerned reliance on card and phone payments rather than cash. One participant wrote, ``I didn't think I did, but as I started navigating a night in my life, I realized that a lot of what I use works on these systems.''

\textbf{Managing over-steering.} Participants valued suggestions that completed a sentence as they intended. In Study 1, $6.7\%$ of success responses praised AIM's predictive ability (one participant said AIM ``seemed to read my mind!''), and some said it helped them ``continue the narrative when I didn't have something relevant to add.'' But unsuitable suggestions were also the largest complaint in Study 1 ($34.6\%$ of failure responses) and the second largest in Study 2 ($30.8\%$). They broke the writer's flow or repeated the scenario.  In Study 1, this appeared related to ``over-steering'': We wanted the AI to tie the story back to the scenario, but this led some participants to feel the AI was too repetitive, restating the prompt and the scenario. We revised the autocomplete prompt for Study 2 to reduce repetition (SI~\autoref{sec:si-study-differences}). Two Study 2 participants also reported ``uncanny personalization'': suggestions that added details that were not true, such as a call to an aunt who had died.

\textbf{Specificity-credibility tradeoff.} Concrete details helped some participants picture the consequences and led others to doubt them. In Study 1, the 14-month doctor wait was both the second most common success theme and the second most common failure theme: ``The wait time seemed exceptionally long in relation with the `problems' I was having in the story.'' In Study 2, implausibility was the largest failure theme ($29$ of $78$ substantive negative responses, $37.2\%$); the story was ``a bit too doomsday to be believable.'' Plausibility complaints were more common in Study 2, where attitude effects were larger, so they did not stop AIM from working. Because participants steer their own stories, a detail that is plausible in one narrative can become implausible in another.

\textbf{The personal crowding out the political.} Personal relevance could also become a failure theme. In Study 1, $12.8\%$ of failure responses described concentrating on a personal predicament and attending less to science funding as a policy issue: ``[Once] the story began developing, I was more focused on science funding in my own life than the overall funding of science.'' A similar Study 2 theme ($16.7\%$ of failure responses) said the story portrayed personal disruption without explaining how an attack caused it or how cybersecurity policy could prevent it.

\section{Consistent yet distinct roles of humans and AI}
\label{sec:linguistic-results}

Humans and AI played distinct roles, largely replicating across studies. We compared human-written text with accepted AI suggestions (Extended Data Figures~\ref{fig:human_ai_log_odds} and~\ref{fig:human_ai_profile}). AIM participants received $23{,}496$ suggestions across both studies and accepted $4{,}068$. We analyzed language through word log-odds, Biber linguistic features~\cite{biber_using_1993}, and Empath word categories~\cite{fast_empath_2016} (similar to LIWC). Using all features, five-fold cross-validated classifiers on all 109 features distinguished the two sources with mean AUCs of .914 in science funding and .940 in satellite cybersecurity---suggesting these features capture differences between human and AI writing well.

Humans and AI performed different narrative work (Extended Data Figure~\ref{fig:human_ai_log_odds}). AI supplied scenario structure and physical detail. Its most distinctive science-funding terms named the scenario---\textit{train}, \textit{fellows}, \textit{NIH}, and \textit{T32}---while its satellite terms established the setting: \textit{street}, \textit{kitchen}, and \textit{nearby}. Participants supplied emotion and interior perspective. Among their twenty most distinctive science-funding terms were \textit{feel}, \textit{know}, \textit{really}, \textit{think}, and \textit{wonder}. Satellite cybersecurity showed a similar pattern, with humans disproportionately using terms like \textit{think}, \textit{worried}, \textit{know}, \textit{feel}, and \textit{panic}.

Register features showed the same division (Extended Data Figure~\ref{fig:human_ai_profile}; $d_z$ is Cohen's $d$ of human--AI differences). The largest difference in both studies was humans' greater use of first-person pronouns ($d_z=0.55$ in science funding and $0.72$ in satellite cybersecurity). Nervousness ($0.38$ and $0.49$) and negative emotion ($0.37$ and $0.46$) ranked among the ten largest human-associated differences in both studies, and humans used more amplifiers such as ``very'' and ``extremely'' ($0.42$ and $0.31$). AI writing was more structural: it used more nouns ($-0.69$ and $-0.72$), more present participial clauses ($-0.70$ and $-0.77$), and longer words ($-0.63$ and $-0.73$). The system supplied scenario description, while participants supplied first-person perspective and emotion.

This division replicated across domains: the 109 feature contrasts correlated $r=.78$ between studies, and 83 of 109 were in the same direction. It was also intentional (Extended Data Table~\ref{best_practices}): we wanted the Autocompleter to avoid being too sentimental, didactic, or prescriptive.

\section{Discussion}
\label{sec:discussion-content}
\label{sec:discussion}

\paragraph{Personalized simulation made two long-term risks feel close.} Across two pre-registered experiments, AIM increased psychological closeness by $0.36$ and $0.38$ SDs relative to an expert-written vignette and by $0.80$ and $0.90$ SDs relative to writing about an unrelated topic. It also shifted policy attitudes: by $0.16$ SDs against the vignette and $0.10$ SDs against the placebo for science funding (the latter was not significant), and by $0.54$ and $0.67$ SDs, respectively, for satellite cybersecurity. The attitude effect was not as dependable.

\paragraph{AIM revealed hidden dependencies.} AIM helped participants project themselves into a specific future episode, a form of episodic future thinking~\citep{atance_episodic_2001,szpunar_taxonomy_2014}. Its strong effects on psychological closeness and the qualitative responses are consistent with this interpretation. In Study 2, $23.1\%$ of success-theme responses described rehearsing what the participant would do, while $34.7\%$ described discovering everyday dependencies, personal reliance, or previously unnoticed vulnerability (Supplementary Table~\ref{tab:qualitative_themes_satellite}). By imagining a day under satellite disruption, participants recognized connections (among satellite infrastructure, payments, navigation, and communication) that they had not considered. These simulations may be especially useful for risks whose consequences are hidden in ordinary life. Infrastructure has this property. It often goes unnoticed until it fails~\cite{star_ethnography_1999}, yet its failures can cascade. Other risks may have this same ``infrastructural'' quality. Democratic norms, for example, can function as social infrastructure:  unnoticed while intact, but consequential when they fail. Because many societal challenges have non-obvious individual consequences, AIM likely has applications beyond the domains studied here.

\paragraph{Humans and AI contributed different parts of the simulation.} AIM supplied scenario structure and physical detail, while participants supplied first-person perspective and emotion (Extended Data Figures~\ref{fig:human_ai_log_odds} and~\ref{fig:human_ai_profile}). This division replicated across domains: The 109 human--AI linguistic contrasts correlated $r = .78$ between studies. AIM did not write a personalized story on the participant's behalf. It supplied a scaffold that participants developed. For designers, this suggests that the value of generative AI co-writing may lie less in producing complete narratives than in providing enough structure for people to imagine an episode on their own terms.

\paragraph{Closeness may change attitudes most when those attitudes are less established.} Closeness effects were nearly identical across studies, but attitude effects were larger in Study 2. One possibility is that science-funding attitudes were more crystallized. SI~\autoref{sec:si-crystallization} gives more details and SI~\autoref{sec:si-attitude-explanations} discusses alternative explanations we ruled out. Each SD of closeness predicted a larger difference in satellite attitudes ($b=0.62$) than in science-funding attitudes ($b=0.41$; interaction $b=0.21$, 95\% CI $[0.13,0.29]$). The interaction held among placebo participants, who did the same task in both studies without issue-relevant information ($b=0.16$, 95\% CI $[0.05,0.28]$). Three other indicators agree with the crystallization account. First, science funding received about 30 times as much news coverage in the preceding two years (Media Cloud~\citep{roberts_media_2021}). Second, ideology explained about six times as much variance in control participants' science-funding attitudes ($R^2=3.9\%$) as in satellite attitudes ($R^2=0.7\%$). Third, participants' open-ended feedback referred to prior views more often in Study 1 than in Study 2 ($13.7\%$ versus $6.5\%$ of coded open-ended responses; $z=4.21$, $p<.001$), including among vignette participants ($14.5\%$ versus $6.3\%$). See SI~\autoref{sec:si-crystallization-validation} for details and validation. By contrast, interface changes, sample composition, story--item similarity, and ceiling effects did not appear to account for the gap (SI~\autoref{sec:si-attitude-explanations}). Differing levels of emotion of the vignette explain at most a fifth of the vignette contrast. Together, the evidence is \textit{consistent} with closeness changing attitudes more when those attitudes are less crystallized. But we did not manipulate crystallization, and other study differences could explain the divergence (SI~\autoref{sec:si-study-differences})---though we find less evidence for these explanations.

\paragraph{AIM did not require favorable attitudes toward AI or a sense of control.} Prior work identifies two factors that could shape the efficacy of AI-assisted interventions: people's attitudes toward algorithms~\citep{dietvorst_algorithm_2015} and their sense of ownership over co-written text~\citep{draxler_ai_2024}. Neither explained our results. AI attitudes did not moderate AIM's effect on either outcome in either study (Figure~\ref{fig:mechanisms}C), and effects held among participants below the scale midpoint (Supplementary Table~\ref{tab:ai_attitude}). Perceived agency was unrelated to attitudes ($r=.04$, $p=.28$) and weakly related to closeness ($r=.10$, $p=.010$).

\paragraph{Specificity made scenarios vivid, but also contestable.} Perceived realism can increase narrative transportation~\citep{green_transportation_2004}, which would suggest using specific details to make scenarios concrete. However, specificity can backfire when the specificity is deemed non-credible. In Study 1, the fixed 14-month wait helped participants imagine the consequence. But it also gave participants a specific claim to reject. One participant said, ``14 months for a wait time across all places seemed unrealistic, no matter the scenario, especially in life or death situations like cancer.'' Others accepted that specialist waits can be long but rejected the idea that science-funding was the cause. Study 2 produced a different problem: Some participants rejected the scale of the entire event rather than any particular detail. Overall, specifics can increase concreteness but also contestability of simulated counterfactuals.  Moreover, in a co-writing system, the participant can also direct the narrative toward contexts in which a previously plausible detail no longer fits. More specificity is not always more credible.

\paragraph{The personal could overshadow the political.} Framing research distinguishes episodic accounts of individual cases from thematic accounts of broader patterns~\citep{iyengar_is_1994,hart_one_2011}. AIM is deliberately episodic, which made consequences concrete but sometimes left institutions and policy causes abstract. In Study 1, $13\%$ of failure-theme responses described focusing on a personal predicament rather than science funding as policy. Prior research finds overly episodic framing may not change attitudes toward systems~\cite{iyengar_is_1994}, and narratives change social attitudes best when they remain connected to credible and typical causes~\cite{polletta_when_2020}. For AIM scenarios, the personal must therefore remain connected to the political, in a causally believable way.

\paragraph{Misuse potential.} AI can be persuasive, especially when it personalizes content~\cite{matz_potential_2024}. AIM could therefore make real or manufactured risks feel personally relevant to a targeted audience, regardless of accuracy. Still, AIM likely poses less risk than ``passive exposure''~\cite{ashkinaze_how_2024}, in which people see AI-generated content without any role in its creation. AIM's architecture requires active engagement: participants write their own stories and decide whether to accept each suggestion. Our deployment added a second safeguard by disclosing the general scenario beforehand, unlike work in which surreptitiously biased suggestions shifted attitudes~\citep{jakesch_co-writing_2023}. A deployer, however, could omit this disclosure, and the deployer also controls the scenario and how strongly the agents steer. Active engagement therefore reduces the risk of manipulation but does not remove it. Generative AI is dual use: Chatbots can both reduce~\citep{costello_durably_2024} and increase~\citep{costello_large_2026} belief in conspiracy theories.

\paragraph{Limitations and future work.} We measured self-reported outcomes immediately after the task. We measured the mediator and outcome together, so our results are \textit{consistent} with an indirect effect through closeness without identifying it. We evaluated AIM as an assembled intervention and did not isolate personalization, the starting scenario, or co-writing. AIM's personalization means the treatment adds variance compared to a fixed script. Each experiment compared AIM to one specific vignette. Our Prolific samples may be more comfortable with AI-assisted writing than other populations. Both scenarios were negative futures. Future studies can ablate components (e.g., ``turn off'' AI suggestions while keeping personalization), measure attitudes longitudinally, and test positive futures.

\paragraph{Conclusion.} We cannot personally experience every consequence that should inform what we support. To address this fundamental problem, we created the As-If Machine. Its goal is to make distant risks feel closer. Across two pre-registered experiments, we find AIM did make distant risks feel closer. Whether that closeness changed what people supported depended on the domain. Generative AI may be able to approximate something experience ordinarily provides: a concrete sense of what alternative futures would mean, for an individual's own life.

\label{future}
\label{sec:conclusion}


\clearpage
\FloatBarrier
\linespread{1.03}\selectfont
\section{Methods}
\label{sec:methods}
\clubpenalty=10000
\widowpenalty=10000

\subsection{Study design and preregistration}
\label{sec:methods-study-design-and-preregistration}

We randomly assigned participants to AIM, an expert vignette, or an interactive placebo, targeting 300 participants per arm. Each registration specified 80\% power to detect $d=0.23$ in a two-sided two-sample test at $\alpha=.05$. We pre-registered the science-funding experiment at \url{https://osf.io/j7dsy} and the satellite experiment at \url{https://osf.io/e967w}. Both registrations specified hypotheses concerning closeness, attitudes, mediation, and familiarity, including familiarity cutoffs of $\leq2$ and $\leq3$ (SI \autoref{sec:si-cutoff} for cutoff rationales). Study 2 additionally pre-registered familiarity interaction tests, a factor-model comparison, and secondary attitude-subscale analyses. Both registrations also specified exploratory causal forests. The quantile, linguistic, and pooled mechanism analyses were additional exploratory analyses. See SI \ref{sec:si-screening} for pre-registration deviations.

\subsection{Procedure and participant protection}
\label{sec:methods-procedure-and-participant-protection}

The University of Michigan's institutional review board determined that both experiments were exempt from ongoing oversight. Study 1 participants provided informed consent before beginning and received warnings about serious medical content in both the Prolific posting and the consent form. Eligibility required age 18 or older, at least 100 prior Prolific submissions, a 98\% approval rating, and desktop or laptop use. Screening included a commitment to high-quality responses and attention and comprehension checks. Study 2 followed a similar procedure and pre-registered screening out Safari users because of interface errors.

Vignette participants read their assigned story and answered comprehension questions. AIM and placebo participants read instructions for the writing interface and completed comprehension checks before providing personal information and co-writing (six minutes for Study 1, seven minutes for Study 2). They then returned to Qualtrics using a completion code and answered outcome, experience, and demographic questions.  The vignette provided domain-relevant information, and the placebo used the same co-writing interface for an unrelated event.  In both experiments, the placebo condition was cross-randomized to use AIM for either (1) meeting a celebrity or (2) having sounds turn into colors since we anticipated these were plausibly unrelated to science funding or satellite cybersecurity.

\subsection{System implementation}
\label{sec:methods-system-implementation}

In the most current version of the system, each agent calls GPT-4.1-mini through an API using prompts stored in a YAML file. The Autocompleter issues an AJAX request only after the participant has paused mid-sentence for 300 ms, written at least one complete sentence, and typed at least one character since the last accepted suggestion. For each request, the system samples temperature uniformly from $[0.5, 1.2]$ and token length from $[10, 30]$ to produce short, varied completions. Extended Data Figure~\ref{fig:prompt_anatomy} shows an example request and identifies the source of each prompt element.

\subsection{Formative development}
\label{sec:methods-formative-development}

Before the randomized experiments, we refined AIM through small-scale, informal user testing (e.g., with research assistants). This testing revealed two problems that informed the current design.

First, in early versions of the system, users objected when the AI attributed interiority to them. For example, a user would type ``I was\ldots'' and the AI would complete the sentence with an emotional state the user did not feel. Because this feedback was common, we constrained the Autocompleter to continue users' stories rather than describe what users thought or felt (Extended Data Table~\ref{best_practices}; Extended Data Figure~\ref{fig:prompt_anatomy}). The linguistic analysis is consistent with these constraints: participants supplied emotion and first-person perspective, while the AI supplied concrete detail and scenario structure (\autoref{sec:linguistic-results}).

Second, users needed a starting point, for two reasons. First, users found it difficult to imagine how the scenario could plausibly begin, so a personalized onset could serve as a plausible entry into the scenario. Second, the task is a creative writing exercise, and many people do not write creatively. Given only an abstract scenario, users asked what exactly they should write about. We therefore display the agent-generated onset, built from each user's intake responses, at the top of the editor (Figure~\ref{fig:asif_diagram}a), so users continue an existing story rather than start from a blank page. Participants rated this prompt as helpful for starting the story (Figure~\ref{fig:mechanisms}E).

\subsection{Outcome measures}
\label{sec:methods-outcome-measures}

\textbf{Science funding: closeness.} Two items assessed closeness to the issue: ``I can clearly imagine how cuts to scientific funding could affect me, personally'' and ``Levels of scientific funding feel relevant to my life'' (both 1--9). We coded items such that higher values mean more psychological closeness (i.e., less psychological distance) and averaged them into a z-scored index.

\textbf{Science funding: attitudes.} Two items assessed attitudes toward science funding: perceived personal benefit of U.S. science over the past decade~\citep[1--5; adapted from][]{ophir_effects_2021} and preferred NIH share of federal spending (0--100\%, similar to \cite{goldfarb_building_2017}). We averaged items into a z-scored index.

\textbf{Satellite cybersecurity: closeness.} Four items measured whether a deliberate satellite attack felt abstract or concrete, whether participants could see effects on their daily lives, whether they could imagine its effects, and whether satellite cybersecurity felt relevant to their lives (1--9).

\textbf{Satellite cybersecurity: attitudes.} Seven attitude items (split into three sub-scales) used 0--100 sliders. Importance comprised personal and comparative national importance. Spending comprised the preferred direction of satellite cybersecurity spending and the priority of cybersecurity grants to state and local governments. Action comprised intentions to vote against a candidate or ballot measure that reduced satellite cybersecurity funding and to oppose a politician proposing such cuts.

We standardized items within each study, averaged them, and standardized the resulting indices, so treatment contrasts are in standard-deviation units. For satellite attitudes, we first constructed the three standardized subscales, then averaged and re-standardized them with equal subscale weights. The three-item action subscale therefore received the same weight as each two-item subscale. All analyzed participants had complete main outcome indices. We coded higher scores as greater closeness or support, following the item wording.

For the pre-registered factor comparison, we fit one-factor and correlated three-factor models to the seven standardized attitude items using semopy's maximum-likelihood discrepancy objective. We retained the pre-registered equal-subscale index rather than estimating scoring weights from the fitted model. Supplementary Table~\ref{tab:satellite-cfa} reports the fit statistics.

\subsection{Pretreatment and posttreatment measures}
\label{sec:methods-pretreatment-and-experience-measures}

\textbf{Pretreatment.} Before treatment, we measured familiarity with the study topic (1--7), creative writing and journaling experience, vividness of mental imagery~\citep[5 items;][]{andrade_assessing_2014}, fantasy, perspective taking, and personal distress (using the 12-item Brief Interpersonal Reactivity Index)~\citep{ingoglia_development_2016}, and AI attitudes~\citep[AIAS-4;][]{grassini_development_2023}.

\textbf{Posttreatment.} In addition to closeness and attitudes, we asked several other questions. For treatment and placebo participants, we assessed system usability and helpfulness with several items. We measured ease of use with a single item from the System Usability Scale~\citep{lewis_system_2018} --- ``I thought the system was easy to use'' --- which accounts for 90\% of the full scale's variance~\citep{sauro_can_2018}. We measured enjoyment with ``I thought the system was fun to use.'' Two items captured helpfulness of system components: ``The writing assistant suggestions helped me continue the story'' and ``The prompt at the top of the page helped me start the story.'' Two items captured perceived story quality: personalization (``The story I wrote felt personalized to aspects of my life'') and realism (``\textit{Assuming the scenario were to happen to me}, it would play out like in the story I wrote''). Finally, we measured perceived agency with ``Who did you feel was in control of the story you wrote?'' ($-3 =$ writing assistant had total control; $+3 =$ I had total control). We measured narrative transportation using four post-task narrative-transportation items from the short-form narrative transportation scale~\cite{appel2015transportation} (SI \autoref{sec:si-narrative-items} for details) assessing picturing oneself in the scenario, mental involvement, wanting to learn more, and emotional impact. Additionally, for both the vignette and treatment participants, we asked what about the activity was (1) effective and (2) ineffective for changing attitudes or making the topic feel more relevant. At the end of these questions, we asked about basic demographics.

\subsection{Exclusions}
\label{sec:methods-exclusions-and-implementation-deviations}

Our pre-registered screening criteria excluded participants who failed attention checks, failed comprehension checks, completed the study faster than three scaled median absolute deviations below their condition's median, or reported technical errors (AIM and placebo arms, which used the writing app). All participants had completed at least 100 prior studies with a $98\%$ approval rating. We quota-sampled participants to match the U.S. population on sex, political party (Democrat, Republican, independent), and age (18--34, 35--54, and 55+ in Study 1; 18--54 and 55 or older in Study 2). See the \hyperref[sec:si-screening]{Supplementary Information} for details.

\subsection{Primary effects and familiarity}
\label{sec:methods-primary-effects-and-familiarity}

For each study and outcome, we fit an OLS model with HC1 heteroscedasticity-robust standard errors:
\begin{equation}
Y_i=\alpha+\gamma_V\,\mathrm{Vignette}_i+\gamma_P\,\mathrm{Placebo}_i+\varepsilon_i,
\label{eq:main}
\end{equation}
with AIM as the reference. This specification was pre-registered. We report $-\gamma_V$ and $-\gamma_P$ as AIM-minus-control effects with $95\%$ confidence intervals. Figure~\ref{fig:forest_2} also reports sharpened false-discovery-rate $q$ values for 12 contrasts per study: two outcomes, two controls, and three samples (full, low familiarity, high familiarity).

As pre-registered for both Studies 1 and 2, we analyzed H1 and H2 separately at two familiarity thresholds ($\leq 2$ and $\leq 3$). We fit Equation~\ref{eq:main} within the low- and high-familiarity groups at each cutoff. We also formally tested whether these cutpoints moderated the effect by interacting treatment-arm indicators with a dummy variable for IsHighFamiliarity.

\subsection{Mediation and exploratory models}
\label{sec:methods-mediation-and-exploratory-models}

We used Python's \texttt{pingouin} mediation package with bootstrap CIs for our simple pre-registered mediation model where the treatment's effect on attitudes is mediated via closeness only. We computed sensitivity using the R function \texttt{medsens} from \texttt{mediation}. Our more complex mediation model in \autoref{fig:mechanisms} (serial multiple-mediator model in which closeness predicts transportation, and both predict attitudes) was fit in the \texttt{lavaan} R package using bootstrap CIs and entering study number as a covariate. We constructed the adoption index by taking an inverse-covariance weighted~\cite{anderson_multiple_2008} index of acceptance rate, number of accepted suggestions, and proportion of AI-origin words; this results in an index of mean 0 and SD 1.

\subsection{Causal forests and treatment-effect heterogeneity}
\label{sec:methods-causal-forests}
We fit separate causal forests for each outcome and AIM--control comparison using the R package \texttt{grf} 2.6.1, with 2,000 honest trees and automatic tuning of all supported hyperparameters~\citep{wager_estimation_2018,athey_estimating_2019}. Honesty uses separate observations to choose splits and estimate effects within leaves. We standardized continuous moderators.
We summarized heterogeneity with best linear projection (BLP) coefficients estimated from doubly robust scores rather than by regressing fitted CATEs directly on the moderators~\citep{semenova_debiased_2021,chernozhukov_doubledebiased_2018}. We used HC3 standard errors. We visualized out-of-bag CATE predictions with LOESS smooths. The Supplementary Information presents all CATE panels and BLP estimates, along with sharpened q values that correct for multiple hypotheses within each forest.

\subsection{Qualitative and linguistic analysis}
\label{sec:methods-qualitative-and-linguistic-analysis}

We asked treatment participants which aspects of the co-writing experience were effective and ineffective. We identified themes using BERTopic\footnote{\url{https://maartengr.github.io/BERTopic/index.html}}---a transformer-based topic modeling pipeline. The pipeline embeds documents, reduces dimensionality, and then clusters resultant document vectors. We embedded responses with \texttt{hkunlp/instructor-large}~\citep{su_one_2023}, reduced dimensionality with UMAP~\citep{mcinnes_umap_2020}, and clustered with HDBSCAN~\citep{mcinnes_hdbscan_2017} (which automatically picks the number of clusters). We excluded filler clusters (e.g., ``nothing was ineffective''), and an author manually interpreted the remaining clusters. Table~\ref{tab:qualitative_themes} reports the substantive themes with representative quotes.

We compared human-written text with accepted AI suggestions using identical lexical and linguistic analyses in both studies. First, we fit a log-odds model with a Dirichlet prior to identify tokens associated with each source. Second, we extracted Biber's 67 lexico-grammatical register features~\cite{biber_using_1993} with PyBiber 0.3.1, which characterize text type and have been used in human--AI comparisons~\cite{sourati_shrinking_2026}. We also extracted 42 Empath semantic category scores~\citep{fast_empath_2016}. Because each participant contributed one human and one AI document, we measured each feature's difference with the paired Cohen's $d_z$: the mean of the within-participant human-minus-AI differences divided by their standard deviation. The figures rank features by $d_z$. We retained a human--AI pair only when both halves exceeded five words, so that neither side was too short to characterize. This yielded 285 pairs in Study 1 and 259 in Study 2, each with 67 Biber features and 42 Empath categories per document.

To test whether these features distinguished human from AI text, we trained an L2-regularized logistic regression classifier ($C=0.5$; maximum 1,000 iterations). We used 5-fold cross-validation (GroupKFold), keeping each participant's paired documents in the same fold. Supplementary Table~\ref{tab:linguistic-classifier} reports mean AUC and its variation across folds. The log-odds analysis used a stopword list, unigrams and bigrams, and a symmetric Dirichlet pseudocount of 0.001.

We constructed AI-authored text by concatenating all accepted suggestions. We constructed human-authored text by removing one exact occurrence of each accepted suggestion from the final narrative. We retained a narrative only when both the AI text and the remaining human text exceeded five words. Because participants could edit or delete AI suggestions after accepting them, the final narrative did not always contain the accepted suggestion verbatim, and we remove the first exact match. We therefore reconstruct the human and AI text approximately.

\subsection{Research use of generative AI}
\label{sec:methods-research-use-of-generative-ai}

We used generative AI for code generation, writing assistance (concision, clarity, and grammar), and stimulus refinement.

\section{Author contributions}
\label{sec:methods-author-contributions}

Joshua Ashkinaze conceived the initial system and experiments, programmed the system and experiments, ran the experiments, conducted the analyses, and wrote the manuscript. Yubo Zhou contributed to experimental design, system programming, analysis, and manuscript writing. Pengyu Chen contributed to experimental design, system programming, and analysis. Avi Bhattacharya contributed to system design and system programming. Hyebin Park and Yanman Zhang contributed to system design. Stephanie Preston, Eric Gilbert, and Ceren Budak supervised the research and provided feedback on the manuscript and figures.

\section{Open Source}
\label{sec:methods-code-availability}

An open-source version of the system is available at \url{https://github.com/josh-ashkinaze/asif-machine}.

\bibliographystyle{unsrtnat}
\begingroup
\raggedright
\bibliography{references}
\endgroup

\subsection{Acknowledgments}
We thank Scott Page, David Rand, Misha Teplitskiy, Lavinia Dunagan, and James Dumlao for feedback. We also received helpful feedback from the Political Communication Working Group at the University of Michigan.

\subsection{Funding}
\label{sec:methods-funding}

Both studies received funding from the Cooperative AI Foundation and the University of Michigan's Propelling Original Data Science program.

\clearpage
\section{Extended Data}
\label{sec:extended-data}
\label{sec:extended}
\setcounter{figure}{0}
\setcounter{table}{0}
\renewcommand{\thefigure}{\arabic{figure}}
\renewcommand{\thetable}{\arabic{table}}
\captionsetup[figure]{name=Extended Data Fig.}
\captionsetup[table]{name=Extended Data Table}
\renewcommand{\theHfigure}{ED.\arabic{figure}}
\renewcommand{\theHtable}{ED.\arabic{table}}
\floatplacement{table}{H}

\subsection{Outcome distributions and primary effects}
\label{sec:extended-data-outcome-distributions-and-primary-effects}
\begingroup
\footnotesize
\setlength{\tabcolsep}{3pt}
\input{study_comparison/tables/descriptive.tex}
\input{study_comparison/tables/primary.tex}
\endgroup
\clearpage
\subsection{Pre-registered familiarity subgroups}
\label{sec:extended-data-pre-registered-familiarity-splits}
\begingroup
\footnotesize
\setlength{\tabcolsep}{3pt}
\input{study_comparison/tables/familiarity_2.tex}
\input{study_comparison/tables/familiarity_3.tex}
\endgroup
\clearpage
\subsection{Study-specific mediation and satellite attitude subscales}
\label{sec:extended-data-study-specific-mediation}
\begingroup
\small
\centering
\input{study_comparison/tables/mediation.tex}
\input{satellite/tables/satellite_subscales.tex}
\endgroup
\clearpage
\subsection{Interface and scenario configuration}
\label{sec:extended-interface-configuration}
\begin{figure}[H]
\centering
\includegraphics[width=\linewidth,height=0.68\textheight,keepaspectratio]{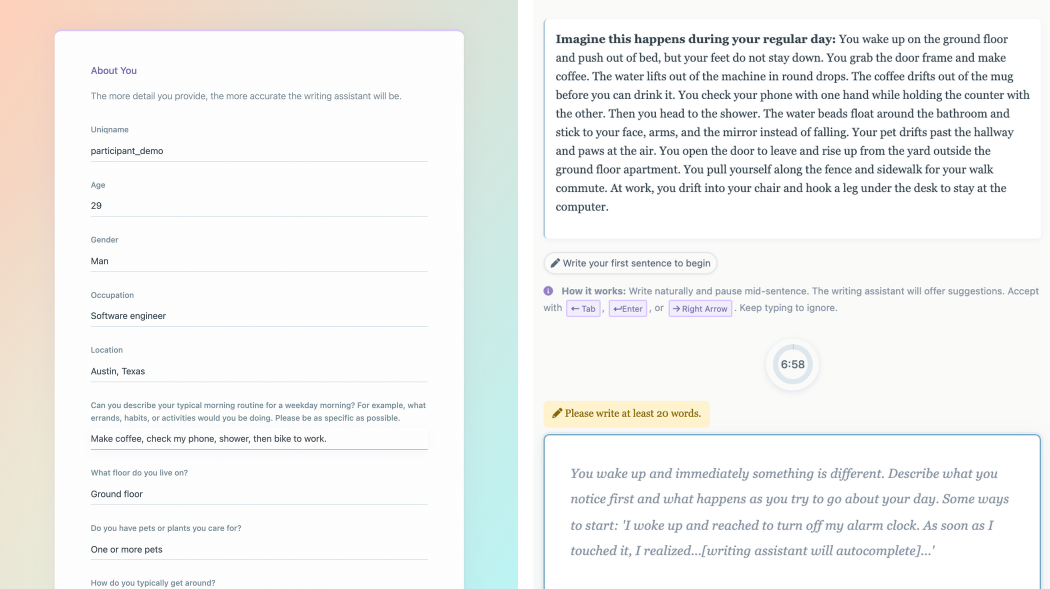}
\caption{AIM interface for an illustrative scenario in which gravity disappears. The next panel shows the configuration file that generates it; neither experiment used this scenario.}
\label{fig:asif_screens}
\end{figure}
\clearpage
\begin{figure}[H]

\centering
\includegraphics[width=\linewidth,height=0.83\textheight,keepaspectratio]{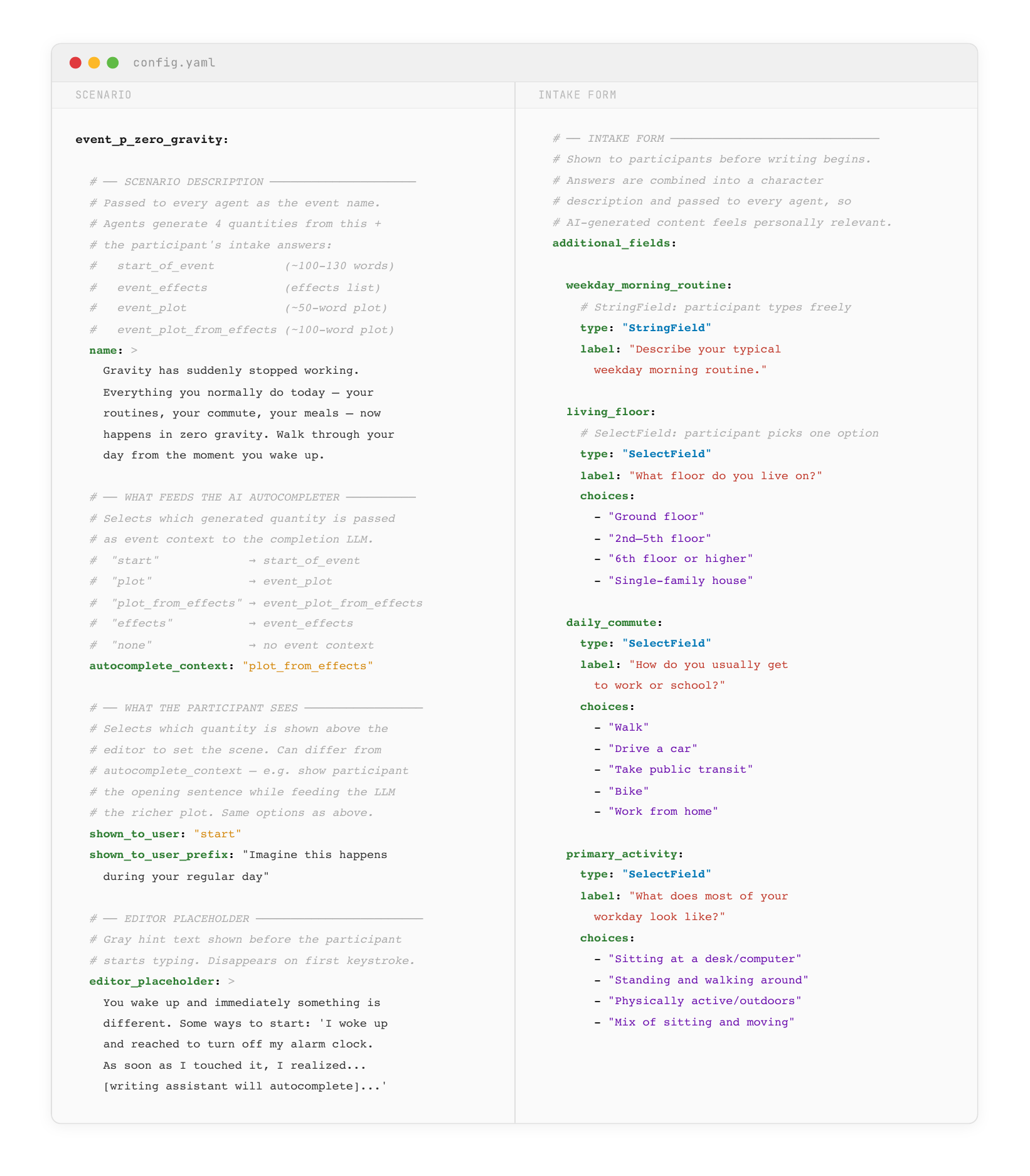}
\caption{AIM interface and scenario configuration, continued. Example YAML configuration for the gravity-disappearance scenario shown in the preceding panel.}
\label{fig:yaml_config}
\end{figure}
\clearpage
\subsection{Satellite interface and autocomplete prompt}
\label{sec:extended-data-science-funding-interface-and-autocomplete-prompt}
\begin{figure}[H]
\centering
\includegraphics[width=\linewidth,height=0.37\textheight,keepaspectratio]{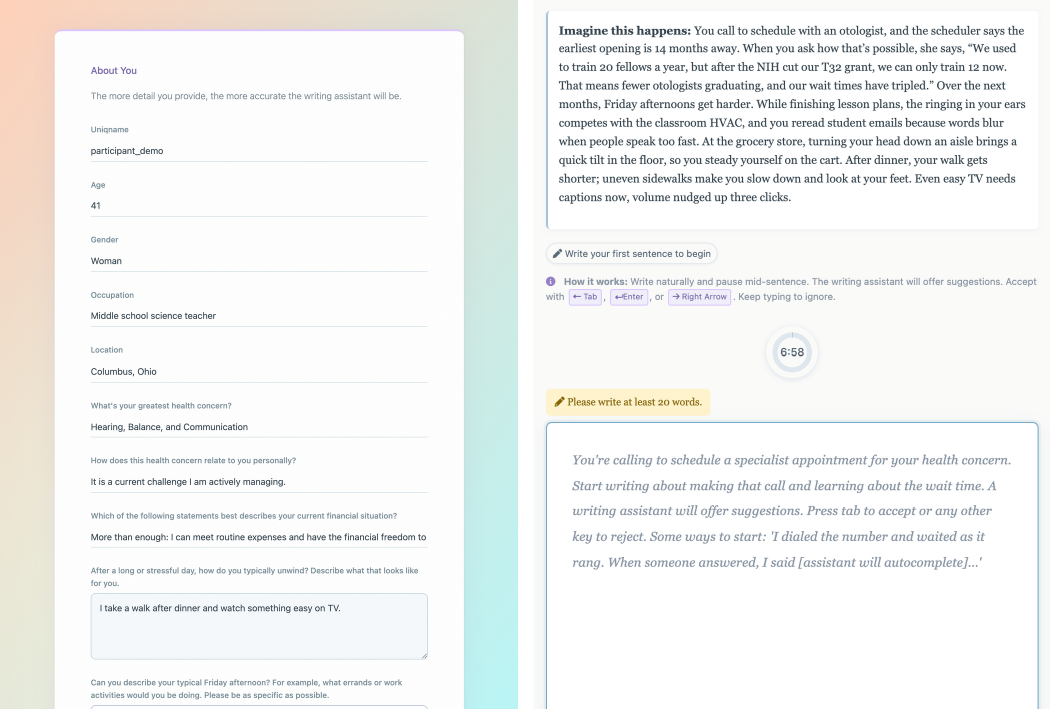}

\medskip
\includegraphics[width=\linewidth,height=0.38\textheight,keepaspectratio]{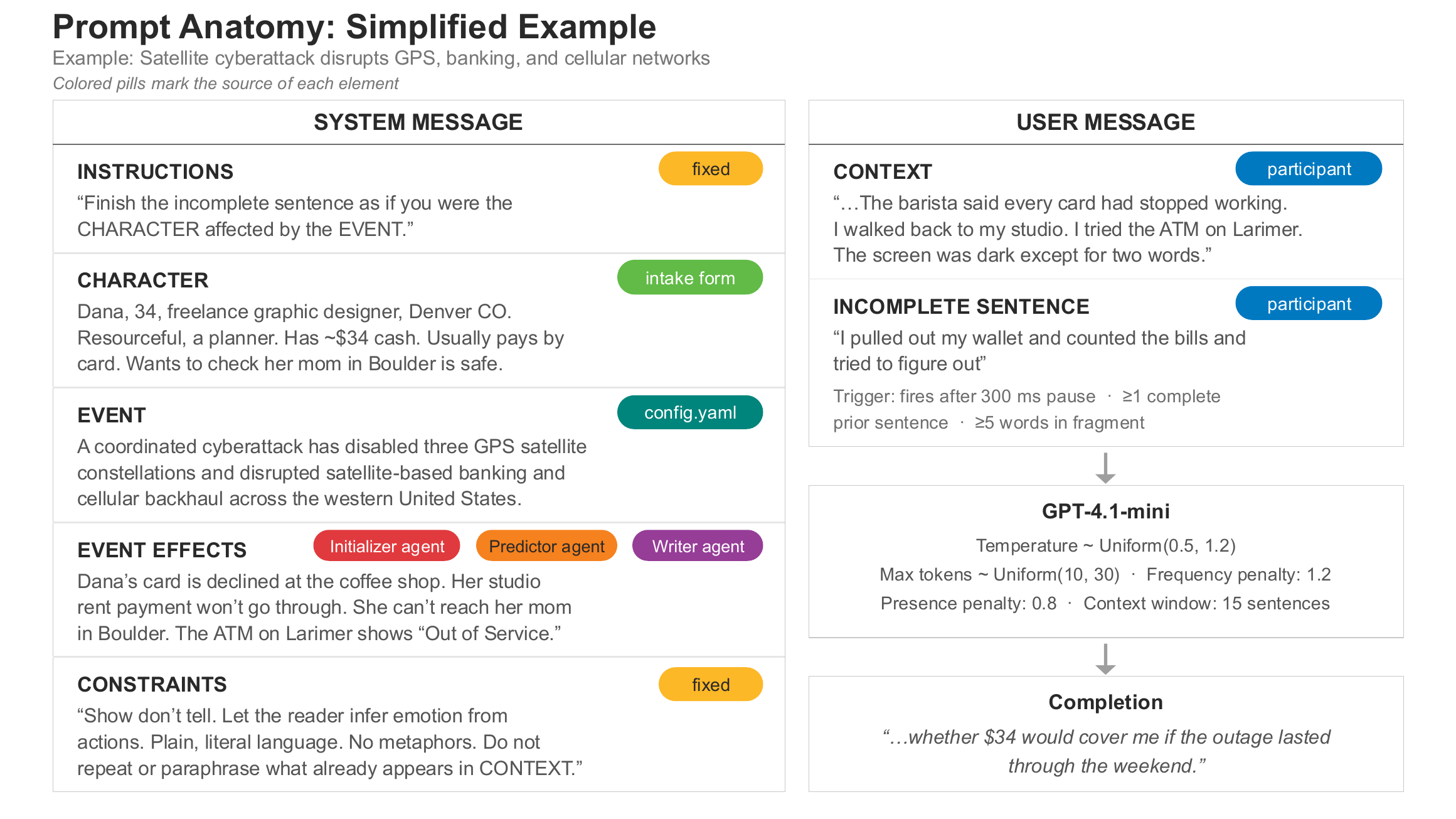}
\caption{Satellite interface and autocomplete prompt. Top, the co-writing interface; bottom, the autocomplete request from step 5 of Figure~\ref{fig:asif_diagram}, with colors marking the source of each element.}
\label{fig:prompt_anatomy}
\end{figure}
\clearpage
\subsection{Human and AI lexical differences}
\label{sec:extended-human-ai-words}
\begin{figure}[H]
\centering
\includegraphics[width=\linewidth]{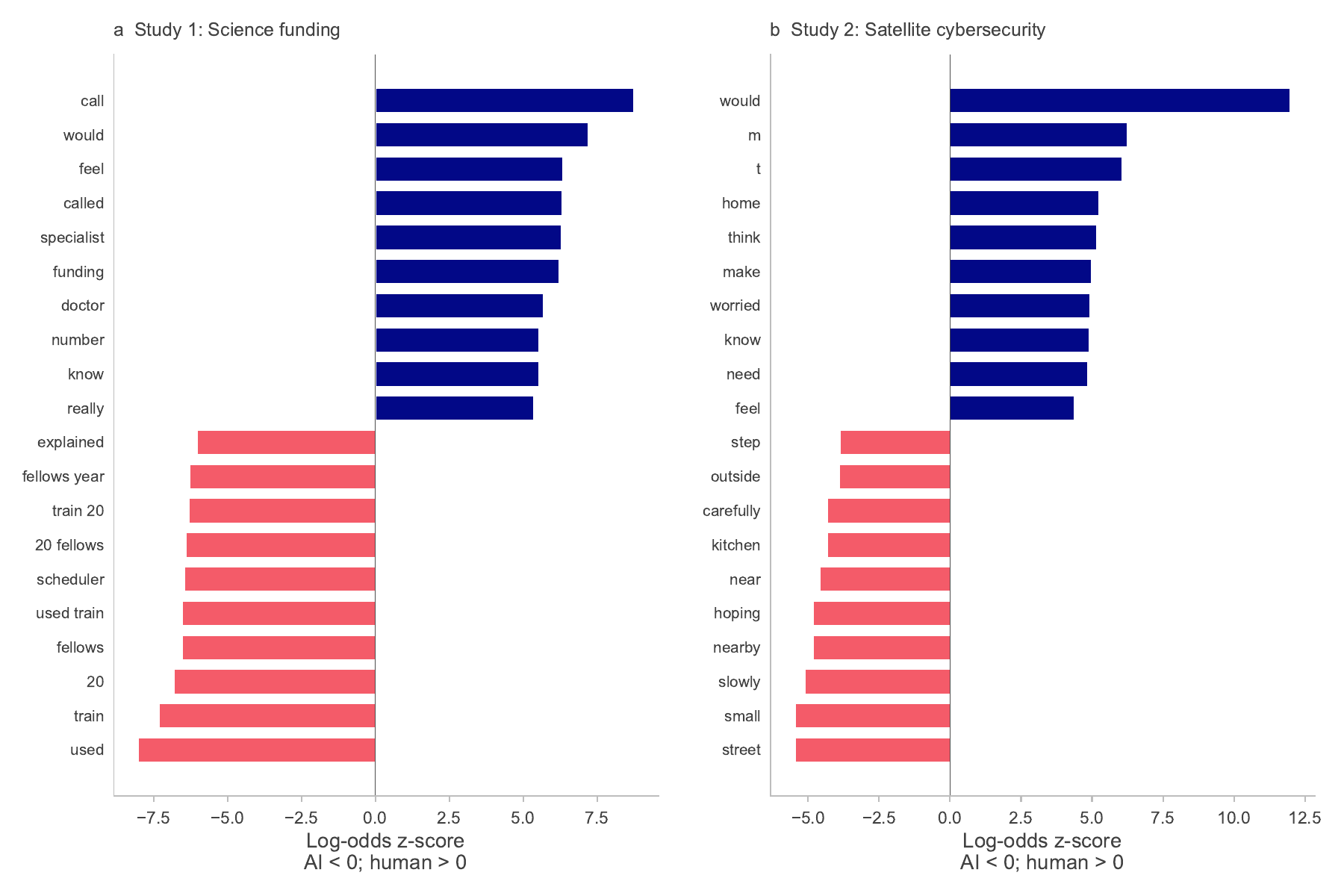}
\caption{Human and AI lexical differences. The ten terms most associated with each source in each study, ranked by log-odds $z$ score; positive values indicate participant-written text. Study 1, 285 paired narratives; Study 2, 259.}
\label{fig:human_ai_log_odds}
\end{figure}
\clearpage
\subsection{Human and AI linguistic features}
\label{sec:extended-human-ai-features}
\begin{figure}[H]
\centering
\includegraphics[width=\linewidth]{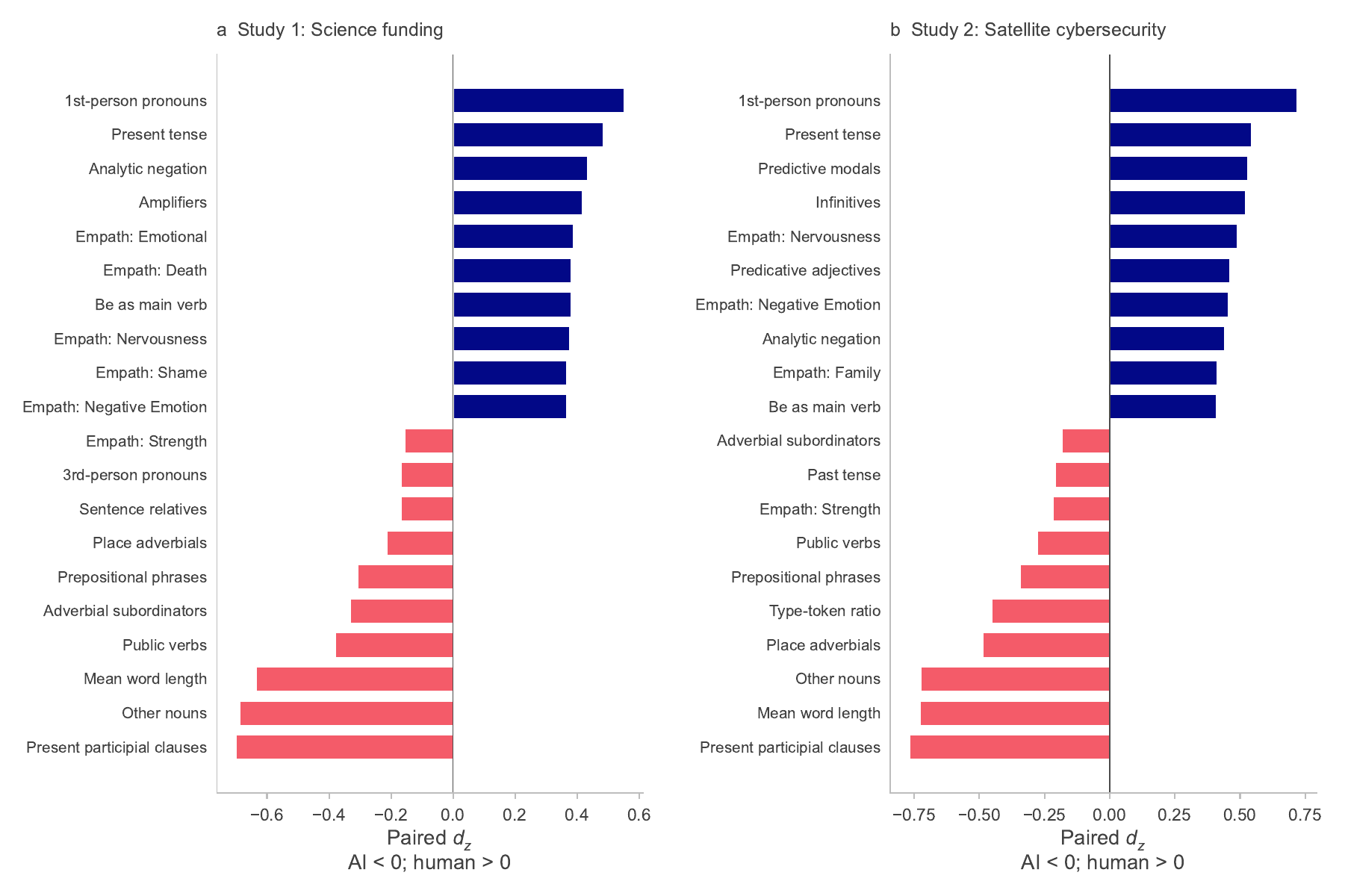}
\caption{Human and AI linguistic features. The ten largest differences in each direction within each study; positive $d_z$ indicates a higher value in participant-written text. Study 1, 285 document pairs; Study 2, 259.}
\label{fig:human_ai_profile}
\end{figure}
\clearpage

\subsection{Narrative Transportation Practices}
\label{sec:extended-narrative-transportation}
Extended Data Table~\ref{best_practices} shows how AIM translates each principle into system components.

\begin{longtable}{@{}p{0.4\textwidth}p{0.58\textwidth}@{}}
\caption{Embedding narrative transportation best practices~\citep{green_transportation_2021} into the system.}
\label{best_practices}\\
\toprule
\raggedright\arraybackslash\textbf{Best Practices for Narrative Transportation} &
\raggedright\arraybackslash\textbf{Our System} \\
\midrule
\endfirsthead

\multicolumn{2}{c}{\bfseries Table \thetable\ continued from previous page} \\
\toprule
\raggedright\arraybackslash\textbf{Best Practices for Narrative Transportation} &
\raggedright\arraybackslash\textbf{Our System} \\
\midrule
\endhead

\midrule
\multicolumn{2}{r}{Continued on next page} \\
\endfoot

\bottomrule
\endlastfoot

\raggedright\arraybackslash Connect the character to the user's life &
\raggedright\arraybackslash AIM uses participants' intake responses to personalize the character and generate a realistic scenario onset grounded in their circumstances. \\
\midrule

\raggedright\arraybackslash Use material familiar to the reader &
\raggedright\arraybackslash Intake questions capture participants' routines and daily lives, giving the agents familiar material to incorporate into each story. \\
\midrule

\raggedright\arraybackslash Cue ``remindings'' of the reader's own experiences &
\raggedright\arraybackslash Intake questions ask participants to describe their own lives, and the scenario is constructed from these responses. \\
\midrule

\raggedright\arraybackslash Use rich mental imagery &
\raggedright\arraybackslash Agent prompts instruct AIM to describe concrete surroundings, actions, and consequences. \\
\midrule

\raggedright\arraybackslash Maintain psychological realism &
\raggedright\arraybackslash The Initializer and Predictor Agents are instructed to be ``very specific and very realistic'' and not to exaggerate, while the scenario onset is constrained to follow from the participant's stated circumstances. \\
\midrule

\raggedright\arraybackslash Keep stories cohesive &
\raggedright\arraybackslash The agent pipeline builds each plot from a scenario onset and its predicted effects, preserving continuity across the narrative. \\
\midrule

\raggedright\arraybackslash Build a clear cause-and-effect structure &
\raggedright\arraybackslash The Predictor and Storyteller Agents construct a plot that connects the scenario to its consequences rather than presenting isolated effects. \\
\midrule

\raggedright\arraybackslash Keep the persuasive message central to the plot &
\raggedright\arraybackslash The Storyteller Agent conditions the narrative arc on the Predictor's forecast effects, so the scenario's consequences drive the plot rather than appearing as commentary on it. \\
\midrule

\raggedright\arraybackslash Show rather than tell &
\raggedright\arraybackslash Agent prompts instruct AIM to convey effects through events and actions rather than stating the persuasive message directly. \\
\midrule

\raggedright\arraybackslash Let the reader supply the emotion &
\raggedright\arraybackslash Autocompleter constraints direct the model to let emotion be inferred from actions and events rather than stated outright and to avoid sentimentality. \\
\midrule

\raggedright\arraybackslash Use plain, accessible language &
\raggedright\arraybackslash Autocompleter constraints require plain, literal language and prohibit metaphor and figurative writing, keeping suggestions close to everyday language. \\
\midrule

\raggedright\arraybackslash Reduce counterarguing by avoiding didactic framing &
\raggedright\arraybackslash AIM never states the policy argument directly. Participants author the story themselves, so its implications are inferred from events rather than asserted by the system. \\

\end{longtable}

\clearpage

\section{Supplementary Information}
\label{sec:supplementary}
\setcounter{table}{0}
\renewcommand{\thetable}{S\arabic{table}}
\renewcommand{\theHtable}{SI.\arabic{table}}
\captionsetup[table]{name=Supplementary Table}
\setcounter{figure}{0}
\renewcommand{\thefigure}{S\arabic{figure}}
\renewcommand{\theHfigure}{SI.\arabic{figure}}
\captionsetup[figure]{name=Supplementary Fig.}

\subsection{Cutoff rationale}
\label{sec:si-familiarity-cutoffs}
\label{sec:si-cutoff}

We pre-registered two cutoffs for low familiarity, $\leq 2$ and $\leq 3$ on the 1--7 familiarity scale, for three reasons.

First, we planned to run multiple studies, so we wanted the same cutoffs on the same 1--7 scale in every study rather than cutoffs chosen for a single dataset. We pre-registered identical cutoffs for Study 1 and Study 2. Second, 2 and 3 were a priori reasonable. A cutoff of $\leq 1$ risks leaving too few participants in the low-familiarity group, and a cutoff of $\leq 4$ places participants at the scale midpoint in the low-familiarity group. That leaves $\leq 2$ and $\leq 3$, so we pre-registered both as sensitivity checks. Both cutoffs approximate common conventions: $\leq 3$ is a midpoint split (all responses below 4), and $\leq 2$ approximates the bottom third of the scale. We did not exclude middle responses, so that every participant entered the analysis. Third, a small pilot on science funding suggested that a cutoff of 3 would leave enough participants in the low-familiarity group.

Note that we also pre-registered a test of familiarity that does not depend on any cutoff. The causal forests treat familiarity as continuous, and we summarize its moderation with the best linear projection (Methods, \autoref{sec:methods-causal-forests}).

\subsection{Narrative Transportation Items}
\label{sec:si-narrative-items}
We adapted the short-form narrative transportation scale~\citep{appel2015transportation} to our setup. Participants rated the extent to which four statements applied to them (1 = not at all, 7 = very much):
``I could picture myself in the scene of the events described in the narrative,''
``I was mentally involved in the narrative while writing it,''
``I wanted to learn how the narrative progressed,'' and
``The narrative affected me emotionally.''
We made three changes to the original scale. First, AIM and placebo participants saw ``writing'' in the second item, while vignette participants saw the original ``reading.'' Second, we replaced ``ended'' with ``progressed'' in the third item. Participants wrote their own stories, so the stories had no fixed ending, but AIM's suggestions could still introduce events participants did not plan. Third, we omitted the scale's two items about named characters, since personalized stories share no common characters. We averaged the four items into a transportation index. 

\subsection{Screening procedure, deviations, complications}
\label{sec:si-screening}

\textbf{Exclusion criteria.} As pre-registered, across both studies we excluded participants who reported serious technical errors, failed the commitment check or attention check, used a mobile device, or completed the study unusually quickly. Serious errors included never receiving an autocomplete, encountering an app error message, or being unable to type; these appeared as predefined response options for participants who reported experiencing a problem. We defined unusually short completion as a duration below the condition median minus three scaled median absolute deviations, using a scaling factor of 1.4826. We also required completion of a pre-task CAPTCHA. We pre-registered a minimum sample size of 900 and recruited additional participants to accommodate anticipated exclusions.

Study 2 included several additional pre-registered criteria. Participants had to meet a minimum writing threshold before proceeding beyond the writing task. Post-task exclusions additionally covered Safari use and reCAPTCHA scores below 0.5 when available.

\textbf{Deviations.} In Study 1, participants who reported a technical problem could select from predefined non-serious and serious errors or describe an unlisted error in an ``Other'' field. We pre-registered manually classifying these custom reports as serious or non-serious. We skipped this review to save time, preserve power, and avoid subjective classifications that would be hard to apply consistently in follow-up studies, so we excluded only participants who selected a predefined serious error. A survey programming error had two consequences. First, we applied the first attention-check exclusion after data collection rather than screening participants out immediately. Second, although both studies required passing a CAPTCHA to advance, we did not record Qualtrics's continuous reCAPTCHA \emph{scores} in Study 1 or for most of Study 2. We therefore applied the reCAPTCHA-score exclusion only to the subset of Study 2 participants for whom scores were recorded. Unlike the pre-task CAPTCHA itself, the reCAPTCHA score did not determine study entry; it provided a continuous estimate of the likelihood that a respondent was a bot. In Study 2, four participants messaged us on Prolific to report significant errors, and we excluded them as well. Finally, the pre-registration discussed pooling Studies 1 and 2 in one causal random forest. We did not pool them because the moderator coefficients correlated only weakly and inconsistently across the two studies ($r$ ranged from $-0.41$ to $0.54$ across the four forests; Supplementary Information~\ref{sec:si-blp-consistency}), so a pooled forest would mislead.

\textbf{Record linkage in Study 1.} Participants entered their Prolific ID in Qualtrics, completed the app intake form (which included again entering their Prolific ID) and writing task, and then returned to Qualtrics. We linked app and survey records using Prolific IDs, assuming participants would enter the same Prolific ID for both the survey and app. However, IDs failed to match for 27 participants in the AIM and placebo conditions. We recovered 22 records by comparing the stories participants copied into Qualtrics, a procedure included prospectively to resolve matching errors. Of the remaining five, we recovered one through manual verification. We retained these participants without matching journal entries in the main analysis but excluded them from analyses requiring journal records (causal random forests, AI-vs-human language).

\textbf{Prolific supplementary data.} Prolific lets researchers download quota-sampling variables that the survey did not ask. We asked age, sex, and ideology (1--7) in Qualtrics but took party ID from this supplementary data. Party ID was unavailable for 6 of 894 participants in Study 2 (0.7\%) and 5 of 1{,}007 in Study 1 (0.5\%). No analysis uses party ID (the causal forests use ideology), so this affects only the party composition in the descriptives.

\textbf{Study 2 interface.} Early in Study 2 we recognized that the longer writing period we had adopted risked badly interacting with the timer's position in the user interface: participants who wrote more may scroll the timer out of view. On 2026-08-22 we changed the timer to stay visible as participants scrolled. The $146$ participants who began on 08-21 wrote under the original timer and the $748$ who began later wrote under the revised one. Dropping the earlier participants left the main results unchanged and no effect $\times$ period interaction was significant (smallest $p = .28$; Supplementary Table~\ref{tab:ruled_out_sticky_timer}).

{\footnotesize
\input{explain_divergence/tables/ruled_out_sticky_timer.tex}
}

\subsection{Participant descriptives by arm}
\label{sec:si-descriptives}
Supplementary Tables~\ref{tab:balance_science} and~\ref{tab:balance_satellite} report pretreatment covariates by arm in each study, with tests of differences across arms and sharpened $q$ values.

\begingroup
\footnotesize
\input{sci_fund_real/tables/balance.tex}
\input{satellite/tables/balance.tex}
\endgroup

\subsection{Differences between the two studies}
\label{sec:si-study-differences}

The two experiments shared the same system, three experimental arms, pretreatment measures, narrative-transportation items, and analysis plan. They differed in the following ways.

\textbf{Topic and scenario.} Study 1 personalized a delay in medical care resulting from cuts to science funding. Study 2 personalized a deliberate attack on satellites that disrupted everyday infrastructure.

\textbf{Outcome measures.} Study 1 measured closeness with two items and attitudes with two items using different response scales. Study 2 used four closeness items and seven attitude sliders organized into three subscales (Methods). We used sliders to reduce ceiling effects and multiple subscales to capture potentially distinct forms of attitude change.

\textbf{Timing.} Study 1 was fielded in April 2026, and Study 2 in August 2026.

\textbf{User experience changes.} In Study 2, we excluded Safari users because internal testing suggested poorer app performance in Safari, consistent with the higher error rate among Safari users in Study 1. We also increased the writing period from six to seven minutes. We also changed the timer so that it remained visible as participants scrolled through the writing interface rather than staying fixed at the top of the page (see SI \autoref{tab:ruled_out_sticky_timer}).

\textbf{System changes.} First, because OpenAI was deprecating GPT-4o, which we used in Study 1, we changed the primary autocomplete model in Study 2 to GPT-4.1-mini. Second, we increased the frequency penalty from 1 to 1.2 and added a presence penalty of 0.8. Both changes were based on feedback from Study 1 of repetition. Third, we changed the generated start of the event to be a minimum of 100 words rather than 30. Fourth, we changed the generated event plot from effects to be 100 words rather than 50. Fifth, we added a constraint to the autocomplete prompt to not be so repetitive: ``Never repeat or paraphrase phrases that already appear in the CONTEXT. Do not re-describe objects, scenes, or actions already mentioned. Each completion must introduce a new action, detail, or event that moves the story forward.''

\subsection{Crystallization}
\label{sec:si-crystallization}

Four measures suggest that attitudes toward science funding were more crystallized than attitudes toward satellite cybersecurity.

\subsubsection{Closeness differences}
\label{sec:si-closeness-differences}

We estimate $\text{Attitudes}_i = \beta_0 + \beta_1\,\text{Closeness}_i + \varepsilon_i$ within each study and arm, and pool the two studies to test the difference:
\begin{equation*}
\text{Attitudes}_i = \beta_0 + \beta_1\,\text{Closeness}_i + \beta_2\,\text{IsStudy2}_i
+ \beta_3\,(\text{Closeness}_i \times \text{IsStudy2}_i) + \varepsilon_i,
\end{equation*}
so $\beta_3$ is the reported interaction.

Closeness converted into attitudes less in Study 1. This held among placebo participants, the cleanest contrast because they wrote about the same unrelated scenarios in both experiments: each standard-deviation increase in closeness predicted a $0.42$ SD difference in attitudes in Study 1 and a $0.59$ SD difference in Study 2 (interaction: $b=0.16$, $95\%$ CI $[0.05, 0.28]$, $p=.007$). Supplementary Table~\ref{tab:conversion} reports every arm.

{\footnotesize
\input{explain_divergence/tables/conversion.tex}
}

\subsubsection{Media Cloud procedure}
\label{sec:si-media-cloud}

We counted U.S. national news stories that mentioned each issue using Media Cloud, an open-source archive of online news~\citep{roberts_media_2021}. We queried its U.S. national collection for stories published between 1 September 2024 and 1 September 2026 that contained the exact phrase ``science funding,'' or either of the exact phrases ``satellite cybersecurity'' and ``space cybersecurity.'' The collection held 8,050,883 stories in that window. Of these, 493 mentioned science funding ($0.61$ per 10,000 stories) and 17 mentioned satellite or space cybersecurity ($0.02$ per 10,000). Adding related phrases did not narrow the gap: including ``research funding,'' ``NIH funding,'' and ``scientific research funding'' raised the science-funding count to 6,579 stories, whereas adding ``satellite hacking'' and ``satellite cyberattack'' found no stories beyond the original 17.

\subsubsection{Ideology differences}
\label{sec:si-ideology-differences}

We coded ideological direction as $-1$ (liberal), $0$ (moderate), or $+1$ (conservative) and ideological strength as the distance from the midpoint of the 7-point ideology scale (0--3). We estimate, separately within each study and sample,
\begin{equation*}
\text{Attitudes}_i = \beta_0 + \beta_1\,\text{IdeoDirection}_i + \beta_2\,\text{IdeoStrength}_i
+ \beta_3\,(\text{IdeoDirection}_i \times \text{IdeoStrength}_i) + \varepsilon_i,
\end{equation*}
and we report $R^2$ with intervals from $5{,}000$ bootstrap resamples drawn within arm. Direction, strength, and their interaction explained more of the variance in science-funding attitudes than in satellite attitudes, among control participants and among all participants (Supplementary Table~\ref{tab:ideology_r2}).

{\footnotesize
\input{explain_divergence/tables/ideology_r2.tex}
}

\subsubsection{Open-ended responses}
\label{sec:si-crystallization-coding}

In the main manuscript, we report how we asked AIM participants what aspect of the treatment was effective or ineffective, though we also asked this of vignette participants. We coded the open-ended responses of AIM and vignette participants for signs of a crystallized attitude; placebo participants did not answer these questions. For each participant we combined the answers to what was effective and what was ineffective, and excluded combined responses of three words or fewer, leaving 664 participants in Study 1 and 600 in Study 2. We sent each response to gpt-4.1-mini at temperature 0, one request per participant, with the prompt below, replacing \texttt{\{topic\}} with ``science funding cuts'' or ``satellite cybersecurity'' and \texttt{\{text\}} with the response. The model judged the two types of crystallization, prior familiarity and
reinforcement, separately, and we coded a response as crystallized if the
model flagged either type. Supplementary Table~\ref{tab:crystallization_types}
reports the share of responses showing each type by study and arm, and
``\nameref{sec:si-crystallization-validation}'' (below) reports the model's
agreement with two human raters.

\begin{figure}[H]
{\footnotesize
\begin{verbatim}
INSTRUCTIONS
You are coding open-ended survey responses from a study about {topic}.
Participants wrote about what was effective and ineffective in a task
meant to make the issue feel personally close. Decide whether the
response shows that the participant's attitude about {topic} was
CRYSTALLIZED: settled before the task, so the task had little room to
move it.

Judge each of the two types below separately. A response can show
neither, one, or both.

Type 1: Prior familiarity
The participant held a settled or strong view on {topic} BEFORE the
study. The view can be supportive or opposed. This type also includes
participants who say they were already aware of or informed about
{topic} or its risks, so the task told them nothing new.
Example signals: "I've always supported this", "I was already
convinced", "my mind was made up", "I've long believed", "I already
cared a lot about this", "I already knew about this", "I was already
aware of these risks", "nothing here was new to me", "I work in this
field, so I knew".

Type 2: Reinforced
The participant explicitly says the task confirmed, reaffirmed, or
strengthened a belief they already held, rather than changing it. If
they describe that belief as vague, weak, or uncertain, Type 2 does
not apply.
Example signals: "it just reinforced what I believed", "it confirmed
what I thought", "it reminded me why I care", "it only strengthened
my view".

Neither type applies when the participant only:
- says "it didn't change my attitude" with no reason, or with a
  reason about the story, the interface, the AI suggestions, or the
  scenario's plausibility rather than a view on {topic}
- doubts the scenario or its plausibility ("this would never happen")
- describes personal EXPERIENCE with the issue without a settled view
  or prior awareness ("I have chronic pain")
- discusses the story, the interface, or the AI suggestions

Return JSON:
{"prior_familiarity": 0 or 1,
 "prior_familiarity_evidence": "<exact quote supporting Type 1, or empty>",
 "reinforced": 0 or 1,
 "reinforced_evidence": "<exact quote supporting Type 2, or empty>"}

Response: """{text}"""
\end{verbatim}
}
\caption{Prompt for LLM coding of crystallized attitudes. We replaced \texttt{\{topic\}} with ``science funding cuts'' or ``satellite cybersecurity'' and \texttt{\{text\}} with the participant's combined open-ended response.}
\label{fig:crystallization-prompt}
\end{figure}

\input{explain_divergence/tables/crystallization_types.tex}

\subsubsection{Validation of the LLM coding}
\label{sec:si-crystallization-validation}

We validated the LLM labels against two human raters. The raters first drafted the codebook together and calibrated it over three rounds before the final validation.

\paragraph{Calibration.} We gave the draft codebook to the LLM and had it code every response, which let us find positive cases for calibration. In each of three rounds, we drew a new set of 20 responses: 10 at random and 10 from responses the LLM had coded as crystallized, so that each set contained enough positive cases to discuss. The two raters coded each set independently, compared their labels, discussed each disagreement, and revised the codebook. After the third round, the raters no longer disagreed on how to apply the codebook, and we fixed it. Rater A then adapted the final codebook into the LLM prompt (Supplementary Figure~\ref{fig:crystallization-prompt}), drawing on suggestions from an AI assistant to add instructions specific to LLM coding. We did not revise the codebook or the prompt after this point. None of the 60 calibration responses appears in the validation sample.

\paragraph{Validation sample.} We drew 200 new responses for validation. We drew 100 at random from all responses. To ensure enough positive cases, we stratified the other 100 on the labels from the draft prompt: 55 responses the draft prompt had coded as crystallized and 45 drawn at random from those it had coded as not crystallized. We shuffled these 100 together so that their position did not reveal how they were drawn. We enriched the sample because crystallized responses are rare: with the final prompt, the LLM coded 10.3\% of all responses as crystallized, so a purely random sample of 200 would contain about 20 positive cases, too few to estimate agreement precisely. Each rater coded all 200 responses for prior familiarity and reinforcement, and we counted a response as crystallized if the rater coded either type. The raters coded independently, did not see the LLM labels, and did not discuss responses during coding.

\paragraph{Agreement.} Supplementary Table~\ref{tab:annotator_agreement} reports Cohen's $\kappa$ for each pair of raters. On all 200 responses, the two human raters agreed with each other ($\kappa=0.83$, 95\% CI $[0.74, 0.90]$) about as well as the LLM agreed with each of them ($\kappa=0.91$, 95\% CI $[0.83, 0.96]$ with Rater A; $\kappa=0.80$, 95\% CI $[0.71, 0.89]$ with Rater B). The LLM agreed more closely with Rater A, who adapted the codebook into the LLM prompt.

Because we selected the stratified cases on an LLM label, which can inflate agreement with the LLM, we also report agreement on the 100 responses drawn at random from all responses, whose class distribution matches the full data. On the 100 random responses, agreement was lower: $\kappa=0.80$, 95\% CI $[0.56, 0.96]$ between the human raters, and $\kappa=0.71$, 95\% CI $[0.39, 0.93]$ and $\kappa=0.66$, 95\% CI $[0.39, 0.86]$ between the LLM and Raters A and B. Raw agreement remained between 93\% and 96\%. Cohen's $\kappa$ falls when one category is rare, and this subset contained only 9 to 13 positive cases per rater, so its intervals are wide and overlap with those on the full sample.

{\footnotesize
\input{explain_divergence/tables/annotator_agreement.tex}
}

\subsection{Alternative explanations for Study 1 and Study 2 attitude differences}
\label{sec:si-attitude-explanations}

We tested five alternative explanations for the larger attitude effects in Study 2.

\subsubsection{UX differences}
\label{sec:si-ux-differences}

Study 2 used an updated interface: a different autocomplete model, stronger repetition penalties, a longer generated scenario, a seven- rather than six-minute writing period, and a minimum word count. The placebo task was identical in the two studies, so placebo participants' system ratings isolate the interface change from the change of scenario; the AIM arm cannot, because its ratings move with both. No rating differed after correction, and four of six were directionally lower in Study 2, the largest being perceived realism ($b=-0.30$, 95\% CI $[-0.57,-0.04]$, $q=.17$; Supplementary Table~\ref{tab:ruled_out_ux}). Participants did not experience the Study 2 system as better, so an improved interface cannot explain its larger attitude effects.

{\footnotesize
\input{explain_divergence/tables/ruled_out_ux.tex}
}

\subsubsection{Composition differences}
\label{sec:si-composition-differences}

If the studies' samples differed in ways that moderate the effect, or if study restrictions mattered, adjusting for them should shrink the difference in attitude effects. We estimate
\begin{multline*}
\text{Attitudes}_i = \beta_0 + \beta_1\,\text{IsVignette}_i + \beta_2\,\text{IsPlacebo}_i
+ \beta_3\,\text{IsStudy2}_i \\
+ \beta_4\,(\text{IsVignette}_i \times \text{IsStudy2}_i)
+ \beta_5\,(\text{IsPlacebo}_i \times \text{IsStudy2}_i)
+ \text{Covariates}_i^{\prime}\gamma + \varepsilon_i,
\end{multline*}
with the AIM arm omitted, so $-\beta_4$ and $-\beta_5$ are the reported gaps. $\text{Covariates}_i$ is empty, the 13 pretreatment covariates, Safari use and story length, or all of these. Adding 13 pretreatment covariates, Safari use, and story length left the difference essentially unchanged against both controls (Supplementary Table~\ref{tab:ruled_out_composition}).

{\footnotesize
\input{explain_divergence/tables/ruled_out_composition.tex}
}

\subsubsection{Measure overlap differences}
\label{sec:si-measure-overlap}

If Study 2 stories drew closer to the content of the attitude items, the attitude items might have been easier to move. We estimate, among AIM participants within each study,
\begin{equation*}
\text{Attitudes}_i = \beta_0 + \beta_1\,\text{ItemSimilarity}_i + \varepsilon_i,
\end{equation*}
where $\text{ItemSimilarity}_i$ is the z-scored cosine similarity between participant $i$'s story and the mean embedding of that study's attitude items. In fact Study 1 stories were more similar to the attitude items than Study 2 stories. Within each study, however, a story's similarity to the items did not significantly predict attitudes (Supplementary Table~\ref{tab:ruled_out_similarity}).

{\footnotesize
\input{explain_divergence/tables/ruled_out_similarity.tex}
}

\subsubsection{Ceiling effects}
\label{sec:si-ceiling-effects}

If Study 1 attitudes were near the top of their scales, ceiling effects could have obscured treatment-induced increases. For example, consider a person who selects ``5'' on a 1-5 scale. They may actually be a 5 or the respondent's true position could be a stronger position that the scale cannot express. OLS cannot distinguish between these cases. Consequently, we refit each attitude item using a two-sided Tobit model, treating responses at either bound as censored. That is, we estimate
\begin{equation*}
\text{Item}^*_i = \beta_0 + \beta_1\,\text{IsAIM}_i + \varepsilon_i, \quad \varepsilon_i \sim N(0, \sigma^2), \quad
\text{Item}_i = \min\{\max(\text{Item}^*_i, L), U\},
\end{equation*}
where $\text{Item}^*_i$ is the unobserved latent response and $L$ and $U$ are the scale bounds (Supplementary Table~\ref{tab:ruled_out_ceiling}). We then z-score the latent outcome and report the treatment effect in standard-deviation units to make it comparable to the OLS estimate.

The results do not provide evidence that ceiling effects suppressed attitude treatment effects in Study 1. For perceived personal benefit of science, the only top-heavy item, the share of participants at the maximum was nearly identical across conditions ($27.4\%$ in placebo and $27.6\%$ in AIM), and the Tobit estimate was identical to the least-squares estimate ($b=0.02$). The preferred NIH budget item was rarely at the upper bound ($0.3\%$ in placebo and $0.6\%$ in AIM), and its estimate changed only from $b=0.12$ under least squares to $b=0.13$ under Tobit. In Study 2, by contrast, AIM increased the share of participants at the scale maximum by $7.0$--$11.2$ percentage points across all seven attitude items, and the Tobit estimates exceeded the least-squares estimates. Thus, ceiling effects do not explain the smaller attitude effects observed in Study 1.

{\footnotesize
\input{explain_divergence/tables/ruled_out_ceiling.tex}
}

\subsubsection{Vignette difference (some support)}
\label{sec:si-vignette-difference}

The two vignettes differed. Narrative transportation predicted attitudes, and against the vignette AIM raised transportation less in Study 1, mostly through the item on being emotionally affected. The difference has to do with the vignettes: Study 1's vignette was more emotionally affecting than Study 2's, whereas AIM stories were about equally emotionally affecting. Giving Study 1 Study 2's effect on emotional engagement closes about a fifth of the difference in the vignette contrast (Supplementary Table~\ref{tab:ruled_out_vignette}). This account applies only to the vignette contrast and does not explain the larger difference against the placebo.

{\footnotesize
\input{explain_divergence/tables/ruled_out_vignette.tex}
}

\subsection{Prior attitudes toward AI}
\label{sec:si-ai-attitudes}

A causal-forest projection shows only that AIM's effect does not \emph{vary} with prior attitude toward AI, not that AIM worked for participants who dislike AI. We therefore estimated the effect directly on each side of the AI-attitude scale midpoint, alongside an explicit arm $\times$ AI-attitude interaction (Supplementary Table~\ref{tab:ai_attitude}). No interaction was significant, and every closeness contrast held on both sides. Co-writing with AIM moved closeness even among participants holding unfavourable views of AI.

{\footnotesize
\input{study_comparison/tables/ai_attitude.tex}
}

\subsection{Mediation Model Selection}
\label{sec:si-mediation-model-selection}

Our model selection procedure---deriving a mediation model that describes how treatment relates to downstream attitudes---combined prior work with consistency with the data. Because we measured transportation, closeness, and attitudes at the same time, we treat our mediation model as a descriptive decomposition, not a causal one.

Unlike the simple closeness-only model, which shows results against both the placebo and the vignette control, for these dual-mediator models we present only the vignette contrast. This is because when placebo participants rated narrative transportation, it was for a topic (meeting a celebrity or having sounds turn into colors) unrelated to the closeness and attitude items, so its relation to downstream attitudes is not as meaningful. That is: $Treatment -> Transportation -> Attitudes$ for placebo participants would be exploring whether narrative transportation into a story \textit{unrelated} to the topic we elicit attitudes about has effects on that topic.

\subsubsection{Serial mediation model}
\label{sec:si-serial-model}

Our primary specification is a serial multiple-mediator model. AIM affects closeness, closeness affects transportation, and both mediators predict attitudes. AIM can therefore reach attitudes through closeness alone, through transportation alone, or through closeness and then transportation. We chose this ordering for two reasons. First, prior studies that measured both psychological distance and narrative transportation also treated psychological distance as preceding transportation~\cite{liu_narrative_2023,morales_riech_role_2025}. Second, the data are consistent with this ordering and inconsistent with the reverse (next section). Because we measured both mediators concurrently, the ordering remains an assumption, and the model is a descriptive decomposition rather than evidence that participants experienced closeness before transportation.

\subsubsection{Considering Directionality}
\label{sec:si-mediation-directionality}
We can use the data to ask which ordering of closeness and transportation is more consistent with the observed associations.

Let $X$ be treatment, $T$ be transportation, and $C$ be closeness. We first consider which of two paths is more consistent with the data:

\textbf{Proposed Path 1}. $X \to T \to C$. Consider that if the true path is $X \to T \to C$, then the effect of $X$ on $C$ should be mediated by $T$---meaning, after controlling for $T$, the effect of $X$ on $C$ should be close to zero.

\textbf{Proposed Path 2}. Conversely, if the true path is $X \to C \to T$, then the effect of $X$ on $T$ should be mediated by $C$: After controlling for $C$, the effect of $X$ on $T$ should be close to zero.

Supplementary Table~\ref{tab:mediation_direction_tests} reports the unadjusted and adjusted treatment coefficient for both proposed chains. Here we find that the coefficient of $X$ on $C$ is large, and controlling for $T$ does not reduce it to zero. That is inconsistent with Path 1. However, the coefficient of $X$ on $T$ goes to zero once we include $C$. That is consistent with Path 2. Thus, the data are inconsistent with the path $X \to T \to C$ and consistent with the path $X \to C \to T$, the ordering of the serial model.

{\tiny
\input{study_comparison/tables/tab_mediation_direction_tests.tex}
}

\subsubsection{Parallel mediation model}
\label{sec:si-parallel-model}

\begin{figure}[H]
\centering
\includegraphics[width=\linewidth]{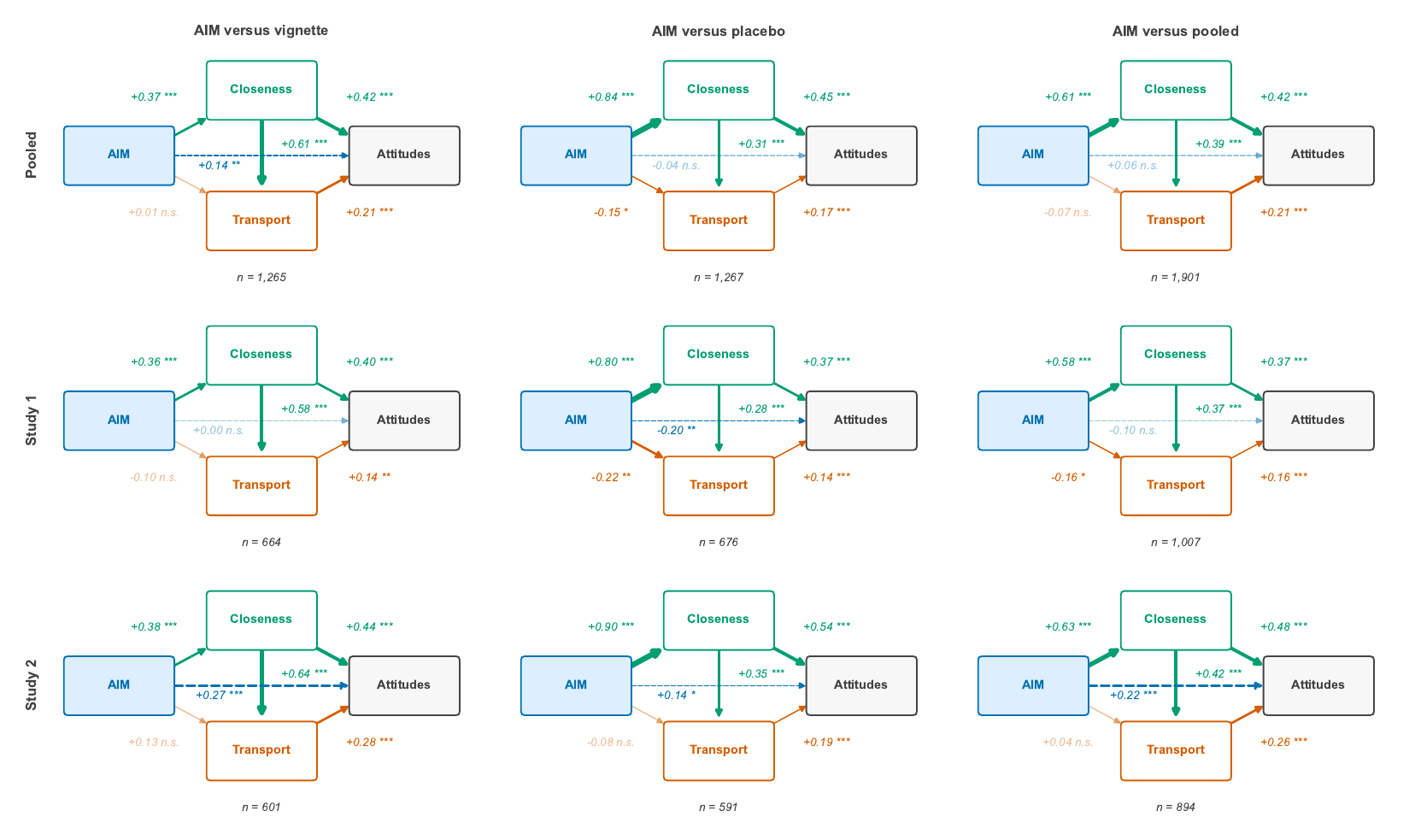}
\caption{Serial multiple-mediator model (primary specification), every study and control comparison. Closeness predicts transportation; see~\S\ref{sec:si-serial-model} for why.}
\label{fig:mediation-grid-serial}
\end{figure}

\begin{figure}[H]
\centering
\includegraphics[width=\linewidth]{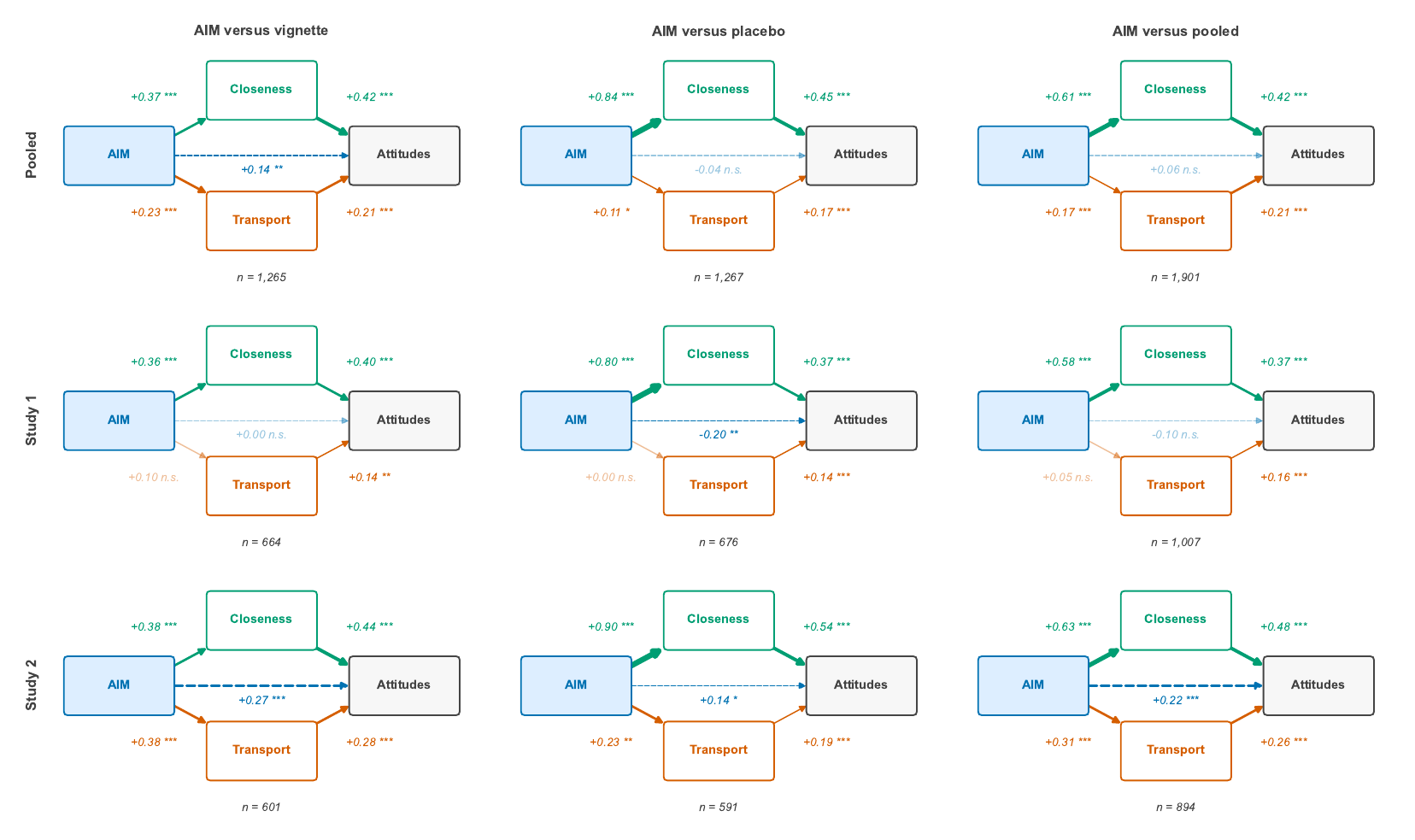}
\caption{Parallel multiple-mediator model (alternative), every study and control comparison.}
\label{fig:mediation-grid-parallel}
\end{figure}

Because the two mediators were measured concurrently, we also fit a parallel model as an alternative. It treats closeness and transportation as correlated mediators: AIM may affect each, each may independently predict attitudes, and their residuals covary freely. The parallel model does not attribute that correlation to a causal path between the mediators. A complication is that both models fit exactly as many parameters as there are observed moments, so fit statistics alone cannot adjudicate between the parallel model (with residuals free to covary) and the serial model. However, the central conclusion holds under either model, in every study, and in every comparison: closeness carried more of the treatment's effect on attitudes than narrative transportation (\autoref{fig:mediation-grid-serial} and \autoref{fig:mediation-grid-parallel}).

\subsection{Qualitative themes}
\label{sec:si-qualitative-themes}
\begingroup
\footnotesize
\setlength{\tabcolsep}{3pt}
\input{sci_fund_real/tables/tab_qualitative_themes.tex}
\input{satellite/tables/tab_qualitative_themes_satellite.tex}
\endgroup
\subsection{Linguistic classification performance}
\label{sec:si-linguistic-classification}
\input{study_comparison/tables/linguistic_classifier.tex}
\clearpage
\subsection{Satellite attitude measurement}
\label{sec:si-satellite-cfa}
\input{satellite/tables/satellite_cfa.tex}

\subsection{Experiment event configurations}
\label{sec:si-event-configs}
The two boxes below reproduce the treatment events from the YAML configuration files we deployed in each experiment. Each \texttt{additional\_fields} entry adds an intake question; AIM folds every answer into the participant's character description. Square-bracket tokens such as \texttt{[state]} refer to these answers. Text inside \texttt{[Tip: ...]} instructs the agents.

\begin{lstlisting}[style=yaml,title={Study 1 (science funding) treatment event.}]
event:
  event_101:
    name: "You need to see a specialist for your health concern, but the earliest appointment is 14 months away. When you ask why, the scheduler explains that their hospital's fellowship training program just lost its NIH T32 training grant—they're now training 40% fewer specialists than before. This is creating a severe shortage, and wait times keep getting longer. Your symptoms are progressing while you wait. [Tip: CRITICAL - You MUST explicitly name the specific specialist type based on health_concern (e.g., 'neurologists' for Neurological Disorders, 'oncologists' for cancer, 'cardiologists' for heart conditions). The story should start with the character calling to schedule an appointment and learning about the 14-month wait. Show the scheduler explicitly explaining the connection: 'We used to train 20 fellows a year, but after the NIH cut our T32 grant, we can only train 12 now. That means fewer specialists graduating, and our wait times have tripled.' Then show the concrete impact as symptoms worsen over the following weeks/months—specific symptoms getting harder to manage, daily activities becoming more difficult, the character trying to cope while waiting. Ground the progression in their typical_friday_afternoon routine and show how worsening symptoms conflict with their stress_management practices. Focus on small, specific, concrete physical changes that tangibly affect daily life—tremors making it harder to write, fatigue making stairs exhausting, pain disrupting sleep. Show the character's attempts to manage and adapt. Do NOT describe emotions directly—show them through actions, dialogue, and physical details.]"
    autocomplete_context: "plot_from_effects"
    shown_to_user: "start"
    shown_to_user_prefix: "Imagine this happens"
    editor_placeholder: "You're calling to schedule a specialist appointment for your health concern. Start writing about making that call and learning about the wait time. A writing assistant will offer suggestions. Press tab to accept or any other key to reject. Some ways to start: 'I dialed the number and waited as it rang. When someone answered, I said [assistant will autocomplete]...'"
    effects:
    additional_fields:
      health_concern:
        type: "SelectField"
        label: "What's your greatest health concern?"
        choices: [
          "Aging and Alzheimer's Disease",
          "Allergies and Infectious Diseases",
          "Arthritis, Musculoskeletal, and Skin Conditions",
          "Cancer",
          "Child and Maternal Health",
          "Dental and Oral Health",
          "Diabetes, Digestive, and Kidney Diseases",
          "Environmental Health",
          "Genetic and Genomic Conditions",
          "Hearing, Balance, and Communication",
          "Heart, Lung, and Blood Conditions",
          "Mental Health",
          "Neurological Disorders and Stroke",
          "Substance Use and Addiction",
          "Vision and Eye Health",
        ]
        validators: [ "DataRequired" ]
      health_relationship:
        type: "SelectField"
        label: "How does this health concern relate to you personally?"
        choices: [
          "It is a current challenge I am actively managing.",
          "It is a potential future challenge I worry about.",
          "It is a past experience I have dealt with."
        ]
        validators: [ "DataRequired" ]

      financial_situation:
        type: "SelectField"
        label: "Which of the following statements best describes your current financial situation?"
        choices: [
          "More than enough: I can meet routine expenses and have the financial freedom to handle unexpected expenses with ease.",
          "Just enough: I can meet routine expenses but feel I do not have the financial freedom to handle unexpected expenses with ease.",
          "Strained: I often worry about making ends meet."
        ]
        validators: [ "DataRequired" ]
      stress_management:
        type: "TextAreaField"
        label: "After a long or stressful day, how do you typically unwind? Describe what that looks like for you."
        validators: [ "DataRequired" ]
      typical_friday_afternoon:
        type: "TextAreaField"
        label: "Can you describe your typical Friday afternoon? For example, what errands or work activities would you be doing. Please be as specific as possible."
        validators: [ "DataRequired" ]
\end{lstlisting}

\begin{lstlisting}[style=yaml,title={Study 2 (satellite cybersecurity) treatment event.}]
event:
  event_satellite_hack:
    name: >
      You live in [metro_type] in [state]. A coordinated cyberattack
      breaches the ground control systems of a major satellite network,
      disabling GPS-linked services, satellite internet, and
      satellite-relayed communications across much of the country.
      [daily_dependency] stops working without warning. Officials say
      it will take days, not hours, to restore service.
      [Tip: 1) THE FIRST HOURS: Open on an ordinary day as
      [daily_dependency] suddenly stops working. Ground it in the
      moment -- what they were doing, what breaks first. Do not
      foreshadow. 2) THE CAUSE (state once): Via a news alert or
      official statement, reveal that attackers breached a
      satellite ground station through a VPN vulnerability and deployed
      malware that wiped the connected systems. Make clear the damage
      is more than lost GPS: satellites also broadcast the precise
      timing signals that keep banks and power grids synchronized, so
      payment systems and utilities are failing too, not just
      navigation. State this once and never repeat it. Not a solar
      storm, not routine maintenance. 3) DAY ONE: Show [cash_situation]
      becoming a real constraint -- card readers rejecting transactions
      because payment networks cannot synchronize, ATMs down. Show them
      trying to reach [person_to_reach] about [reason_to_reach] and
      failing -- dropped calls, no signal. Let wider chaos arrive only
      as overheard fragments: a radio mentioning stalled traffic, a
      headline about banks limiting withdrawals. Do not narrate the
      national picture directly. 4) THE UNRESOLVED STRETCH: Move into
      days two through five. Show active adaptation -- rationing cash,
      navigating without GPS, checking on [person_to_reach] again,
      still unable to resolve [reason_to_reach]. Things escelate in background:
      rolling power outages, empty shelves --- glimpsed,
      not explained. Officials say restoration will take longer than
      first stated. End on a complete sentence with the disruption
      ongoing -- no fix, no rescue, no hopeful pivot. RULES: never
      name emotions -- reveal them through action; no metaphors; do not
      let money alone solve it, since the systems themselves are down;
      do not deliver the cascading effects as a news-anchor summary --
      only as things the character personally overhears or is told.]
    effects:
    autocomplete_context: "plot_from_effects"
    shown_to_user: "start"
    shown_to_user_prefix: "Imagine this is happening to you right now"
    editor_placeholder: >
      Something you rely on every day just stopped working, and it is not a normal outage -- a cyberattack has
      knocked out satellite systems, and because payment networks cannot verify timestamps, card readers and ATMs are
      failing along with GPS and  internet service. Describe the first thing you notice, or the first thing you try. A
      writing assistant will offer suggestions when you pause -- press tab to accept or keep typing to ignore. Continue
      from where the story left off. What happens next?

    additional_fields:
      state:
        type: "StringField"
        label: "What U.S. state do you live in?"
        validators: [ "DataRequired" ]

      metro_type:
        type: "SelectField"
        label: "Where do you live?"
        choices:
          - "a big city"
          - "a suburb"
          - "a small town"
          - "a rural area"
        validators: [ "DataRequired" ]

      daily_dependency:
        type: "SelectField"
        label: "Which of these do you rely on most in a normal day?"
        choices:
          - "using GPS or a maps app to get somewhere"
          - "paying with a card or getting cash from an ATM"
          - "checking weather or traffic before heading out"
          - "a work task that depends on GPS or satellite data"
          - "staying in touch with someone through my phone or the internet"
        validators: [ "DataRequired" ]

      cash_situation:
        type: "SelectField"
        label: "If card readers and ATMs stopped working for several days, how much cash would you have on hand?"
        choices:
          - "enough for several days"
          - "enough for about a day"
          - "almost none"
        validators: [ "DataRequired" ]

      person_to_reach:
        type: "SelectField"
        label: "Who would you most want to reach if phone and internet service became unreliable for days?"
        choices:
          - "a parent or older relative"
          - "my partner or spouse"
          - "my child or someone I care for"
          - "a close friend"
          - "a coworker or boss"
        validators: [ "DataRequired" ]

      reason_to_reach:
        type: "StringField"
        label: "What's something you'd urgently need to coordinate with them about? Answer as a short phrase (e.g., 'making sure they have supplies', 'picking up my kids', 'checking they're safe', 'covering a shift at work')."
        validators: [ "DataRequired", { "Length": { "max": 70 } } ]
\end{lstlisting}
\clearpage
\subsection{Causal-forest heterogeneity analyses}
\label{sec:si-moderators}
We report causal-forest results for both outcomes and both control comparisons in
each study: four tables of best linear projections and two figures for each
combination. The conventions below apply to every table and figure in this
section. Each table reports the best linear projection of the conditional average
treatment-effect function onto the moderator set. These estimates are
exploratory. We used two-sided tests and HC3 standard errors, and standardized
continuous moderators within each sample.

\textbf{Moderators and samples.} Both comparisons included familiarity, age,
mental imagery, three IRI subscales, journaling and creative-writing frequency,
education, income, ideology, AI attitudes, and a male-gender
indicator. The placebo comparison added completions received, story length,
suggestion acceptance rate, and proportion of AI-origin words. In Study 1, the
complete-case samples comprised $n=644$ participants for AIM versus placebo and
$638$ for AIM versus vignette; in Study 2, $n=564$ and $579$. We used
complete cases, requiring all covariates (and, for the placebo comparison, journal entries) to be available, and excluded participants outside the male and female
categories we used to code gender. These restrictions applied only to the causal
forests, not to the primary analyses.

\textbf{Figures.} Best linear projection figures plot the projection
coefficients with their $95\%$ intervals. Red marks $p<.05$ and blue marks
$p\geq.05$. A positive coefficient indicates a larger AIM-minus-control effect as
the moderator increases. Conditional effect figures plot out-of-bag CATE
predictions against each standardized numeric moderator, with a LOESS smooth of
span $0.75$. The shaded bands describe uncertainty in the smooth, not confidence
intervals for individual treatment effects. Moderators taking fewer than eight
values appear as box plots with the median, the interquartile range, and whiskers
extending to $1.5$ times that range.


\newcommand{\forestfigure}[5]{
\begin{figure}[H]\centering
\includegraphics[width=\linewidth,height=0.76\textheight,keepaspectratio]{#1/#2_treat_vs_#3_#4.pdf}
\caption{#5}
\end{figure}}
\newcommand{\forestblock}[4]{
\clearpage
\input{#2_#4.tex}
\forestfigure{#1}{blp}{#4}{closeness_index}{Best linear projections: #3, psychological closeness, AIM versus #4.}
\forestfigure{#1}{blp}{#4}{attitude_index}{Best linear projections: #3, policy attitudes, AIM versus #4.}
\forestfigure{#1}{loess}{#4}{closeness_index}{Conditional effects: #3, psychological closeness, AIM versus #4.}
\forestfigure{#1}{loess}{#4}{attitude_index}{Conditional effects: #3, policy attitudes, AIM versus #4.}}
\clearpage
\input{study_comparison/tables/grf_sci_fund_real_placebo.tex}
\forestfigure{sci_fund_real/plots}{blp}{placebo}{closeness_index}{Best linear projections: Study 1: Science funding, psychological closeness, AIM versus placebo.}
\forestfigure{sci_fund_real/plots}{blp}{placebo}{attitude_index}{Best linear projections: Study 1: Science funding, policy attitudes, AIM versus placebo.}
\forestfigure{sci_fund_real/plots}{loess}{placebo}{closeness_index}{Conditional effects: Study 1: Science funding, psychological closeness, AIM versus placebo.}
\forestfigure{sci_fund_real/plots}{loess}{placebo}{attitude_index}{Conditional effects: Study 1: Science funding, policy attitudes, AIM versus placebo.}
\clearpage
\input{study_comparison/tables/grf_sci_fund_real_vignette.tex}
\forestfigure{sci_fund_real/plots}{blp}{vignette}{closeness_index}{Best linear projections: Study 1: Science funding, psychological closeness, AIM versus vignette.}
\forestfigure{sci_fund_real/plots}{blp}{vignette}{attitude_index}{Best linear projections: Study 1: Science funding, policy attitudes, AIM versus vignette.}
\forestfigure{sci_fund_real/plots}{loess}{vignette}{closeness_index}{Conditional effects: Study 1: Science funding, psychological closeness, AIM versus vignette.}
\forestfigure{sci_fund_real/plots}{loess}{vignette}{attitude_index}{Conditional effects: Study 1: Science funding, policy attitudes, AIM versus vignette.}
\clearpage
\input{study_comparison/tables/grf_satellite_placebo.tex}
\forestfigure{satellite/plots}{blp}{placebo}{closeness_index}{Best linear projections: Study 2: Satellite cybersecurity, psychological closeness, AIM versus placebo.}
\forestfigure{satellite/plots}{blp}{placebo}{attitude_index}{Best linear projections: Study 2: Satellite cybersecurity, policy attitudes, AIM versus placebo.}
\forestfigure{satellite/plots}{loess}{placebo}{closeness_index}{Conditional effects: Study 2: Satellite cybersecurity, psychological closeness, AIM versus placebo.}
\forestfigure{satellite/plots}{loess}{placebo}{attitude_index}{Conditional effects: Study 2: Satellite cybersecurity, policy attitudes, AIM versus placebo.}
\clearpage
\input{study_comparison/tables/grf_satellite_vignette.tex}
\forestfigure{satellite/plots}{blp}{vignette}{closeness_index}{Best linear projections: Study 2: Satellite cybersecurity, psychological closeness, AIM versus vignette.}
\forestfigure{satellite/plots}{blp}{vignette}{attitude_index}{Best linear projections: Study 2: Satellite cybersecurity, policy attitudes, AIM versus vignette.}
\forestfigure{satellite/plots}{loess}{vignette}{closeness_index}{Conditional effects: Study 2: Satellite cybersecurity, psychological closeness, AIM versus vignette.}
\forestfigure{satellite/plots}{loess}{vignette}{attitude_index}{Conditional effects: Study 2: Satellite cybersecurity, policy attitudes, AIM versus vignette.}
\clearpage

\subsection{Consistency of moderators across studies}
\label{sec:si-blp-consistency}
\begin{figure}[H]
\centering
\includegraphics[width=\linewidth]{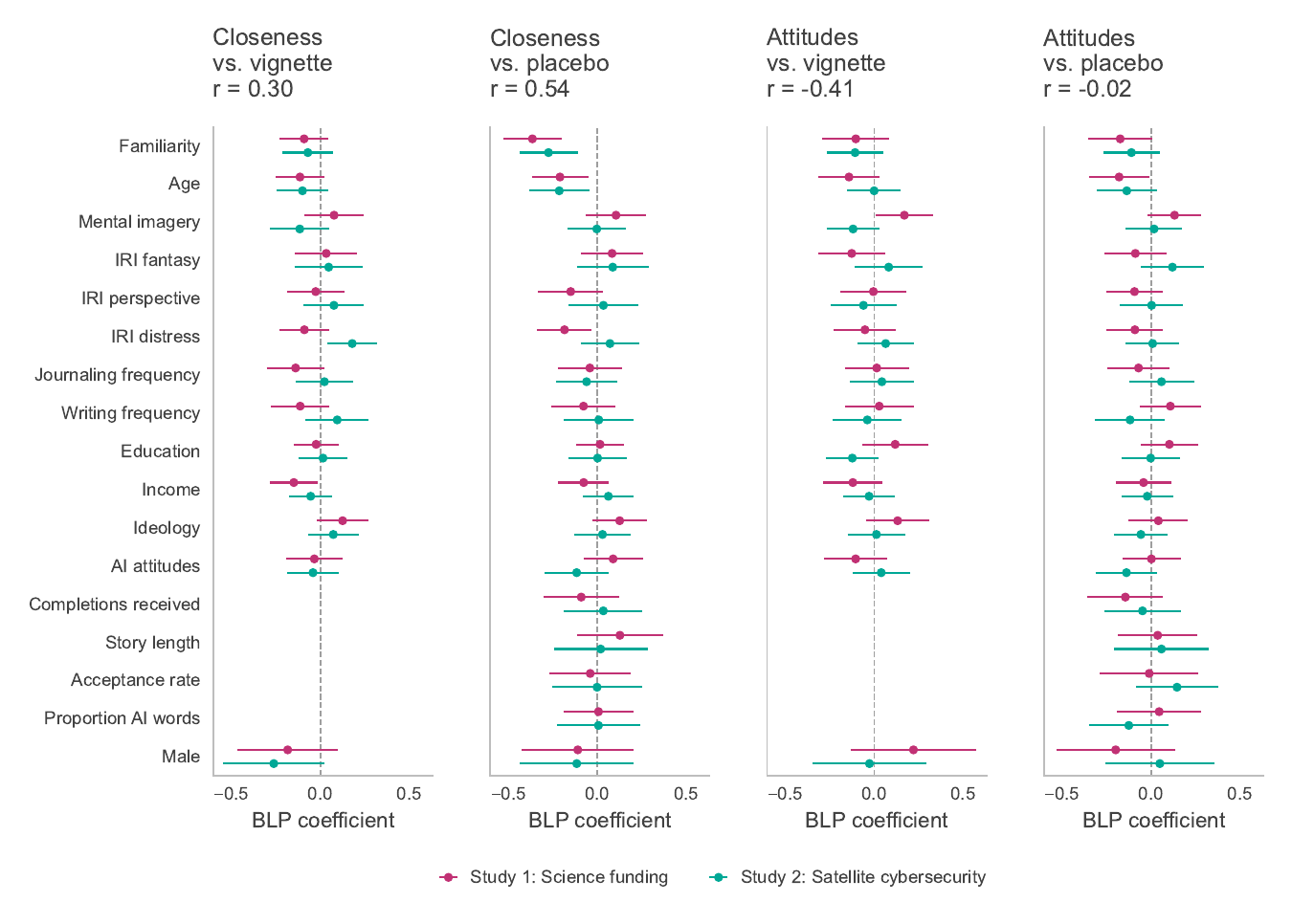}
\caption{Best linear projection coefficients in both studies. Points show each moderator's coefficient with its $95\%$ confidence interval for Study 1 (science funding) and Study 2 (satellite cybersecurity). Moderators are standardized within study. Each panel is one outcome and AIM--control comparison, and $r$ is the correlation between the two studies' coefficients across that panel's moderators. The vignette comparison omits the four app-usage moderators, which only AIM and placebo participants have.}
\label{fig:blp-cross-study}
\end{figure}

To check whether the same moderators shaped AIM's effects in both studies, we plotted each moderator's best linear projection coefficient from Study 1 next to its coefficient from Study 2 (Supplementary Figure~\ref{fig:blp-cross-study}). We standardized moderators within each study, so the coefficients are comparable. For each of the four forests, we also correlated the two studies' coefficients across moderators.

\clearpage

\subsection{Sensitivity of the mediation analyses}
\label{sec:si-mediation-sensitivity}
\begin{figure}[H]
\centering
\includegraphics[width=0.82\linewidth]{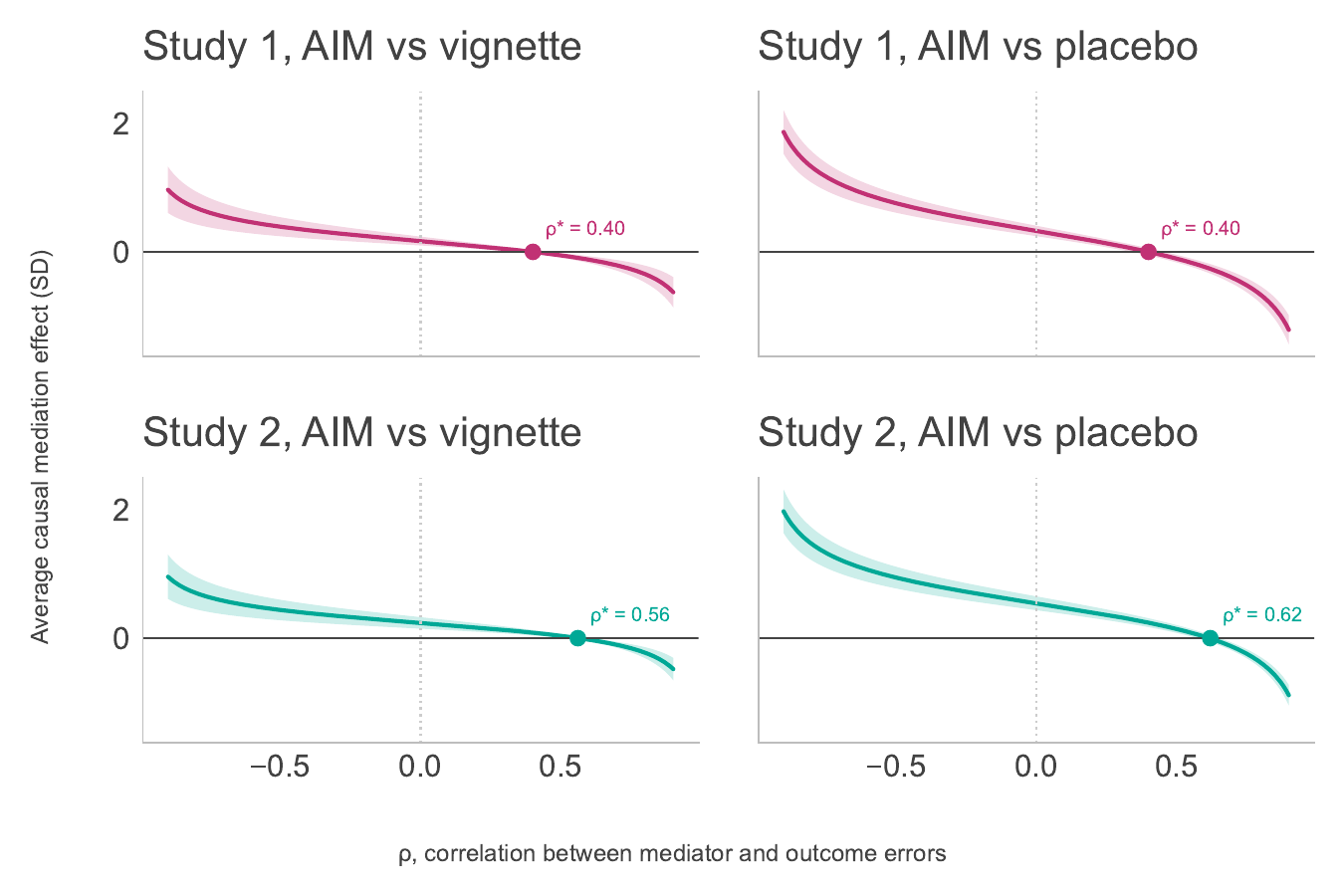}
\caption{Sensitivity of the pre-registered mediation analyses to unmeasured confounding. Curves show the indirect effect as a function of $\rho$, the mediator--outcome error correlation~\citep{imai_identification_2010}, with $95\%$ intervals; points mark $\rho^{*}$, where it crosses zero. Estimated with the R package \texttt{mediation} and 1,000 simulations}
\label{fig:mediation-sensitivity}
\end{figure}

\end{document}

%% file: study_comparison/tables/descriptive.tex
\begin{table}
\caption{Outcome means (SD) by study and condition. Bold rows are the two outcome indices, z-scored within study (mean 0, SD 1). Indented rows are the items that form each index, in their original units; the response scale follows each item. Study 2's attitude index averages three subscales (importance, spending, action).}
\label{tab:main_effects_desc}
\begin{tabular}{lrrr}
\toprule
Measure & AIM & Vignette & Placebo \\
\midrule
\addlinespace[4pt]\textbf{Study 1: Science funding} & \textit{n}=333 & \textit{n}=331 & \textit{n}=343 \\
\textbf{Closeness index} & 0.40 (0.79) & 0.04 (0.95) & -0.40 (1.09) \\
\hspace{1em}Effects on me (1--9) & 7.55 (1.63) & 6.66 (2.03) & 5.73 (2.37) \\
\hspace{1em}Relevance to my life (1--9) & 7.32 (1.83) & 6.74 (2.02) & 5.84 (2.29) \\
\textbf{Attitude index} & 0.07 (1.00) & -0.09 (1.06) & -0.02 (0.95) \\
\hspace{1em}Personal benefit of U.S. science (1--5) & 3.84 (0.97) & 3.73 (1.01) & 3.83 (0.97) \\
\hspace{1em}Preferred NIH budget share (0--100\%) & 19.42 (22.48) & 17.16 (21.99) & 16.77 (20.21) \\
\addlinespace[4pt]\textbf{Study 2: Satellite cybersecurity} & \textit{n}=298 & \textit{n}=303 & \textit{n}=293 \\
\textbf{Closeness index} & 0.42 (0.78) & 0.04 (0.92) & -0.47 (1.08) \\
\hspace{1em}Concreteness (1--9) & 6.53 (2.04) & 6.17 (2.04) & 5.02 (2.30) \\
\hspace{1em}Effects on daily life (1--9) & 7.94 (1.37) & 6.96 (1.91) & 6.03 (2.22) \\
\hspace{1em}Imagining effects (1--9) & 7.49 (1.70) & 6.85 (1.95) & 5.88 (2.25) \\
\hspace{1em}Relevance to my life (1--9) & 7.39 (1.80) & 6.57 (2.08) & 5.76 (2.30) \\
\textbf{Attitude index} & 0.40 (0.86) & -0.14 (1.02) & -0.26 (0.98) \\
\hspace{1em}\textit{Importance subscale} &  &  &  \\
\hspace{2em}Personal importance (0--100) & 83.08 (18.89) & 71.94 (23.84) & 68.59 (25.44) \\
\hspace{2em}National importance (0--100) & 75.65 (21.73) & 63.83 (26.47) & 62.26 (26.59) \\
\hspace{1em}\textit{Spending subscale} &  &  &  \\
\hspace{2em}Spending direction (0--100) & 74.30 (16.37) & 67.11 (18.50) & 65.67 (17.60) \\
\hspace{2em}State and local grants (0--100) & 74.15 (18.47) & 62.32 (22.29) & 61.88 (21.25) \\
\hspace{1em}\textit{Action subscale} &  &  &  \\
\hspace{2em}Vote against candidate (0--100) & 71.58 (24.45) & 62.09 (26.81) & 57.65 (24.25) \\
\hspace{2em}Vote against ballot measure (0--100) & 73.78 (24.63) & 65.25 (27.37) & 60.73 (25.70) \\
\hspace{2em}Oppose politician (0--100) & 69.59 (25.73) & 58.67 (30.03) & 56.68 (26.71) \\
\bottomrule
\end{tabular}
\end{table}

%% file: study_comparison/tables/primary.tex
\begin{table}
\caption{Primary treatment effects. AIM-minus-control estimates from condition-only OLS models. Outcomes are the closeness and attitude indices, z-scored within study, so $b$ is in standard deviations; Extended Data Table~\ref{tab:main_effects_desc} lists the items in each index.}
\label{tab:primary}
\begin{tabular}{lllrrrrrr}
\toprule
Study & Outcome & Control & $n$ & $b$ & $SE$ & $z$ & 95\% CI & $p$ \\
\midrule
Science & Closeness & vignette & 1007 & 0.356 & 0.068 & 5.271 & [0.224, 0.489] & $1.36\times10^{-7}$ \\
Science & Closeness & placebo & 1007 & 0.801 & 0.073 & 10.956 & [0.657, 0.944] & $6.25\times10^{-28}$ \\
Science & Attitudes & vignette & 1007 & 0.160 & 0.080 & 2.000 & [0.003, 0.317] & $.0455$ \\
Science & Attitudes & placebo & 1007 & 0.098 & 0.075 & 1.308 & [-0.049, 0.244] & $.191$ \\
Satellite & Closeness & vignette & 894 & 0.379 & 0.069 & 5.468 & [0.243, 0.515] & $4.55\times10^{-8}$ \\
Satellite & Closeness & placebo & 894 & 0.896 & 0.078 & 11.545 & [0.744, 1.048] & $7.8\times10^{-31}$ \\
Satellite & Attitudes & vignette & 894 & 0.544 & 0.077 & 7.056 & [0.393, 0.695] & $1.72\times10^{-12}$ \\
Satellite & Attitudes & placebo & 894 & 0.666 & 0.076 & 8.748 & [0.517, 0.816] & $2.18\times10^{-18}$ \\
\bottomrule
\end{tabular}
\end{table}

%% file: study_comparison/tables/familiarity_2.tex
\begin{table}
\caption{Familiarity split at 2. AIM-minus-control estimates with $95\%$ confidence intervals.}
\label{tab:familiarity2}
\begin{tabular}{llllrrr}
\toprule
study & outcome & sample & control & n & Effect [95\% CI] & $p$ \\
\midrule
Science & Closeness & Low & vignette & 488 & 0.55 [0.35, 0.76] & $2.00\times 10^{-7}$ \\
Science & Closeness & Low & placebo & 488 & 1.09 [0.87, 1.31] & $1.65\times 10^{-22}$ \\
Science & Closeness & High & vignette & 519 & 0.16 [0.00, 0.31] & $.049$ \\
Science & Closeness & High & placebo & 519 & 0.52 [0.35, 0.70] & $4.44\times 10^{-9}$ \\
Science & Attitudes & Low & vignette & 488 & 0.32 [0.11, 0.54] & $.003$ \\
Science & Attitudes & Low & placebo & 488 & 0.21 [0.00, 0.42] & $.047$ \\
Science & Attitudes & High & vignette & 519 & -0.00 [-0.23, 0.22] & $.969$ \\
Science & Attitudes & High & placebo & 519 & -0.01 [-0.22, 0.19] & $.900$ \\
Satellite & Closeness & Low & vignette & 425 & 0.40 [0.19, 0.62] & $1.92\times 10^{-4}$ \\
Satellite & Closeness & Low & placebo & 425 & 1.22 [0.98, 1.46] & $4.03\times 10^{-24}$ \\
Satellite & Closeness & High & vignette & 469 & 0.37 [0.19, 0.55] & $5.99\times 10^{-5}$ \\
Satellite & Closeness & High & placebo & 469 & 0.62 [0.43, 0.80] & $7.26\times 10^{-11}$ \\
Satellite & Attitudes & Low & vignette & 425 & 0.68 [0.46, 0.90] & $2.01\times 10^{-9}$ \\
Satellite & Attitudes & Low & placebo & 425 & 0.84 [0.63, 1.06] & $3.49\times 10^{-14}$ \\
Satellite & Attitudes & High & vignette & 469 & 0.47 [0.27, 0.68] & $4.30\times 10^{-6}$ \\
Satellite & Attitudes & High & placebo & 469 & 0.53 [0.34, 0.73] & $9.62\times 10^{-8}$ \\
\bottomrule
\end{tabular}
\end{table}

%% file: study_comparison/tables/familiarity_3.tex
\begin{table}
\caption{Familiarity split at 3. AIM-minus-control estimates with $95\%$ confidence intervals.}
\label{tab:familiarity3}
\begin{tabular}{llllrrr}
\toprule
study & outcome & sample & control & n & Effect [95\% CI] & $p$ \\
\midrule
Science & Closeness & Low & vignette & 662 & 0.47 [0.30, 0.65] & $1.30\times 10^{-7}$ \\
Science & Closeness & Low & placebo & 662 & 1.00 [0.82, 1.19] & $5.50\times 10^{-27}$ \\
Science & Closeness & High & vignette & 345 & 0.15 [-0.03, 0.34] & $.100$ \\
Science & Closeness & High & placebo & 345 & 0.36 [0.17, 0.56] & $2.81\times 10^{-4}$ \\
Science & Attitudes & Low & vignette & 662 & 0.23 [0.05, 0.41] & $.012$ \\
Science & Attitudes & Low & placebo & 662 & 0.15 [-0.02, 0.31] & $.087$ \\
Science & Attitudes & High & vignette & 345 & 0.05 [-0.25, 0.34] & $.763$ \\
Science & Attitudes & High & placebo & 345 & -0.02 [-0.30, 0.26] & $.884$ \\
Satellite & Closeness & Low & vignette & 584 & 0.39 [0.21, 0.57] & $1.90\times 10^{-5}$ \\
Satellite & Closeness & Low & placebo & 584 & 1.05 [0.86, 1.24] & $9.21\times 10^{-28}$ \\
Satellite & Closeness & High & vignette & 310 & 0.39 [0.19, 0.60] & $2.12\times 10^{-4}$ \\
Satellite & Closeness & High & placebo & 310 & 0.58 [0.35, 0.80] & $8.39\times 10^{-7}$ \\
Satellite & Attitudes & Low & vignette & 584 & 0.63 [0.44, 0.81] & $7.44\times 10^{-11}$ \\
Satellite & Attitudes & Low & placebo & 584 & 0.74 [0.56, 0.92] & $1.48\times 10^{-15}$ \\
Satellite & Attitudes & High & vignette & 310 & 0.49 [0.25, 0.73] & $7.10\times 10^{-5}$ \\
Satellite & Attitudes & High & placebo & 310 & 0.53 [0.29, 0.77] & $1.50\times 10^{-5}$ \\
\bottomrule
\end{tabular}
\end{table}

%% file: study_comparison/tables/mediation.tex
\begin{table}
\caption{Single-mediator models by study and control. Indirect effects use bias-corrected bootstrap intervals.}
\label{tab:mediation}
\begin{tabular}{llrlr}
\toprule
study & control & n & path & Effect [95\% CI] \\
\midrule
Science & vignette & 664 & Total & 0.160 [0.003, 0.317] \\
Science & vignette & 664 & Direct & -0.009 [-0.156, 0.138] \\
Science & vignette & 664 & Indirect & 0.169 [0.100, 0.247] \\
Science & placebo & 676 & Total & 0.098 [-0.049, 0.245] \\
Science & placebo & 676 & Direct & -0.229 [-0.375, -0.083] \\
Science & placebo & 676 & Indirect & 0.327 [0.247, 0.421] \\
Satellite & vignette & 601 & Total & 0.544 [0.392, 0.695] \\
Satellite & vignette & 601 & Direct & 0.307 [0.178, 0.435] \\
Satellite & vignette & 601 & Indirect & 0.237 [0.147, 0.336] \\
Satellite & placebo & 591 & Total & 0.666 [0.517, 0.816] \\
Satellite & placebo & 591 & Direct & 0.123 [-0.007, 0.254] \\
Satellite & placebo & 591 & Indirect & 0.543 [0.431, 0.660] \\
\bottomrule
\end{tabular}
\end{table}

%% file: satellite/tables/satellite_subscales.tex
\begin{table}
\caption{Secondary satellite attitude subscales. AIM-minus-control estimates from three-arm OLS models with HC1-robust intervals and two-sided tests.}
\label{tab:subscales}
\begin{tabular}{llrrrrr}
\toprule
Subscale & Control & $n$ & $b$ & 95\% CI & $z$ & $p$ \\
\midrule
Importance & Vignette & 894 & 0.495 & [0.349, 0.642] & 6.636 & $3.22\times10^{-11}$ \\
Importance & Placebo & 894 & 0.603 & [0.450, 0.756] & 7.732 & $1.06\times10^{-14}$ \\
Spending & Vignette & 894 & 0.508 & [0.354, 0.663] & 6.453 & $1.09\times10^{-10}$ \\
Spending & Placebo & 894 & 0.562 & [0.412, 0.713] & 7.306 & $2.75\times10^{-13}$ \\
Action & Vignette & 894 & 0.407 & [0.249, 0.566] & 5.030 & $4.89\times10^{-7}$ \\
Action & Placebo & 894 & 0.563 & [0.414, 0.712] & 7.399 & $1.37\times10^{-13}$ \\
\bottomrule
\end{tabular}
\end{table}

%% file: explain_divergence/tables/ruled_out_sticky_timer.tex
\begin{table}[H]
\caption{Sensitivity of the Study 2 effects to the timer change of 2026-08-22, shown on all participants and again excluding the 146 who began before it. Estimates are AIM-minus-control differences in SDs with HC1 confidence intervals; the last column is the effect $\times$ period interaction, the formal test of whether the effect differs across the change.}
\label{tab:ruled_out_sticky_timer}
\begin{tabular}{llllll}
\toprule
Outcome & Control & \shortstack{All participants\\$b$ [95\% CI]} & \shortstack{Excluding 08-21\\$b$ [95\% CI]} & \shortstack{Effect $\times$ period\\$b$ [95\% CI]} & $p$ \\
\midrule
Closeness & Placebo & 0.90 [0.74, 1.05] & 0.86 [0.70, 1.03] & -0.23 [-0.63, 0.18] & .280 \\
Closeness & Vignette & 0.38 [0.24, 0.51] & 0.42 [0.27, 0.57] & 0.20 [-0.16, 0.55] & .276 \\
Attitudes & Placebo & 0.67 [0.52, 0.82] & 0.66 [0.49, 0.82] & -0.09 [-0.45, 0.28] & .645 \\
Attitudes & Vignette & 0.54 [0.39, 0.69] & 0.55 [0.38, 0.72] & 0.00 [-0.37, 0.37] & .989 \\
\bottomrule
\end{tabular}
\end{table}

%% file: sci_fund_real/tables/balance.tex
\begin{table}
\caption{Study 1 (science funding) participant descriptives by arm: pre-treatment covariates (M (SD) for continuous; \% for binary). One-way ANOVA $F$-tests for continuous variables; $\chi^2$ for binary. Party ID (Democrat / Independent / Republican) is a $\chi^2$ test on participants with Prolific party data, missing for 5 participants. $q$ is the sharpened false-discovery-rate $q$ value, correcting for multiple comparisons across covariates.}
\label{tab:balance_science}
\begin{tabular}{llllll}
\toprule
Variable & Treat & Placebo & Vignette & Test & q \\
\midrule
Age & 45.62 (15.43) & 45.39 (15.36) & 45.70 (15.69) & F=0.04 & 1.000 \\
Male (\%) & 45.9\% & 47.2\% & 49.8\% & $\chi^2$=1.05 & 1.000 \\
Education (Recoded 1–8) & 3.15 (0.98) & 3.16 (1.14) & 3.25 (1.29) & F=0.81 & 1.000 \\
Income (Recoded 1–7) & 4.06 (1.90) & 4.22 (1.93) & 4.38 (1.82) & F=2.38 & 1.000 \\
Political Ideology (+ = conservative) & 3.63 (1.79) & 3.71 (1.75) & 3.83 (1.83) & F=1.06 & 1.000 \\
Familiarity w/ sci. funding & 2.87 (1.54) & 2.76 (1.49) & 2.82 (1.54) & F=0.42 & 1.000 \\
Mental imagery & 8.69 (1.25) & 8.56 (1.42) & 8.60 (1.40) & F=0.77 & 1.000 \\
IRI Fantasy & 3.69 (0.95) & 3.71 (0.93) & 3.69 (0.90) & F=0.10 & 1.000 \\
IRI Perspective-taking & 3.90 (0.83) & 3.86 (0.87) & 3.90 (0.82) & F=0.30 & 1.000 \\
IRI Personal distress & 2.59 (0.92) & 2.57 (0.97) & 2.50 (0.94) & F=0.89 & 1.000 \\
Creative writing exp. & 3.08 (1.10) & 3.12 (1.08) & 2.99 (1.16) & F=1.07 & 1.000 \\
Journaling freq. & 2.88 (1.14) & 2.81 (1.13) & 2.80 (1.18) & F=0.42 & 1.000 \\
AI attitudes & 6.51 (2.38) & 6.80 (2.15) & 6.62 (2.29) & F=1.47 & 1.000 \\
Party ID: D / I / R (\%) & 27.5 / 46.2 / 26.3 & 27.9 / 44.4 / 27.6 & 27.8 / 43.8 / 28.4 & $\chi^2$=0.52 & 1.000 \\
\bottomrule
\end{tabular}
\end{table}

%% file: satellite/tables/balance.tex
\begin{table}
\caption{Study 2 (satellite cybersecurity) participant descriptives by arm: pre-treatment covariates (M (SD) for continuous; \% for binary). One-way ANOVA $F$-tests for continuous variables; $\chi^2$ for binary. Party ID (Democrat / Independent / Republican) is a $\chi^2$ test on participants with Prolific party data, missing for 6 participants. $q$ is the sharpened false-discovery-rate $q$ value, correcting for multiple comparisons across covariates.}
\label{tab:balance_satellite}
\begin{tabular}{llllll}
\toprule
Variable & Treat & Placebo & Vignette & Test & q \\
\midrule
Age & 45.81 (15.23) & 45.61 (14.65) & 47.29 (14.36) & F=1.17 & 0.329 \\
Male (\%) & 40.9\% & 46.8\% & 49.2\% & $\chi^2$=4.33 & 0.221 \\
Education (Recoded 1–8) & 3.27 (1.13) & 3.22 (1.10) & 3.18 (1.04) & F=0.48 & 0.567 \\
Income (Recoded 1–7) & 4.29 (1.89) & 4.10 (1.81) & 3.87 (1.86) & F=3.78 & 0.133 \\
Political Ideology (+ = conservative) & 3.56 (1.72) & 3.83 (1.76) & 3.82 (1.85) & F=2.21 & 0.221 \\
Familiarity w/ satellite cybersecurity & 2.72 (1.60) & 2.80 (1.51) & 3.05 (1.55) & F=3.72 & 0.133 \\
Mental imagery & 8.58 (1.42) & 8.57 (1.49) & 8.65 (1.39) & F=0.26 & 0.712 \\
IRI Fantasy & 3.80 (0.87) & 3.69 (0.86) & 3.73 (0.94) & F=1.15 & 0.329 \\
IRI Perspective-taking & 3.86 (0.87) & 3.92 (0.83) & 3.95 (0.86) & F=0.87 & 0.416 \\
IRI Personal distress & 2.61 (0.96) & 2.44 (0.93) & 2.53 (0.95) & F=2.40 & 0.221 \\
Creative writing exp. & 3.37 (1.06) & 3.12 (1.05) & 3.29 (1.07) & F=4.02 & 0.133 \\
Journaling freq. & 3.08 (1.13) & 2.89 (1.15) & 2.91 (1.16) & F=2.51 & 0.221 \\
AI attitudes & 6.69 (2.30) & 6.76 (2.25) & 6.73 (2.29) & F=0.08 & 0.856 \\
Party ID: D / I / R (\%) & 31.4 / 44.3 / 24.3 & 26.1 / 44.0 / 29.9 & 27.2 / 43.5 / 29.2 & $\chi^2$=3.62 & 0.416 \\
\bottomrule
\end{tabular}
\end{table}

%% file: explain_divergence/tables/conversion.tex
\begin{table}[H]
\caption{Conversion of mediators into attitudes. Each slope is the change in attitudes (SD) per SD of the mediator, by study. Study 2 minus Study 1 rows report the study x mediator interaction from models pooling both studies within each sample. HC1 intervals.}
\label{tab:conversion}
\begin{tabular}{lllllr}
\toprule
Sample & Study & Mediator & Slope [95\% CI] & p & n \\
\midrule
All participants & Study 1 & Closeness & 0.41 [0.36, 0.47] & <.001 & 1007 \\
All participants & Study 2 & Closeness & 0.62 [0.57, 0.68] & <.001 & 894 \\
All participants & Study 2 minus Study 1 & Closeness & 0.21 [0.13, 0.29] & <.001 & 1901 \\
AIM Treatment & Study 1 & Closeness & 0.38 [0.22, 0.54] & <.001 & 333 \\
AIM Treatment & Study 2 & Closeness & 0.65 [0.51, 0.78] & <.001 & 298 \\
AIM Treatment & Study 2 minus Study 1 & Closeness & 0.27 [0.06, 0.47] & .011 & 631 \\
Placebo & Study 1 & Closeness & 0.42 [0.34, 0.50] & <.001 & 343 \\
Placebo & Study 2 & Closeness & 0.59 [0.50, 0.67] & <.001 & 293 \\
Placebo & Study 2 minus Study 1 & Closeness & 0.16 [0.05, 0.28] & .007 & 636 \\
Vignette & Study 1 & Closeness & 0.54 [0.44, 0.64] & <.001 & 331 \\
Vignette & Study 2 & Closeness & 0.61 [0.50, 0.73] & <.001 & 303 \\
Vignette & Study 2 minus Study 1 & Closeness & 0.07 [-0.08, 0.23] & .341 & 634 \\
All controls & Study 1 & Closeness & 0.44 [0.38, 0.51] & <.001 & 674 \\
All controls & Study 2 & Closeness & 0.57 [0.51, 0.64] & <.001 & 596 \\
All controls & Study 2 minus Study 1 & Closeness & 0.13 [0.04, 0.22] & .005 & 1270 \\
\bottomrule
\end{tabular}
\end{table}

%% file: explain_divergence/tables/ideology_r2.tex
\begin{table}[H]
\caption{Variance in attitudes explained by ideology: the R$^2$ of attitudes on ideological direction, strength, and their interaction, among control participants and among all participants, by study. Difference is Study 1 minus Study 2 in percentage points. Brackets give 95\% percentile intervals from 5,000 bootstrap draws, resampling participants within arm.}
\label{tab:ideology_r2}
\begin{tabular}{lllll}
\toprule
Sample & Study 1 R$^2$ (\%) & Study 2 R$^2$ (\%) & Difference (points) & n (Study 1, Study 2) \\
\midrule
Both controls & 3.9 [2.0, 7.7] & 0.7 [0.1, 3.2] & 3.3 [0.2, 6.6] & 674, 596 \\
All arms & 3.3 [1.8, 5.9] & 0.4 [0.1, 2.2] & 2.9 [0.5, 5.2] & 1007, 894 \\
\bottomrule
\end{tabular}
\end{table}

%% file: explain_divergence/tables/crystallization_types.tex
\begin{table}[H]
\caption{Share of participants whose open-ended responses showed a crystallized attitude, by type, with two-proportion z-tests comparing the studies. Labels from gpt-4.1-mini, using the prompt above.}
\label{tab:crystallization_types}
\begin{tabular}{lllllll}
\toprule
Sample & Label & Study 1 (\%) & Study 2 (\%) & n (Study 1, Study 2) & z & p \\
\midrule
AIM and vignette & Crystallized (either type) & 13.7 & 6.5 & 664, 600 & 4.21 & <.001 \\
AIM and vignette & Type 1: prior familiarity & 13.1 & 6.2 & 664, 600 & 4.14 & <.001 \\
AIM and vignette & Type 2: reinforced & 2.0 & 0.7 & 664, 600 & 1.99 & .047 \\
Vignette only & Crystallized (either type) & 14.5 & 6.3 & 331, 302 & 3.35 & <.001 \\
Vignette only & Type 1: prior familiarity & 13.9 & 6.0 & 331, 302 & 3.31 & <.001 \\
Vignette only & Type 2: reinforced & 1.8 & 1.0 & 331, 302 & 0.87 & .384 \\
AIM only & Crystallized (either type) & 12.9 & 6.7 & 333, 298 & 2.59 & .009 \\
AIM only & Type 1: prior familiarity & 12.3 & 6.4 & 333, 298 & 2.54 & .011 \\
AIM only & Type 2: reinforced & 2.1 & 0.3 & 333, 298 & 1.98 & .048 \\
\bottomrule
\end{tabular}
\end{table}

%% file: explain_divergence/tables/annotator_agreement.tex
\begin{table}[H]
\caption{Inter-annotator agreement. Two human raters independently coded 200 open-ended responses as crystallized or not, and the LLM coded the same responses with a codebook fixed before human coding began. A response counts as crystallized if it shows prior familiarity or reinforcement, judged separately as subtypes. The first 100 responses were drawn at random; the remaining 100 combine random responses with responses the LLM coded as crystallized, shuffled together, to ensure enough positive cases. The table reports agreement between the two human raters and between the LLM and each human rater, on all responses and on the random subset alone. $\kappa$ is Cohen's kappa with a 95\% percentile bootstrap interval (2,000 resamples of responses). Positives gives the number of responses each rater coded 1, in the order the raters are listed.}
\label{tab:annotator_agreement}
\begin{tabular}{lllllrl}
\toprule
Sample & Raters & $\kappa$ & 95\% CI & Agreement (\%) & Disagreements & Positives \\
\midrule
All ($N=200$) & Rater A vs Rater B & 0.83 & [0.74, 0.90] & 92.5 & 15 & 63 / 62 \\
All ($N=200$) & LLM vs Rater A & 0.91 & [0.83, 0.96] & 96.0 & 8 & 61 / 63 \\
All ($N=200$) & LLM vs Rater B & 0.80 & [0.71, 0.89] & 91.5 & 17 & 61 / 62 \\
Random ($N=100$) & Rater A vs Rater B & 0.80 & [0.56, 0.96] & 96.0 & 4 & 9 / 13 \\
Random ($N=100$) & LLM vs Rater A & 0.71 & [0.39, 0.93] & 95.0 & 5 & 10 / 9 \\
Random ($N=100$) & LLM vs Rater B & 0.66 & [0.39, 0.86] & 93.0 & 7 & 10 / 13 \\
\bottomrule
\end{tabular}
\end{table}

%% file: explain_divergence/tables/ruled_out_ux.tex
\begin{table}[H]
\caption{UX changes. Each row is one system rating, on its administered scale, among placebo participants only. The placebo task was identical in the two studies, so these differences isolate the interface change from the change of scenario. Estimates are Study 2 minus Study 1 with $95\%$ HC1 confidence intervals, $p$ values, and sharpened FDR $q$-values across the six ratings. n = 636 placebo participants.}
\label{tab:ruled_out_ux}
\begin{tabular}{llll}
\toprule
Rating & Study 2 $-$ Study 1 [95\% CI] & $p$ & $q$ \\
\midrule
Fun & -0.20 [-0.45, 0.05] & .111 & .384 \\
Tailored & -0.15 [-0.40, 0.11] & .257 & .522 \\
Realistic & -0.30 [-0.57, -0.04] & .024 & .166 \\
Helped & -0.08 [-0.33, 0.18] & .556 & .589 \\
Easy & 0.06 [-0.07, 0.20] & .369 & .536 \\
Agency & 0.09 [-0.14, 0.32] & .436 & .536 \\
\bottomrule
\end{tabular}
\end{table}

%% file: explain_divergence/tables/ruled_out_composition.tex
\begin{table}[H]
\caption{Composition effects and study restrictions. Each row models attitudes on arm, study, and their interactions, with the AIM arm omitted. Composition adds 13 pretreatment covariates; Restrictions adds Safari use (Study 2 excluded Safari users) and the number of words written, measured in hundreds. Gap vs vignette is Study 2's AIM effect against the vignette minus Study 1's corresponding effect; gap vs placebo is defined analogously. Brackets give 95\% confidence intervals (HC1). If participant composition or the study restrictions explained the difference between studies, these gaps would shrink toward zero after adjustment.}
\label{tab:ruled_out_composition}
\begin{tabular}{lll}
\toprule
Model & Gap vs vignette & Gap vs placebo \\
\midrule
Simple & 0.40 [0.18, 0.62] & 0.57 [0.36, 0.79] \\
Composition & 0.44 [0.24, 0.65] & 0.55 [0.35, 0.75] \\
Restrictions & 0.40 [0.18, 0.62] & 0.58 [0.36, 0.79] \\
Composition + restrictions & 0.44 [0.24, 0.65] & 0.55 [0.35, 0.75] \\
\bottomrule
\end{tabular}
\end{table}

%% file: explain_divergence/tables/ruled_out_similarity.tex
\begin{table}[H]
\caption{Tighter connection to the DVs. Similarity is the mean cosine similarity between an AIM participant's story and the mean embedding of that study's attitude items (SBERT all-mpnet-base-v2). Slope is the change in attitudes (SD) per SD of similarity among AIM participants, with a $95\%$ confidence interval (HC1) and $p$ value. Study 1 stories were the more similar of the two to their own attitude items (Welch $t(591)=3.04$, $p=.002$), the opposite of what an explanation running through item overlap would require.}
\label{tab:ruled_out_similarity}
\begin{tabular}{llllr}
\toprule
Study & Mean similarity & Slope [95\% CI] & $p$ & $n$ \\
\midrule
Study 1 & 0.150 & -0.08 [-0.20, 0.04] & .176 & 330 \\
Study 2 & 0.127 & 0.09 [-0.01, 0.19] & .079 & 298 \\
\bottomrule
\end{tabular}
\end{table}

%% file: explain_divergence/tables/ruled_out_ceiling.tex
\begin{table}[H]
\caption{Ceiling effects on the attitude items. Each item is z-scored within its study, and effects are AIM minus placebo in item standard deviations. The two-sided Tobit treats values as censored and estimates the effect on the latent scale, rescaled by the observed standard deviation so OLS and Tobit are comparable. comparable. Two columns show percentage of each condition's participants at the maximum item.}
\label{tab:ruled_out_ceiling}
\begin{tabular}{lllll}
\toprule
Study & Item & \shortstack{At scale max\\placebo $\to$ AIM} & OLS [95\% CI] & Tobit [95\% CI] \\
\midrule
Study 1 & Preferred NIH budget share & 0.3\% $\to$ 0.6\% & 0.12 [-0.03, 0.27] & 0.13 [-0.03, 0.29] \\
Study 1 & Science benefits people like me & 27.4\% $\to$ 27.6\% & 0.02 [-0.13, 0.17] & 0.02 [-0.19, 0.23] \\
Study 2 & Personal importance & 13.7\% $\to$ 24.8\% & 0.62 [0.46, 0.77] & 0.72 [0.54, 0.90] \\
Study 2 & Importance vs other issues & 6.5\% $\to$ 14.8\% & 0.53 [0.38, 0.69] & 0.59 [0.42, 0.76] \\
Study 2 & Spending: reduce risk & 3.8\% $\to$ 10.7\% & 0.49 [0.34, 0.65] & 0.53 [0.36, 0.70] \\
Study 2 & Spending: priority & 2.7\% $\to$ 12.8\% & 0.59 [0.44, 0.74] & 0.65 [0.47, 0.82] \\
Study 2 & Action: take action & 5.8\% $\to$ 15.8\% & 0.55 [0.40, 0.71] & 0.62 [0.45, 0.79] \\
Study 2 & Action: support bill & 8.5\% $\to$ 19.1\% & 0.50 [0.35, 0.66] & 0.58 [0.40, 0.76] \\
Study 2 & Action: oppose cuts & 5.8\% $\to$ 16.8\% & 0.48 [0.32, 0.64] & 0.55 [0.37, 0.74] \\
\bottomrule
\end{tabular}
\end{table}

%% file: explain_divergence/tables/ruled_out_vignette.tex
\begin{table}[H]
\caption{Whether greater emotional engagement in Study 2 can account for its larger AIM attitude effect. Path a is AIM's effect on emotional engagement relative to the vignette. Path b is the association between emotional engagement and attitudes, adjusting for treatment and closeness. Path c is AIM's total effect on attitudes relative to the vignette. The predicted additional attitude effect is (a2-a1)b1, and the share accounted for is 100(a2-a1)b1/(c2-c1).}
\label{tab:ruled_out_vignette}
\begin{tabular}{lll}
\toprule
Quantity & Study 1 & Study 2 \\
\midrule
Vignette emotion & 4.57 & 3.65 \\
AIM emotion & 4.02 & 4.13 \\
a: AIM effect on emotion & -0.55 & 0.48 \\
b: Emotion -> attitudes & 0.073 & 0.136 \\
c: AIM effect on attitudes & 0.16 & 0.54 \\
Extra AIM effect on emotion in Study 2 & 1.024 &  \\
Predicted extra attitude effect from emotion & 0.074 &  \\
Observed attitude-effect gap & 0.384 &  \\
Gap accounted for by emotion (\%) & 19.4 &  \\
\bottomrule
\end{tabular}
\end{table}

%% file: study_comparison/tables/ai_attitude.tex
\begin{table}[H]
\caption{AIM's effects by prior attitude toward AI, split at the midpoint of the 1--10 AI-attitude scale. Estimates are AIM-minus-control differences in outcome SDs with HC1 confidence intervals. Interaction is the arm by AI-attitude term, in SDs per SD of AI attitude.}
\label{tab:ai_attitude}
\begin{tabular}{lllcccc}
\toprule
Study & Outcome & vs. & \shortstack{Unfavourable\\$b$ [95\% CI]} & \shortstack{Favourable\\$b$ [95\% CI]} & \shortstack{Interaction\\$b$ [95\% CI]} & $p$ \\
\midrule
Study 1 & Closeness & Vignette & +0.53 [+0.24, +0.81] & +0.30 [+0.15, +0.45] & -0.08 [-0.22, +0.05] & .229 \\
Study 1 & Closeness & Placebo & +0.76 [+0.44, +1.07] & +0.81 [+0.65, +0.97] & +0.05 [-0.10, +0.21] & .494 \\
Study 1 & Attitudes & Vignette & +0.25 [-0.02, +0.51] & +0.13 [-0.06, +0.32] & -0.10 [-0.26, +0.06] & .209 \\
Study 1 & Attitudes & Placebo & +0.03 [-0.24, +0.31] & +0.12 [-0.05, +0.29] & -0.02 [-0.17, +0.13] & .785 \\
Study 2 & Closeness & Vignette & +0.41 [+0.13, +0.70] & +0.36 [+0.21, +0.52] & -0.05 [-0.19, +0.09] & .475 \\
Study 2 & Closeness & Placebo & +0.99 [+0.67, +1.31] & +0.86 [+0.69, +1.03] & -0.14 [-0.30, +0.02] & .093 \\
Study 2 & Attitudes & Vignette & +0.42 [+0.11, +0.73] & +0.58 [+0.41, +0.75] & -0.01 [-0.15, +0.14] & .945 \\
Study 2 & Attitudes & Placebo & +0.76 [+0.45, +1.06] & +0.62 [+0.46, +0.79] & -0.15 [-0.30, +0.00] & .053 \\
\bottomrule
\end{tabular}
\end{table}

%% file: study_comparison/tables/tab_mediation_direction_tests.tex
\begin{longtable}{lllrlll}
\caption{Treatment coefficients before and after adjusting for the proposed upstream mediator. Coefficients are in SD units with 95\% HC1 confidence intervals.} \label{tab:mediation_direction_tests} \\
\toprule
Study & Control & Proposed chain & $n$ & Unadjusted treatment coefficient & Adjusted treatment coefficient & Tolerated \\
\midrule
\endfirsthead
\caption[]{Treatment coefficients before and after adjusting for the proposed upstream mediator. Coefficients are in SD units with 95\% HC1 confidence intervals.} \\
\toprule
Study & Control & Proposed chain & $n$ & Unadjusted treatment coefficient & Adjusted treatment coefficient & Tolerated \\
\midrule
\endhead
\midrule
\multicolumn{7}{r}{Continued on next page} \\
\midrule
\endfoot
\bottomrule
\endlastfoot
Pooled & Vignette & AIM $\rightarrow$ transportation $\rightarrow$ closeness & 1265 & +0.37 [+0.27, +0.46] & +0.26 [+0.18, +0.35] & No \\
Pooled & Vignette & AIM $\rightarrow$ closeness $\rightarrow$ transportation & 1265 & +0.23 [+0.12, +0.35] & +0.01 [-0.09, +0.11] & Yes \\
Study 1 & Vignette & AIM $\rightarrow$ transportation $\rightarrow$ closeness & 664 & +0.36 [+0.22, +0.49] & +0.31 [+0.20, +0.43] & No \\
Study 1 & Vignette & AIM $\rightarrow$ closeness $\rightarrow$ transportation & 664 & +0.10 [-0.05, +0.26] & -0.10 [-0.24, +0.04] & Yes \\
Study 2 & Vignette & AIM $\rightarrow$ transportation $\rightarrow$ closeness & 601 & +0.38 [+0.24, +0.51] & +0.21 [+0.09, +0.32] & No \\
Study 2 & Vignette & AIM $\rightarrow$ closeness $\rightarrow$ transportation & 601 & +0.38 [+0.22, +0.54] & +0.13 [-0.01, +0.28] & Yes \\
\end{longtable}

%% file: sci_fund_real/tables/tab_qualitative_themes.tex
\begin{table}[htbp]
\centering
\small
\setlength{\tabcolsep}{5pt}
\begin{tabularx}{\textwidth}{l >{\raggedright\arraybackslash}p{3.8cm} r X}
\toprule
\textbf{Type} & \textbf{Theme} & \textbf{\%} & \textbf{Representative quote} \\
\midrule
Success Mode & Made issue more personal & 30.1\% & {\footnotesize\textit{Having to contemplate a real-life scenario for myself made it personal and, thus, more effective.}} \\[1pt]
Success Mode & Scenario specifics were effective & 29.0\% & {\footnotesize\textit{Simply the idea of having to wait 1 year + for an appointment}} \\[1pt]
Success Mode & The experience caused reflection & 16.6\% & {\footnotesize\textit{Thinking about what could happen and how it could affect my life was effective.}} \\[1pt]
Success Mode & The activity made the issue feel real & 9.3\% & {\footnotesize\textit{Writing the story showed me how funding cuts can directly affect people's health, which made the issue feel more real.}} \\[1pt]
Success Mode & Writing was emotionally charged & 8.3\% & {\footnotesize\textit{I felt that it was totally effecting [sic] my own life in that I was suffering and unable to find a solution, so that everyday went by while I seemed to be on hold.}} \\[1pt]
Success Mode & AI predictive ability & 6.7\% & {\footnotesize\textit{The responses after I typed a couple of words were effective and captured what I was trying to say.}} \\[1pt]
\addlinespace[3pt]\midrule\addlinespace[3pt]
Failure Mode & AI added things the user didn't intend & 34.6\% & {\footnotesize\textit{[S]ome of the writing suggestions were not where I was trying to go with the story.}} \\[1pt]
Failure Mode & Wait time was unrealistic & 21.8\% & {\footnotesize\textit{The wait time seemed exceptionally long in relation with the 'problems' I was having in the story.}} \\[1pt]
Failure Mode & AI was too scenario-driven & 21.1\% & {\footnotesize\textit{The suggestions of the AI were narrowly focused on just feelings of the original prompt rather than my own inclination which was to find a way to resolve the problem.}} \\[1pt]
Failure Mode & Story became self-focused & 12.8\% & {\footnotesize\textit{Once I the story began developing, I was more focused on science funding in my own life than the overall funding of science.}} \\[1pt]
Failure Mode & Felt hypothetical & 9.8\% & {\footnotesize\textit{It still felt somewhat hypothetical since I wasn't actually experiencing the situation.}} \\[1pt]
\bottomrule
\end{tabularx}
\caption{Qualitative themes in Study 1 (science funding) open-text responses from the AIM arm. Participants described what was effective and what was ineffective about the co-writing experience. We clustered responses with BERTopic on INSTRUCTOR embeddings and excluded clusters that restated the question or gave no substantive answer (e.g., ``nothing was ineffective''). Percentages use the 193 substantive responses about what was effective and the 133 about what was ineffective.}
\label{tab:qualitative_themes}
\end{table}

%% file: satellite/tables/tab_qualitative_themes_satellite.tex
\begin{table}[htbp]
\centering
\small
\setlength{\tabcolsep}{5pt}
\begin{tabularx}{\textwidth}{l >{\raggedright\arraybackslash}p{3.8cm} r X}
\toprule
\textbf{Type} & \textbf{Theme} & \textbf{\%} & \textbf{Representative quote} \\
\midrule
Success Mode & Realizing the scale of disruption & 31.4\% & {\footnotesize\textit{It made me realize how big a threat this is to my everyday life. My job, communication, and navigating abilities would be severely disrupted.}} \\[1pt]
Success Mode & Rehearsing what I would do & 23.1\% & {\footnotesize\textit{It made me feel like I was in the situation and forced me to think about what I would actually do.}} \\[1pt]
Success Mode & Everyday dependencies made it concrete & 13.2\% & {\footnotesize\textit{Writing out a first person scenario grounded the abstract idea of satellite systems into immediate, everyday consequences like failed card payment, loss of phone service and bank outages.}} \\[1pt]
Success Mode & Topic had never been considered & 10.7\% & {\footnotesize\textit{I haven't really pondered cybersecurity satellites in the past so it was a bit eye opening.}} \\[1pt]
Success Mode & Recognizing personal reliance & 10.7\% & {\footnotesize\textit{Making me more aware of what I use in my life that would be effected by this. I didn't think I did, but as I started navigating a night in my life, I realized that a lot of what I use works on these systems.}} \\[1pt]
Success Mode & Surfaced an unnoticed vulnerability---not having cash on hand & 10.7\% & {\footnotesize\textit{Making it very obvious that my habit of never having cash on hand probably isn't a good idea.}} \\[1pt]
\addlinespace[3pt]\midrule\addlinespace[3pt]
Failure Mode & Scenario felt implausible & 37.2\% & {\footnotesize\textit{The story was a bit too doomsday to be believable. Things would be bad, but it wouldn't fall apart within hours.}} \\[1pt]
Failure Mode & AI suggestions missed the mark & 30.8\% & {\footnotesize\textit{The AI suggestions occasionally broke my train of thought while trying to write.}} \\[1pt]
Failure Mode & Story omitted the causal mechanism & 16.7\% & {\footnotesize\textit{The story focused on personal difficulties but did not explain how satellite cybersecurity works or how attacks could be prevented.}} \\[1pt]
Failure Mode & AI took over story & 15.4\% & {\footnotesize\textit{The stuff that the AI kept trying to add made it feel just like some general story and not my story. It made me feel disconnected and a little annoyed, if anything.}} \\[1pt]
\bottomrule
\end{tabularx}
\caption{Qualitative themes in Study 2 (satellite cybersecurity) open-text responses from the AIM arm. Participants described what was effective and what was ineffective about the co-writing experience. We clustered responses with BERTopic on INSTRUCTOR embeddings and excluded clusters that restated the question or gave no substantive answer (e.g., ``nothing was ineffective''). Percentages use the 121 substantive responses about what was effective and the 78 about what was ineffective.}
\label{tab:qualitative_themes_satellite}
\end{table}

%% file: study_comparison/tables/linguistic_classifier.tex
\begin{table}
\caption{Human--AI text classification with all 109 linguistic features (Biber and Empath). Mean AUC, with its SD across five participant-grouped folds.}
\label{tab:linguistic-classifier}
\begin{tabular}{lrl}
\toprule
Study & Pairs & Mean AUC (SD) \\
\midrule
Science funding & 285 & 0.914 (0.026) \\
Satellite & 259 & 0.940 (0.028) \\
\bottomrule
\end{tabular}
\end{table}

%% file: satellite/tables/satellite_cfa.tex
\begin{table}
\caption{Satellite attitude measurement models. Confirmatory factor analysis of the seven attitude items ($n=894$).}
\label{tab:satellite-cfa}
\begin{tabular}{lrrrrrrr}
\toprule
Model & $n$ & $\chi^2$ & df & CFI & TLI & RMSEA & $p$ \\
\midrule
Three factors & 894 & 122.584 & 11 & 0.973 & 0.948 & 0.107 & $5.47\times10^{-21}$ \\
One factor & 894 & 932.935 & 14 & 0.775 & 0.662 & 0.271 & $3.77\times10^{-190}$ \\
\bottomrule
\end{tabular}
\end{table}

%% file: study_comparison/tables/grf_sci_fund_real_placebo.tex
\begin{table}
\footnotesize
\setlength{\tabcolsep}{3pt}
\caption{Study 1: Science funding, AIM versus placebo: causal-forest best linear projections ($n=644$). Gender uses female as the reference. $q$ is the sharpened false-discovery-rate q-value, with the moderator coefficients of each outcome as one family.}
\label{tab:grf-sci_fund_real-placebo}
\begin{tabular}{>{\raggedright\arraybackslash}p{3.0cm}rrrrrr}
\toprule
Moderator & Closeness $B$ [95\% CI] & Closeness $p$ & Closeness $q$ & Attitudes $B$ [95\% CI] & Attitudes $p$ & Attitudes $q$ \\
\midrule
Intercept & 0.90 [0.71, 1.09] & $7.2\times10^{-20}$ &  & 0.17 [-0.05, 0.39] & $.123$ &  \\
Familiarity & -0.36 [-0.53, -0.20] & $1.4\times10^{-5}$ & $\leq.001$ & -0.17 [-0.35, 0.01] & $.0611$ & $.913$ \\
Age & -0.21 [-0.37, -0.05] & $.00967$ & $.084$ & -0.18 [-0.35, -0.01] & $.039$ & $.913$ \\
Mental imagery & 0.11 [-0.07, 0.28] & $.229$ & $.684$ & 0.13 [-0.02, 0.28] & $.0842$ & $.913$ \\
IRI fantasy & 0.08 [-0.09, 0.26] & $.346$ & $.687$ & -0.09 [-0.26, 0.09] & $.322$ & $.913$ \\
IRI perspective & -0.15 [-0.33, 0.03] & $.109$ & $.444$ & -0.09 [-0.25, 0.07] & $.253$ & $.913$ \\
IRI distress & -0.18 [-0.34, -0.03] & $.018$ & $.099$ & -0.09 [-0.25, 0.07] & $.262$ & $.913$ \\
Journaling frequency & -0.04 [-0.22, 0.14] & $.65$ & $1.000$ & -0.07 [-0.25, 0.10] & $.43$ & $.979$ \\
Writing frequency & -0.08 [-0.26, 0.10] & $.403$ & $.687$ & 0.11 [-0.06, 0.28] & $.219$ & $.913$ \\
Education & 0.02 [-0.12, 0.15] & $.823$ & $1.000$ & 0.10 [-0.06, 0.26] & $.21$ & $.913$ \\
Income & -0.08 [-0.22, 0.07] & $.293$ & $.684$ & -0.04 [-0.20, 0.11] & $.594$ & $.979$ \\
Ideology & 0.13 [-0.03, 0.28] & $.11$ & $.444$ & 0.04 [-0.13, 0.21] & $.63$ & $.979$ \\
AI attitudes & 0.09 [-0.08, 0.26] & $.297$ & $.684$ & 0.00 [-0.16, 0.17] & $.983$ & $.979$ \\
Completions received & -0.09 [-0.30, 0.12] & $.407$ & $.687$ & -0.14 [-0.36, 0.07] & $.184$ & $.913$ \\
Story length & 0.13 [-0.12, 0.37] & $.305$ & $.684$ & 0.04 [-0.19, 0.26] & $.746$ & $.979$ \\
Acceptance rate & -0.04 [-0.27, 0.19] & $.733$ & $1.000$ & -0.01 [-0.29, 0.27] & $.94$ & $.979$ \\
Proportion AI words & 0.01 [-0.19, 0.20] & $.946$ & $1.000$ & 0.05 [-0.19, 0.28] & $.707$ & $.979$ \\
Male & -0.11 [-0.42, 0.20] & $.493$ & $.837$ & -0.20 [-0.53, 0.14] & $.244$ & $.913$ \\
\bottomrule
\end{tabular}
\end{table}

%% file: study_comparison/tables/grf_sci_fund_real_vignette.tex
\begin{table}
\footnotesize
\setlength{\tabcolsep}{3pt}
\caption{Study 1: Science funding, AIM versus vignette: causal-forest best linear projections ($n=638$). Gender uses female as the reference. $q$ is the sharpened false-discovery-rate q-value, with the moderator coefficients of each outcome as one family.}
\label{tab:grf-sci_fund_real-vignette}
\begin{tabular}{>{\raggedright\arraybackslash}p{3.0cm}rrrrrr}
\toprule
Moderator & Closeness $B$ [95\% CI] & Closeness $p$ & Closeness $q$ & Attitudes $B$ [95\% CI] & Attitudes $p$ & Attitudes $q$ \\
\midrule
Intercept & 0.45 [0.26, 0.64] & $2.7\times10^{-6}$ &  & 0.04 [-0.21, 0.28] & $.777$ &  \\
Familiarity & -0.09 [-0.23, 0.05] & $.195$ & $.451$ & -0.10 [-0.29, 0.08] & $.279$ & $.674$ \\
Age & -0.11 [-0.25, 0.02] & $.0956$ & $.451$ & -0.14 [-0.31, 0.03] & $.109$ & $.674$ \\
Mental imagery & 0.08 [-0.09, 0.24] & $.369$ & $.528$ & 0.17 [0.01, 0.33] & $.0425$ & $.674$ \\
IRI fantasy & 0.03 [-0.14, 0.21] & $.713$ & $.528$ & -0.13 [-0.32, 0.06] & $.19$ & $.674$ \\
IRI perspective & -0.03 [-0.19, 0.13] & $.746$ & $.528$ & -0.00 [-0.19, 0.18] & $.961$ & $.674$ \\
IRI distress & -0.09 [-0.23, 0.05] & $.213$ & $.451$ & -0.05 [-0.23, 0.12] & $.563$ & $.674$ \\
Journaling frequency & -0.14 [-0.30, 0.02] & $.0931$ & $.451$ & 0.01 [-0.17, 0.19] & $.876$ & $.674$ \\
Writing frequency & -0.11 [-0.28, 0.05] & $.171$ & $.451$ & 0.03 [-0.16, 0.22] & $.774$ & $.674$ \\
Education & -0.02 [-0.15, 0.10] & $.707$ & $.528$ & 0.12 [-0.07, 0.30] & $.213$ & $.674$ \\
Income & -0.15 [-0.28, -0.01] & $.0302$ & $.451$ & -0.12 [-0.29, 0.05] & $.158$ & $.674$ \\
Ideology & 0.12 [-0.02, 0.27] & $.0917$ & $.451$ & 0.13 [-0.05, 0.31] & $.15$ & $.674$ \\
AI attitudes & -0.03 [-0.19, 0.12] & $.671$ & $.528$ & -0.10 [-0.28, 0.07] & $.252$ & $.674$ \\
Male & -0.18 [-0.47, 0.10] & $.202$ & $.451$ & 0.22 [-0.13, 0.57] & $.222$ & $.674$ \\
\bottomrule
\end{tabular}
\end{table}

%% file: study_comparison/tables/grf_satellite_placebo.tex
\begin{table}
\footnotesize
\setlength{\tabcolsep}{3pt}
\caption{Study 2: Satellite cybersecurity, AIM versus placebo: causal-forest best linear projections ($n=564$). Gender uses female as the reference. $q$ is the sharpened false-discovery-rate q-value, with the moderator coefficients of each outcome as one family.}
\label{tab:grf-satellite-placebo}
\begin{tabular}{>{\raggedright\arraybackslash}p{3.0cm}rrrrrr}
\toprule
Moderator & Closeness $B$ [95\% CI] & Closeness $p$ & Closeness $q$ & Attitudes $B$ [95\% CI] & Attitudes $p$ & Attitudes $q$ \\
\midrule
Intercept & 0.96 [0.76, 1.16] & $8.64\times10^{-19}$ &  & 0.66 [0.46, 0.85] & $1.25\times10^{-10}$ &  \\
Familiarity & -0.27 [-0.44, -0.11] & $.00116$ & $.021$ & -0.11 [-0.27, 0.05] & $.172$ & $1.000$ \\
Age & -0.21 [-0.38, -0.04] & $.0147$ & $.134$ & -0.14 [-0.30, 0.03] & $.111$ & $1.000$ \\
Mental imagery & -0.00 [-0.17, 0.16] & $.973$ & $1.000$ & 0.02 [-0.14, 0.18] & $.837$ & $1.000$ \\
IRI fantasy & 0.09 [-0.12, 0.29] & $.4$ & $1.000$ & 0.12 [-0.06, 0.30] & $.183$ & $1.000$ \\
IRI perspective & 0.03 [-0.16, 0.23] & $.731$ & $1.000$ & 0.00 [-0.17, 0.18] & $.974$ & $1.000$ \\
IRI distress & 0.07 [-0.09, 0.24] & $.396$ & $1.000$ & 0.01 [-0.14, 0.16] & $.907$ & $1.000$ \\
Journaling frequency & -0.06 [-0.23, 0.11] & $.494$ & $1.000$ & 0.06 [-0.12, 0.24] & $.532$ & $1.000$ \\
Writing frequency & 0.01 [-0.19, 0.20] & $.935$ & $1.000$ & -0.12 [-0.31, 0.08] & $.237$ & $1.000$ \\
Education & 0.00 [-0.16, 0.17] & $.983$ & $1.000$ & -0.00 [-0.17, 0.16] & $.979$ & $1.000$ \\
Income & 0.06 [-0.08, 0.20] & $.393$ & $1.000$ & -0.02 [-0.17, 0.12] & $.767$ & $1.000$ \\
Ideology & 0.03 [-0.13, 0.19] & $.722$ & $1.000$ & -0.06 [-0.21, 0.10] & $.462$ & $1.000$ \\
AI attitudes & -0.12 [-0.29, 0.06] & $.203$ & $1.000$ & -0.14 [-0.31, 0.04] & $.12$ & $1.000$ \\
Completions received & 0.03 [-0.19, 0.25] & $.764$ & $1.000$ & -0.05 [-0.26, 0.17] & $.663$ & $1.000$ \\
Story length & 0.02 [-0.24, 0.28] & $.887$ & $1.000$ & 0.06 [-0.21, 0.32] & $.666$ & $1.000$ \\
Acceptance rate & -0.00 [-0.26, 0.25] & $.993$ & $1.000$ & 0.15 [-0.08, 0.38] & $.214$ & $1.000$ \\
Proportion AI words & 0.01 [-0.23, 0.24] & $.956$ & $1.000$ & -0.13 [-0.35, 0.10] & $.278$ & $1.000$ \\
Male & -0.12 [-0.43, 0.20] & $.479$ & $1.000$ & 0.05 [-0.26, 0.36] & $.755$ & $1.000$ \\
\bottomrule
\end{tabular}
\end{table}

%% file: study_comparison/tables/grf_satellite_vignette.tex
\begin{table}
\footnotesize
\setlength{\tabcolsep}{3pt}
\caption{Study 2: Satellite cybersecurity, AIM versus vignette: causal-forest best linear projections ($n=579$). Gender uses female as the reference. $q$ is the sharpened false-discovery-rate q-value, with the moderator coefficients of each outcome as one family.}
\label{tab:grf-satellite-vignette}
\begin{tabular}{>{\raggedright\arraybackslash}p{3.0cm}rrrrrr}
\toprule
Moderator & Closeness $B$ [95\% CI] & Closeness $p$ & Closeness $q$ & Attitudes $B$ [95\% CI] & Attitudes $p$ & Attitudes $q$ \\
\midrule
Intercept & 0.53 [0.35, 0.72] & $3.93\times10^{-8}$ &  & 0.58 [0.37, 0.79] & $1.1\times10^{-7}$ &  \\
Familiarity & -0.07 [-0.21, 0.07] & $.33$ & $.881$ & -0.11 [-0.27, 0.05] & $.186$ & $1.000$ \\
Age & -0.10 [-0.25, 0.05] & $.179$ & $.881$ & -0.00 [-0.15, 0.15] & $.996$ & $1.000$ \\
Mental imagery & -0.12 [-0.28, 0.05] & $.175$ & $.881$ & -0.12 [-0.27, 0.03] & $.122$ & $1.000$ \\
IRI fantasy & 0.05 [-0.15, 0.24] & $.635$ & $1.000$ & 0.08 [-0.11, 0.27] & $.401$ & $1.000$ \\
IRI perspective & 0.08 [-0.09, 0.25] & $.38$ & $.881$ & -0.06 [-0.25, 0.12] & $.523$ & $1.000$ \\
IRI distress & 0.18 [0.04, 0.32] & $.0121$ & $.187$ & 0.06 [-0.10, 0.23] & $.431$ & $1.000$ \\
Journaling frequency & 0.02 [-0.14, 0.18] & $.785$ & $1.000$ & 0.04 [-0.14, 0.22] & $.639$ & $1.000$ \\
Writing frequency & 0.09 [-0.08, 0.27] & $.297$ & $.881$ & -0.04 [-0.23, 0.15] & $.694$ & $1.000$ \\
Education & 0.01 [-0.12, 0.15] & $.838$ & $1.000$ & -0.12 [-0.27, 0.03] & $.105$ & $1.000$ \\
Income & -0.05 [-0.17, 0.06] & $.37$ & $.881$ & -0.03 [-0.18, 0.12] & $.701$ & $1.000$ \\
Ideology & 0.07 [-0.07, 0.21] & $.318$ & $.881$ & 0.01 [-0.15, 0.17] & $.88$ & $1.000$ \\
AI attitudes & -0.04 [-0.19, 0.10] & $.572$ & $1.000$ & 0.04 [-0.12, 0.20] & $.629$ & $1.000$ \\
Male & -0.26 [-0.55, 0.02] & $.072$ & $.761$ & -0.03 [-0.35, 0.29] & $.872$ & $1.000$ \\
\bottomrule
\end{tabular}
\end{table}